\documentclass[12pt]{article}

\usepackage{epsfig,lscape}
\usepackage{amssymb,amsfonts,amsmath,amsthm}
\usepackage{rotating}
\usepackage{bbm}

\usepackage{threeparttable}
\usepackage{booktabs}
\usepackage{subcaption}
\usepackage{times}
\usepackage{xcolor}

\usepackage[left=1in,right=1in,top=.9in,bottom=1.1in]{geometry}

\def\me{\mathrm e}

\def\dif{\mathrm d}

\def\var{\mathrm{var}}

\def\N{\mathrm{N}}

\def\T{ {\mathrm{\scriptscriptstyle T}} }

\def\bbR{\mathbb R}

\def\one{\mathbbm{1}}

\newenvironment{prf}
{\noindent \textbf{Proof.}}{\hfill $\Box$ \vspace{.1in}}

\newtheorem{thm}{Theorem}
\newtheorem{lem}{Lemma}
\newtheorem{pro}{Proposition}

\newtheorem{ass}{Assumption}

\theoremstyle{definition}

\theoremstyle{definition}
\newtheorem{rem}{Remark}

\newcommand{\independent}{\coprod}

\begin{document}

\begin{titlepage}

\begin{center}
{\Large High-dimensional Model-assisted Inference for Local Average Treatment Effects with Instrumental Variables}

\vspace{.1in} {\large Baoluo Sun\footnotemark[1] \& Zhiqiang Tan\footnotemark[1]}

\vspace{.1in}
\today
\end{center}

\footnotetext[1]{Baoluo Sun is Assistant Professor, Department of Statistics and Applied Probability, National University of Singapore, Singapore, SG 117546
(Email: stasb@nus.edu.sg),
and Zhiqiang Tan is Professor, Department of Statistics, Rutgers University,
Piscataway, NJ 08854, USA (Email: ztan@stat.rutgers.edu). }

\paragraph{Abstract.}
Consider the problem of estimating the local average treatment effect with an instrument variable, where the instrument unconfoundedness holds after adjusting for a set of measured covariates.
Several unknown functions of the covariates need to be estimated through regression models, such as instrument propensity score and treatment and outcome regression models.
We develop a computationally tractable method in high-dimensional settings where the numbers of regression terms
are close to or larger than the sample size.
Our method exploits regularized calibrated estimation, which involves Lasso penalties but carefully chosen loss functions
 for estimating coefficient vectors in these regression models, and then employs a doubly robust estimator for the treatment parameter through augmented inverse probability weighting.
We provide rigorous theoretical analysis to show that the resulting Wald confidence intervals are valid for the treatment parameter under
suitable sparsity conditions if the instrument propensity score model is correctly specified, but the treatment and outcome regression models may be misspecified.
For existing high-dimensional methods, valid confidence intervals are obtained for the treatment parameter if all three models are correctly specified.
We evaluate the proposed methods via extensive simulation studies and an empirical application to estimate the returns to education.


\end{titlepage}

\section{Introduction}

A major difficulty in drawing causal inference from observational studies is the possible existence of unobserved background variables that are related to both the treatment status and outcome of interest,
which is usually referred to as unmeasured confounding.
In such settings, biased estimates of the causal effects may be obtained by comparing
observed outcomes between treated and untreated individuals, even after adjusting for measured covariates.
To tackle unmeasured confounding, instrumental variable (IV) methods have been widely used for estimating causal effects.
While conventional IV methods are rooted in econometrics (Wright 1928),
there are modern IV approaches which formulate structural assumptions required to be
satisfied by IVs and provide nonparametric identification results for certain causal contrasts in terms of potential outcomes (Angrist et al.~1996; Robins 1994).
Two basic IV assumptions are called instrument unconfoundedness and exclusion restriction. Intuitively,
a valid IV serves as an exogenous experimental handle, the turning of which may change each individual's treatment status and, through and only through this effect, also change observed outcome.
Then under a monotonicity assumption and other technical conditions,
the local average treatment effect (LATE), defined as the average treatment effect among individuals whose treatment status would be manipulated through the change of the IV,
is identified from observed data nonparametrically (Angrist et al.~1996).

We consider the problem of estimating population LATEs provided that the IV assumptions hold after conditioning on a
set of measured covariates.
While a completely randomized IV is conceptually easy to interpret,
the conditional version of instrument unconfoundedness is more plausible in allowing an IV to be randomized within different values of measured covariates.
In general, it is helpful to think of estimation of LATEs in two stages. First, regression models are built and fitted for certain
unknown functions of the covariates,
such as the instrument propensity score and the treatment and outcome regression functions in Tan (2006a)
or regression functions in Frolich (2007), Uysal (2011), and Ogburn et al.~(2015).
In the second stage, the fitted functions are substituted into various estimators, related to the identification formulas of LATEs.\
For the regression tasks in the first stage, a conventional approach involves an iterative process of model diagnosis, modification, and refitting until some criterion is satisfied,
for example, by inspection of residual plots for outcome regression or covariate balance for propensity score models.
This approach depends on ad hoc choices of how regression terms are added or dropped in model building.
Moreover, uncertainty in the iterative process is complicated and often ignored in subsequent inference (i.e., confidence intervals or hypothesis testing) about LATEs.

In this article, we develop a new method, extending the doubly robust method in Tan (2006a) for estimating population LATEs to high-dimensional settings
where the numbers of regression terms are close to or larger than the sample size in the first-stage estimation described above.
The instrument and treatment are assumed to be binary.
Three regression models are involved: an instrument propensity score model for the conditional probability of the instrument being 1 given the covariates,
a treatment regression model for  the conditional probability of the treatment being 1 given the instrument and covariates,
and an outcome regression model for the conditional mean of the observed outcome given the treatment, instrument and covariates.
The regression terms in all three models are pre-specified, for example, as main effects from a large number of covariates or
additional interaction or nonlinear terms from even a moderate number of covariates.

Our method uses the doubly robust estimator of the LATE in Tan (2006a), in the form of a ratio of two augmented inverse probability weighted (AIPW) estimators.
To tackle high-dimensional data, however, our method employs regularized calibrated estimation for estimating coefficients sequentially in the instrument propensity score
and treatment and outcome regression models. The Lasso penalties (Tibshirani 1996) are introduced to achieve adequate estimation with a large number of
regression terms under sparsity conditions where only a small but unknown subset of regression terms are associated with nonzero coefficients.
The loss functions are carefully chosen for regularized estimation, different from least squares or maximum likelihood,
by leveraging regularized calibrated estimation in Tan (2020b) for estimating average treatment effects (ATEs) under treatment uncounfoundedness in high-dimensional data.
In fact, our estimators for the coefficient vectors in the instrument propensity score and treatment regression models
are directly transferred from Tan (2020b).
Moreover, our estimator for the coefficient vector in the outcome regression model is new as a regularized weighted likelihood estimator,
with a pseudo-response depending on both treatment status and observed outcome. This differs sharply from
maximum quasi-likelihood estimation where the response depends on observed outcome only.

We provide rigorous high-dimensional analysis of the regularized calibrated estimators and the resulting AIPW estimator for the LATE.\
We establish sufficiently fast convergence rates for the regularized calibrated estimators,
in spite of their sequential construction with data-dependent weights and mean functions.
Moreover, we show that  under suitable sparsity conditions, the proposed estimator of LATE achieves a desired asymptotic expansion
in the usual order $O_p(n^{-1/2})$ with a sample size $n$ and then valid Wald confidence intervals can be obtained,
provided that the instrument propensity score model is correctly specified, but the treatment and outcome regression models may be misspecified.
Following the survey literature (Sarndal et al.~2003), our confidence intervals for the LATE
are said to be instrument propensity score model based, and treatment and outcome regression models assisted.
It should be stressed that our method is aimed to be computationally tractable for practical use, with the sequential construction
of the regularized calibrated estimators. In principle, doubly robust confidence intervals for LATEs can also be obtained,
which remain valid if either the instrument propensity score model or the treatment and outcome regression models are correctly specified.
But there are several analytical and computational issues which need to be properly addressed in this direction (see Remarks~\ref{rem:model-assisted} and \ref{rem:asymmetry}).

\vspace{.1in}
{\bf Related work.}
There is an extensive literature on IV and related methods for causal inference (e.g., Baiocchi~et al.~2014; Imbens 2014).
For space limitation, we discuss closely related work only.
Under the IV monotonicity assumption, parametric and semiparametric methods for estimating conditional LATEs given the full vector of covariates
include Little \& Yau (1998), Hirano et al.~(2000), and Abadie (2003). For estimating population or subpopulation LATEs,
doubly robust methods are proposed in Tan (2006a), Uysal (2011), and Ogburn et al.~(2015),
whereas nonparametric smoothing based methods are studied in Frolich (2007).
Alternatively, there are various methods using IVs based on homogeneity assumptions for estimating certain average treatment effects on the treated
(Robins 1994; Vansteelandt \& Goetghebeur 2003; Tan 2010a). Doubly robust estimation of ATEs using IVs is
studied in Okui et al.~(2012) under a partially linear model
and in Wang \& Tchetgen Tchetgen (2018) under suitable identification assumptions.

The foregoing IV methods are developed in low-dimensional settings without regularized estimation. With high-dimensional data,
Chernozhukov et al.~(2018) proposed debiased methods for estimating various treatment parameters, using regularized likelihood-based estimation.
In particular, for estimating population LATEs under the monotonicity assumption, their method exploits the doubly robust estimating function in Tan (2006a) similarly as our method,
but employs regularized likelihood estimation for fitting three regression models as in Uysal (2011).
The Wald confidence intervals for LATEs are shown to be valid under similar sparsity conditions as ours, provided that
all the three regression models are correctly specified (or with negligible biases).
Our main contribution is therefore to provide model-assisted confidence intervals for LATEs using differently configured regularized estimation. See Remark~\ref{rem:ortho-cond} for further discussion.

Similar methods to non-regularized calibrated estimation are proposed in Kim \& Haziza (2014)
for estimating ATEs under treatment unconfoundedness and in Vermeulen \& Vansteelandt (2015) for general doubly robust estimation.
In low-dimensional settings, such methods lead to computationally simpler variance estimation and confidence intervals than likelihood-based estimation for nuisance parameters,
where valid confidence intervals can still be derived using usual influence functions (see Remark~\ref{rem:low-high-dim}).
Similarly as in Tan (2020b), we exploit these ideas in high-dimensional settings to obtain model-assisted confidence intervals for LATEs,
which would not be feasible if using regularized likelihood-based estimation as in Chernozhukov et al.~(2018).

There is a growing literature on confidence intervals and hypothesis testing in high-dimensional settings.
Examples include debiased Lasso
in generalized linear models (Zhang \& Zhang 2014; van de Geer et al.~2014),
and double robustness related methods
(Belloni et al.~2014; Farrell 2015; Avagyan \& Vansteelandt~2017; Smucler et al.~2019; Bradic et al.~2019; Ning et al.~2020),
in addition to Chernozhukov et al.~(2018) and Tan (2020b).
Sample splitting and cross fitting are used in some of these methods, but not pursued here.





\section{Background}

Suppose that $(O_1, \ldots, O_n)$ are independent and identically distributed observations of $O=(Y,D,$ $Z,X)$,
where $Y$ is an outcome variable, $D$ is a binary, treatment variable encoding the presence $(D=1)$ or absence of treatment $(D=0)$, $Z$ is a binary, instrument variable, and $X$ is a vector of measured covariates. We use the potential outcomes notation (Neyman 1990; Rubin 1974) to define quantities of causal interest.
For $z, d\in \{0,1\}$, let $D(z)$ denote the potential treatment status that would be observed if $Z$ were set to level $z$, and let $Y({d,z})$ denote the potential outcome that would be observed if
the treatment and instrument $D$ and $Z$ were set to the levels $d$ and $z$ respectively.
Following Angrist et al.~(1996), the population can be divided into four strata:
the compliers with $D(1)>D(0)$, the always-takers with $D(1)=D(0)=1$, the never-takers with $D(1)=D(0)=0$, and the defiers with $D(1)<D(0)$.

\subsection{Structural assumptions and LATE}

Angrist et al.~(1996) formalized an IV approach with a set of structural assumptions
for identification of the local average treatment effect (LATE), also known as the average treatment effect among the compliers.
Throughout, we make the following conditional versions of the IV assumptions (Abadie 2003; Tan 2006a; Frolich 2007):
\begin{itemize}\addtolength{\itemsep}{-.1in}
\item[(a)] Instrument unconfoundedness: $(Y(d,z), D(z)) \independent Z |X $ for $d,z\in\{0,1\}$, where $\independent$ denotes independence.
\item[(b)] Exclusion restriction: $Y(d,1) = Y(d,0)$, henceforth denoted as $Y(d)$, for $d \in \{0,1\}$.
\item[(c)] Monotonicity: $D({1})\geq D({0})$ with probability 1.
\item[(d)] Instrument overlap: $0< \pi^*(X) <1$ with probability 1, where $\pi^*(X) = P(Z=1|X)$ is called the instrument propensity score (Tan 2006a).
\item[(e)] Instrumentation: $P(D(1)=1)\neq P(D(0)=1)$.
\item[(f)] Consistency: $Y=DY(1)+(1-D)Y(0)$ and $D=ZD(1)+(1-Z)D(0)$.
\end{itemize}

Assumption (a) states that the instrument $Z$ is essentially randomized within levels of the covariate $X$.
Assumptions (a) and (b) together imply an independence condition,
$(Y(d), D(z)) \independent Z | X$ for $d,z\in \{0,1\}$,
which can be technically used in place of Assumptions (a) and (b).
Assumption (c) excludes the existence of defiers in the population.  Vytlacil (2002) and Tan (2006a) showed that the independence and monotonicity
assumptions are equivalent to the assumptions of a nonparametric latent index model: 
\begin{itemize}\addtolength{\itemsep}{-.1in}
\item[\phantom{i}(i)] $D(z)=\one\{\eta(z,X)\geq U\}$ for a function $\eta$ and a random variable $U$, where $\one(\cdot)$ is the indicator function.
\item[(ii)] $(Y(d), U) \independent Z|X$ and $U\independent (Z,X)$.
\end{itemize}
As a result, $U$ can be transformed to be uniformly distributed on $[0,\,1]$ and hence $\eta(z,x)$ equals the treatment propensity score $P(D=1|Z=z,X=x)$.
The preceding representation is helpful for understanding the data-generating process, which is used in our simulation studies.

Assumption (d) ensures that every unit within levels of $X$ has a positive probability of receiving each instrument level $z\in\{0,1\}$.
Assumption (e) requires a non-null causal effect of $Z$ on $D$, in accordance with the concept of $Z$ being an experimental handle; turning of the handle $Z$ needs to change treatment status $D$.
Assumption (f) relates the potential outcomes and treatments to the observed data, under no interference and well-defined intervention conditions.

Under Assumptions (a)--(f), the LATE conditionally on $X=x$, defined as $\mbox{LATE}(x) = E( Y(1) - Y(0) | D(1) > D(0), X=x) $,
can be identified as (Angrist et al.~1996)
\begin{align*}
\frac{ E( Y | Z=1, X=x ) - E(Y| Z=0, X=x)}{ E( D | Z=1, X=x) - E( D | Z=0,X=x) }.
\end{align*}
For high-dimensional $X$, LATE$(x)$ is difficult to interpret, depending on all covariates in $X$.
Moreover, estimation of LATE$(x)$ can be sensitive to modeling assumptions on the conditional expectations above.
Hence it is of interest to consider the population LATE (or in short LATE), defined as $\mbox{LATE}= E( Y(1) - Y(0) | D(1) > D(0))$.
As shown by Tan (2006a) and Frolich (2007), LATE can be identified under Assumptions (a)--(f) in two distinct ways:
\begin{align}
\mbox{LATE } = \frac{ E\{ E( Y | Z=1, X ) - E(Y| Z=0, X) \} }{ E \{ E( D | Z=1, X) - E( D | Z=0,X) \} }, \label{eq:LATE-or}
\end{align}
depending on the regression functions $E(Y | Z=z, X)$ and $E(D|Z=z,X)$ for $z\in\{0,1\}$, or
\begin{align}
\mbox{LATE } = \frac{ E\{ \frac{Z}{\pi^*(X)} Y  \} - E \{\frac{1-Z}{1-\pi^*(X)} Y \} }{ E \{ \frac{Z}{\pi^*(X)} D \} - E \{\frac{1-Z}{1-\pi^*(X)} D \} }, \label{eq:LATE-ps}
\end{align}
depending on the instrument propensity score $\pi^*(X) = P(Z=1|X)$. Both (\ref{eq:LATE-or}) and (\ref{eq:LATE-ps})
are in the form of a ratio of the difference in outcome $Y$ over that in treatment $D$.

A further identification exploited later in our approach is that
the individual expectations $\theta_d = E (Y(d) | D(1) > D(0))$ for $d\in\{0,1\}$, not just the difference LATE $=\theta_1-\theta_0$, can also be identified.
In fact, $\theta_1$ is identified under Assumptions (a)--(f) as
\begin{align}
\theta_1 = \frac{ E\{ E( DY | Z=1, X ) - E(DY | Z=0, X) \} }{ E \{ E( D | Z=1, X) - E( D | Z=0,X) \} }, \label{eq:phi-or}
\end{align}
or equivalently as
\begin{align}
\theta_1 = \frac{ E\{ \frac{Z}{\pi^*(X)} DY  \} - E \{\frac{1-Z}{1-\pi^*(X)} DY \} }{ E \{ \frac{Z}{\pi^*(X)} D \} - E \{\frac{1-Z}{1-\pi^*(X)} D \} }. \label{eq:phi-ps}
\end{align}
Similarly, $\theta_0$ is identified as (\ref{eq:phi-or}) or (\ref{eq:phi-ps}) with $D$ replaced by $1-D$. The difference of the
corresponding identification equations for $\theta_1$ and $\theta_0$ leads back to (\ref{eq:LATE-or}) or (\ref{eq:LATE-ps}).
As shown in Tan (2006a), both (\ref{eq:phi-or}) and (\ref{eq:phi-ps}) can be derived from the following expression of $\theta_1$:
\begin{eqnarray}
\theta_1=\frac{E\{ D(1)Y(1)\} -E\{ D(0)Y(1) \} }{E \{D(1)\} -E\{D(0)\}}, \label{eq:phi1}
\end{eqnarray}
which is a ratio of two differences, depending on potential outcomes and treatments. Because
$Z$ is an experimental handle with $(D,Y)$ as ``outcomes'' under Assumption (a) (instrument unconfoundedness),
each expectation in the numerator and denominator of (\ref{eq:phi1}) can be identified through outcome regression averaging
or inverse probability weighting, so that (\ref{eq:phi-or}) or (\ref{eq:phi-ps}) are obtained.
These results are parallel to related identification results under the assumption of treatment unconfoundedness.
See Tan (2006b, 2010b) and references therein.

\subsection{Modeling assumptions and existing estimators} \label{sec:existing-est}

For estimating $(\theta_1,\theta_0)$ and LATE from sample data,
additional modeling assumptions are required to estimate unknown functions in the identification equations (\ref{eq:LATE-or})--(\ref{eq:LATE-ps})
or (\ref{eq:phi-or})--(\ref{eq:phi-ps}). There are at least two distinct approaches,
depending on models for the instrument propensity score $\pi^*(x) = P(Z=1|X=x)$ or
treatment and outcome regression functions $m^*_z(x) = P(D=1|Z=z,X=x)$
and $m^*_{dz}(x) = E( Y| D=d, Z=z,X=x)$  for $d,z\in\{0,1\}$ (Tan 2006a).
For simplicity, estimation of $\theta_1$ is discussed, whereas that of $\theta_0$ can be similarly handled.
Throughout, $\tilde{E}(\cdot)$ denotes a sample average such that $\tilde{E}\{b(O)\}=n^{-1}\sum_{i=1}^n b(O_i)$ for a function $b(O)$.

\begin{rem}[On modeling choices] \label{rem:modeling}
Consideration of models for $m^*_z(x)$ and $m^*_{dz}(x)$ is conveniently aligned with our interest in
estimating both $(\theta_1,\theta_0)$ and LATE, through identification equations (\ref{eq:phi-or})--(\ref{eq:phi-ps}).
As illustrated in Tan (2006a, Section 5), separate estimates of $\theta_1$ and $\theta_0$ can be informative in applications.
The conditional expectation $E( DY | Z=z, X )$ in (\ref{eq:phi-or}) is decomposed as
$ P(D=1 | Z=z,X) E(Y|D=1, Z=z, X) = m^*_z(x) m^*_{1z}(x)$.
Both models for $m^*_z(x)$ and $m^*_{1z}(x)$ can be specified using
appropriate links functions, as in (\ref{eq:mD})--(\ref{eq:mY}) below. If estimation of LATE is solely of interest
through identification equations (\ref{eq:LATE-or})--(\ref{eq:LATE-ps}), then modeling assumptions can be introduced on
$m^*_z(x) =E(D |Z=z,X) $ and $E(Y |Z=z,x)$ (Froelich 2007; Uysal 2011).
In this case, our methods and theory developed later can be similarly extended.
\end{rem}

First, consider an instrument propensity score model
\begin{eqnarray}
P(Z=1|X=x)=\pi(x; \gamma)=\Pi\{\gamma^\T f(x)\}, \label{eq:pi}
\end{eqnarray}
where $\Pi(\cdot)$ is an inverse link function, $f(x)=\{1,f_1(x),...,f_p(x)\}^\T $ is a vector of known functions such as $( 1,x^\T )^\T $ and $\gamma=(\gamma_0,\gamma_1,...,\gamma_p)$ is a vector of unknown parameters. For concreteness, assume that logistic regression is used such that
$\pi(x; \gamma)=[ 1+\exp\{-\gamma^\T f(x)\}]^{-1}$. By (\ref{eq:phi-ps}), the inverse probability weighted (IPW) estimator of $\theta_1$ is
\begin{eqnarray}
\hat{\theta}_{1,\text{\tiny IPW}}(\hat{\pi})
=\frac{\tilde{E}\left\{\frac{Z}{\hat{\pi}(X)}DY\right\}-\tilde{E}\left\{\frac{1-Z}{1-\hat{\pi}(X)}DY\right\}}{\tilde{E}\left\{\frac{Z}{\hat{\pi}(X)}D\right\}-\tilde{E}\left\{\frac{1-Z}{1-\hat{\pi}(X)}D\right\}}, \label{eq:ipw}
\end{eqnarray}
where $\hat{\pi}(X)=\pi(X; \hat{\gamma})$ is a fitted instrument propensity score.
For low-dimensional $X$, $\hat{\gamma}$ is customarily the maximum likelihood estimator of $\gamma$.
In high-dimensional settings, $\hat{\gamma}$ can be a Lasso penalized maximum likelihood estimator $\hat{\gamma}_{\text{\tiny RML}}$, defined as a minimizer of
$L_{\text{\tiny RML}}(\gamma)= L_{\text{\tiny ML}}(\gamma) + \lambda \|\gamma_{1:p} \|_1$,
where $\|\cdot\|_1$ denotes the $L_1$ norm, $\gamma_{1:p}=(\gamma_1,\ldots,\gamma_p)^\T$,  $\lambda\ge 0$ is a tuning parameter, and
$L_{\text{\tiny ML}}(\gamma)$ is the average negative log-likelihood
\begin{align}
L_{\text{\tiny ML}}(\gamma) = \tilde E \left[ - Z \gamma^\T f(X) + \log \{1 + \me^{\gamma^\T f(X)} \}  \right]. \label{eq:loss-ps-ml}
\end{align}

Alternatively, for $z\in\{0,1\}$, consider treatment and outcome regression models, which can both be called ``outcome regression'' with $(D,Y)$ as ``outcomes'':
\begin{align}
P(D=1|Z=z,X=x)&=m_z(x;\alpha_z)=\psi_{\scriptscriptstyle D}\{\alpha^\T _z g(x)\}, \label{eq:mD} \\
 E(Y|D=1, Z=z, X=x) &=m_{1z}(x;\alpha_{1z})=\psi_{\scriptscriptstyle Y} \{\alpha^\T_{1z} h(x)\}, \label{eq:mY}
\end{align}
where  $\psi_{\scriptscriptstyle D}(\cdot)$ and $\psi_{\scriptscriptstyle Y}(\cdot)$ are inverse link functions,
assumed to be increasing with $\psi_{\scriptscriptstyle D}(-\infty)=0$ and $\psi_{\scriptscriptstyle D}(\infty)=1$, $g(x)=\{1,g_1(x),..., g_{q_1}(x)\}^\T $ and $h(x)=\{1,h_1(x),...,h_{q_2}(x)\}^\T $ are two vectors of known functions, and $\alpha_z$ and $\alpha_{1z}$ are two vectors of unknown parameters of dimensions $1+q_1$ and $1+q_2$ respectively.
By (\ref{eq:phi-or}), the outcome-regression based estimator of $\theta_1$ is
\begin{eqnarray}
\hat{\theta}_{1,\text{\tiny OR}}(\hat{m}_\bullet,\hat{m}_{1\bullet}) = \frac{\tilde{E}\{\hat{m}_{11}(X) \hat{m}_1 (X) \}- \tilde{E}\{ \hat{m}_{10}(X) \hat{m}_0(X) \} }{\tilde{E}\{ \hat{m}_1 (X) \} -\tilde{E}\{ \hat{m}_0 (X) \}}, \label{eq:OR}
\end{eqnarray}
where $\hat{m}_\bullet = (\hat{m}_1, \hat{m}_0)$, $\hat{m}_{1\bullet} = (\hat{m}_{11}, \hat{m}_{10})$,
and, for $z\in\{0,1\}$, $\hat{m}_z (X) = m_z(X ; \hat{\alpha}_z)$ is a fitted treatment regression function and
$\hat{m}_{1z}(X) =m_{1z}(X;\hat{\alpha}_{1z})$ is a fitted outcome regression function.
For low-dimensional $X$, $\hat{\alpha}_z$ and $\hat{\alpha}_{1z}$ are customarily maximum quasi-likelihood estimators of
$\alpha_z$ and $\alpha_{1z}$ or their variants.
In high-dimensional settings, $\hat{\alpha}_z$ and $\hat{\alpha}_{1z}$ can be regularized estimators.
For concreteness, let $\hat{\alpha}_{z,\text{\tiny RML}}$ be a Lasso penalized quasi-likelihood estimator of $\alpha_z$ which is a minimizer of
$L_{\text{\tiny RML}}(\alpha_z ) =  L_{\text{\tiny ML}}(\alpha_z )  + \lambda \| ( \alpha_z)_{1:q_1} \|_1$,
where $( \alpha_z)_{1:q_1}$ is $\alpha_z$ excluding the intercept, $\lambda\ge 0$ is a tuning parameter, and
\begin{align}
L_{\text{\tiny ML}}(\alpha_z ) = \tilde{E}\left( \one\{Z=z\}
[-D\alpha^\T _zg(X)+\Psi_{\scriptscriptstyle D}\{\alpha^\T _zg(X)\}] \right), \label{eq:ML-lossD}
\end{align}
where $\Psi_{\scriptscriptstyle D} (u) = \int_0^u \psi_{\scriptscriptstyle D} (\tilde u)\,\dif \tilde u$.
Let $\hat{\alpha}_{1z,\text{\tiny RML}}$ be a Lasso penalized quasi-likelihood estimator of $\alpha_{1z}$ which is a minimizer of
$L_{\text{\tiny RML}}(\alpha_{1z})= L_{\text{\tiny ML}}(\alpha_{1z}) +
\lambda \| (\alpha_{1z})_{1:q_2} \|_1$,
where $(\alpha_{1z})_{1:q_2}$ is $\alpha_{1z}$ excluding the intercept, $\lambda\ge 0$ is a tuning parameter, and
\begin{align}
L_{\text{\tiny ML}}(\alpha_{1z})= \tilde{E}\left( \one\{Z=z\}  D [-Y \alpha^\T _{1z}h(X)+
 \Psi_{\scriptscriptstyle Y}\{\alpha^\T _{1z}h(X) \}]  \right) , \label{eq:ML-lossY}
\end{align}
where $\Psi_{\scriptscriptstyle Y} (u) = \int_0^u \psi_{\scriptscriptstyle Y} (\tilde u)\,\dif \tilde u$.
The loss function (\ref{eq:ML-lossD}) or (\ref{eq:ML-lossY}) is the average negative log-quasi-likelihood
in the case where model (\ref{eq:mD}) or (\ref{eq:mY}) corresponds to a generalized linear model with a canonical link (McCullagh and Nelder 1989).

\begin{rem}[On outcome regression] \label{rem:model-Y}
We comment on specification of outcome regression model (\ref{eq:mY}).
By the IV assumptions, $E(Y|D=1, Z=z, X=x)=E(Y|Z=z, D(z)=1, X=x)=E(Y(1)|U\leq m^*_z(x),X=x)$ depends on $(z,x)$ only through $m^*_z(x)$ and $x$.
On one hand, this relationship can be incorporated in model (\ref{eq:mY}), by including in $h(x)$ various functions of $x$ and $\hat{m}_z(x)$, as well as their interactions,
for example, $(1,x^\T , \hat{m}_z,x^\T \hat{m}_z)^\T $.
On the other hand, such specification of $h(x)$, depending on an estimator $\hat{\alpha}_z$,
introduces additional variation which needs to be taken account of in theoretical analysis.
For simplicity, this complication is not addressed in our theoretical results later. In fact,
our method is shown to yield valid inference when model (\ref{eq:mY}) may be misspecified, and hence $h(x)$ can be chosen
independently of $\hat{m}_z(x)$.
See Remarks~\ref{rem:model-assisted} and \ref{rem:asymmetry} for related discussions on choices of model-assisted inference.
\end{rem}

Consistency of the estimator $\hat{\theta}_{1,\text{\tiny IPW}}(\hat{\pi})$ relies on correct specification of model (\ref{eq:pi}),
whereas consistency of $ \hat{\theta}_{1,\text{\tiny OR}}(\hat{m}_\bullet,\hat{m}_{1\bullet})$
relies on correct specification of models (\ref{eq:mD})--(\ref{eq:mY}).
The weighting and regression approaches can be combined to obtain doubly robust estimators through augmented IPW estimation (Tan 2006a),
in a similar manner as in the setting of treatment unconfoundedness (Robins et al.~1994; Tan 2007).
The expectations $E \{D(1)\}$ and $E \{D(0)\}$ in (\ref{eq:phi1}) can be estimated by
$\tilde{E} \{\varphi_{\scriptscriptstyle D_1}(O;\hat{\pi},\hat{m}_1) \}$
and $\tilde{E} \{ \varphi_{\scriptscriptstyle D_0}(O;\hat{\pi},\hat{m}_0) \}$ respectively, where
\begin{align}
& \varphi_{\scriptscriptstyle D_1}(O;\hat{\pi},\hat{m}_1) = \frac{Z}{\hat{\pi}(X)} D-\left\{\frac{Z}{\hat{\pi}(X)}-1\right\}\hat{m}_1 (X),  \label{eq:phi-D1}\\
& \varphi_{\scriptscriptstyle D_0}(O;\hat{\pi},\hat{m}_0) = \frac{1-Z}{1-\hat{\pi}(X)} D-\left\{\frac{1-Z}{1-\hat{\pi}(X)}-1\right\}\hat{m}_0 (X). \label{eq:phi-D0}
\end{align}
The expectations $E \{ D(1)Y(1)\}$ and $E\{ D(0)Y(1) \}$ in (\ref{eq:phi1})  can be estimated by
$\tilde{E} \{\varphi_{\scriptscriptstyle D_1Y_{11}}(O;\hat{\pi},$ $\hat{m}_1, \hat{m}_{11}) \}$ and
$\tilde{E} \{\varphi_{\scriptscriptstyle D_0Y_{10}}(O;\hat{\pi},\hat{m}_0, \hat{m}_{10}) \}$ respectively,
where
\begin{align}
&  \varphi_{\scriptscriptstyle D_1Y_{11}}(O;\hat{\pi},\hat{m}_1, \hat{m}_{11} )=
\frac{Z}{\hat{\pi}(X)}  DY-\left\{\frac{Z}{\hat{\pi}(X)}-1\right\} \hat{m}_1 (X)\hat{m}_{11}(X) , \label{eq:phi-D1Y11} \\
& \varphi_{\scriptscriptstyle D_0Y_{10}}(O;\hat{\pi},\hat{m}_0, \hat{m}_{10} ) =
\frac{1-Z}{1-\hat{\pi}(X)} DY -\left\{\frac{1-Z}{1-\hat{\pi}(X)}-1\right\}\hat{m}_0 (X)\hat{m}_{10} (X) . \label{eq:phi-D0Y10}
\end{align}
By (\ref{eq:phi1}), the resulting doubly robust estimator of $\theta_1$ is
\begin{align}
\hat\theta_1 (\hat{\pi}, \hat{m}_\bullet,\hat{m}_{1\bullet} ) =
\frac{\tilde{E}\{\varphi_{\scriptscriptstyle D_1Y_{11}}(O;\hat{\pi},\hat{m}_1, \hat{m}_{11}) - \varphi_{\scriptscriptstyle D_0Y_{10}}(O;\hat{\pi},\hat{m}_0, \hat{m}_{10})\}}
{\tilde{E}\{\varphi_{\scriptscriptstyle D_1}(O;\hat{\pi},\hat{m}_1) - \varphi_{\scriptscriptstyle D_0}(O;\hat{\pi},\hat{m}_0)\}}, \label{eq:dr-phi}
\end{align}
where $\hat{m}_\bullet = (\hat{m}_1, \hat{m}_0)$ and $\hat{m}_{1\bullet} = (\hat{m}_{11}, \hat{m}_{10})$. Consistency of $\hat\theta_1 (\hat{\pi}, \hat{m}_\bullet,\hat{m}_{1\bullet} )$
can be achieved if either model model (\ref{eq:pi}) or  models (\ref{eq:mD})--(\ref{eq:mY}) are correctly specified.

There is potentially a further advantage of doubly robust estimators in high-dimensional settings.
In this case, the estimator $\hat{\theta}_{1,\text{\tiny IPW}}(\hat{\pi})$ or $\hat{\theta}_{1,\text{\tiny OR}}(\hat{m}_\bullet,\hat{m}_{1\bullet})$
in general converges at a slower rate than $O_p(n^{-1/2})$ to the true value $\theta_1$ under correctly specified  model (\ref{eq:pi}) or  models (\ref{eq:mD})--(\ref{eq:mY}) respectively.
Denote $\hat{\pi}_{\text{\tiny RML}} (X) = \pi(X; \hat{\gamma}_{\text{\tiny RML}})$,
$\hat{m}_{z,\text{\tiny RML}} (X) = m_z(X ; \hat{\alpha}_{z,\text{\tiny RML}})$, and
$\hat{m}_{1z,\text{\tiny RML}} (X) =m_{1z}(X;\hat{\alpha}_{1z,\text{\tiny RML}})$,
obtained from Lasso penalized likelihood estimation.
By related results in Chernozhukov et al.~(2018, Section 5.2), it can be shown that
if both models (\ref{eq:pi}) and (\ref{eq:mD})--(\ref{eq:mY}) are correctly specified, then under suitable sparsity conditions,
$\hat{\theta}_{1,\text{\tiny RML}} = \hat\theta_1 (\hat{\pi}_{\text{\tiny RML}}, \hat{m}_{\bullet,\text{\tiny RML}},\hat{m}_{1\bullet,\text{\tiny RML}} )$ converges to $\theta_1$ at rate $O_p(n^{-1/2})$ and admits the asymptotic expansion
\begin{align}
& \hat{\theta}_{1,\text{\tiny RML}}
= \frac{\tilde{E} \{
\varphi_{\scriptscriptstyle D_1Y_{11}}(O;\pi^*,m^*_1, m^*_{11})-
\varphi_{\scriptscriptstyle D_0Y_{10}}(O;\pi^*,m^*_0, m^*_{10}) \} }
{ \tilde{E} \{ \varphi_{\scriptscriptstyle D_1}(O;\pi^*,m^*_1)-\varphi_{\scriptscriptstyle D_0}(O;\pi^*,m^*_0)\} } + o_p(n^{-1/2}), \label{eq:expansion-RML}
\end{align}
where $\pi^* (X) = \pi(X; \gamma^*)$,
$m^*_z(X) = m_z(X ;  \alpha^*_z)$, and
$m^*_{1z}(X) =m_{1z}(X;\alpha^*_{1z})$,
with $(\gamma^*,\alpha^*_z, \alpha^*_{1z})$ the true values in  models (\ref{eq:pi}) and (\ref{eq:mD})--(\ref{eq:mY}).
From this expansion, valid Wald confidence intervals based on $\hat{\theta}_{1,\text{\tiny RML}}$ can be obtained for $\theta_1$.

\section{Methods and theory}

\subsection{Regularized calibrated estimation} \label{sec:methods}

To focus on main ideas, we describe our new method for estimating $\theta_1$. Estimation of $\theta_0$ and LATE is discussed later in this section.
Similarly as in Section~\ref{sec:existing-est}, consider logistic regression model (\ref{eq:pi}),
for estimating the instrument propensity score $\pi^*(x) = P(Z=1|X=x)$,
and models (\ref{eq:mD})--(\ref{eq:mY}) for estimating treatment and outcome regression functions $m^*_z(x) = P(D=1|Z=z,X=x)$
and $m^*_{1z}(x) = E( Y| D=1, Z=z,X=x)$ respectively for $z\in\{0,1\}$.
For technical reasons (see Section \ref{sec:theory}), we require that the ``regressor'' vector $f(x)$ in model (\ref{eq:pi}) is a subvector of $g(x)$ and $h(x)$ in models (\ref{eq:mD})--(\ref{eq:mY})
(hence $p\le q_1$ and $p\le q_2$).
This condition can be satisfied possibly after enlarging models (\ref{eq:mD})--(\ref{eq:mY}) to accommodate $f(x)$.

A class of doubly robust estimators of $\theta_1$, slightly more flexible than (\ref{eq:dr-phi}), is
\begin{align*}
\hat\theta_1  (\hat{\pi}_\bullet, \hat{m}_\bullet,\hat{m}_{1\bullet} ) =
\frac{\tilde{E} \{\tau_{\scriptscriptstyle DY_1} (O;  \hat{\pi}_\bullet, \hat{m}_\bullet,\hat{m}_{1\bullet} ) \} }
{\tilde{E} \{ \tau_{\scriptscriptstyle D} (O;  \hat{\pi}_\bullet, \hat{m}_\bullet )  \} },
\end{align*}
where
$\hat{\pi}_\bullet =(\hat{\pi}_1,\hat{\pi}_0)$ with $\hat{\pi}_1$ and $\hat{\pi}_2$ two possibly different versions of fitted values for $\pi^*$,
$\hat{m}_\bullet = (\hat{m}_1, \hat{m}_0)$ and
$\hat{m}_{1\bullet} = (\hat{m}_{11}, \hat{m}_{10})$
with $(\hat{m}_z, \hat{m}_{1z})$ fitted values for $(m^*_z, m^*_{1z})$ respectively for $z=\{0,1\}$,
and, with $\varphi_{\scriptscriptstyle D_z}$ and $\varphi_{\scriptscriptstyle D_zY_{1z}}$ defined as (\ref{eq:phi-D1})--(\ref{eq:phi-D0Y10}),
\begin{align*}
\tau_{\scriptscriptstyle D} (O; \hat{\pi}_\bullet, \hat{m}_\bullet ) & =
\varphi_{\scriptscriptstyle D_1}(O;\hat{\pi}_1,\hat{m}_1 ) -
\varphi_{\scriptscriptstyle D_0}(O;\hat{\pi}_0,\hat{m}_0 ), \\
\tau_{\scriptscriptstyle DY_1} (O; \hat{\pi}_\bullet, \hat{m}_\bullet,\hat{m}_{1\bullet} ) & =
\varphi_{\scriptscriptstyle D_1Y_{11}}(O;\hat{\pi}_1,\hat{m}_1, \hat{m}_{11}) -
\varphi_{\scriptscriptstyle D_0Y_{10}}(O;\hat{\pi}_0,\hat{m}_0, \hat{m}_{10} ) .
\end{align*}
Our point estimator of $\theta_1$ is
$\hat{\theta}_{1,\text{\tiny RCAL}} = \hat\theta_1 (\hat{\pi}_{\bullet,\text{\tiny RCAL}}, \hat{m}_{\bullet,\text{\tiny RWL}},\hat{m}_{1\bullet,\text{\tiny RWL}} )$,
where, for $z \in \{0,1\}$,
$\hat{\pi}_{z,\text{\tiny RCAL}} (X) = \pi(X; \hat{\gamma}_{z,\text{\tiny RCAL}})$,
$\hat{m}_{z,\text{\tiny RWL}}(X) = m_z (X; \hat{\alpha}_{z,\text{\tiny RWL}})$,
and $\hat{m}_{1z,\text{\tiny RWL}}(X) =m_{1z} (X; \hat{\alpha}_{1z,\text{\tiny RWL}})$ are fitted values,
and $(\hat{\gamma}_{z,\text{\tiny RCAL}}, \hat{\alpha}_{z,\text{\tiny RWL}},  \hat{\alpha}_{1z,\text{\tiny RWL}})$ are
estimators of $(\gamma, \alpha_z, \alpha_{1z})$ defined as follows.

For logistic regression model (\ref{eq:pi}), the estimator $\hat{\gamma}_{z,\text{\tiny RCAL}}$ is a regularized calibrated estimator of
$\gamma$ (Tan 2020a), defined as a minimizer of the Lasso penalized objective function
\begin{eqnarray}
\label{eq:rcal-loss1a} L_{z,\text{\tiny RCAL}}(\gamma)=L_{z,\text{\tiny CAL}}(\gamma) +\lambda \|\gamma_{1:p} \|_1, \quad z \in \{0,1\},
\end{eqnarray}
with the calibration loss functions
\begin{eqnarray}
\label{eq:rcal-loss1b} L_{0,\text{\tiny CAL}}(\gamma)&=\tilde{E}\{(1-Z)e^{\gamma^\T  f(X)}-Z\gamma^\T  f(X)\},\\
\label{eq:rcal-loss1c} L_{1,\text{\tiny CAL}}(\gamma)&=\tilde{E}\{Ze^{-\gamma^\T  f(X)}+(1-Z)\gamma^\T  f(X)\}.
\end{eqnarray}
Minimization of (\ref{eq:rcal-loss1a}) can be implemented using R package \texttt{RCAL} (Tan 2020a).
For treatment regression model (\ref{eq:mD}),  $\hat{\alpha}_{z,\text{\tiny RWL}}$ is a regularized weighted likelihood estimator of $\alpha_z$, defined as a minimizer of
the Lasso penalized objective function
\begin{eqnarray}
\label{eq:rcal-loss2a}L_{z,\text{\tiny RWL}}(\alpha_z;\hat{\gamma}_{z,\text{\tiny RCAL}}) =L_{z,\text{\tiny WL}}(\alpha_z;\hat{\gamma}_{z,\text{\tiny RCAL}}) +\lambda \|(\alpha_z)_{1:q_1} \|_1,
\end{eqnarray}
with the weighted (quasi-)likelihood loss function
\begin{eqnarray}
\label{eq:rcal-loss2b}L_{z,\text{\tiny WL}}(\alpha_z;\hat{\gamma}_z)&=&\tilde{E}\left(\one\{Z=z\} w_z(X;\hat{\gamma}_z)[-D\alpha^\T _zg(X)+\Psi_{\scriptscriptstyle D}\{\alpha^\T _z g(X)\}] \right),
\end{eqnarray}
where the weight function is $w_z(x;\gamma)= [\{1-\pi(X;\gamma)\} /\pi(X;\gamma) ]^{2z-1}$ for $z\in\{0,1\}$. 
For outcome regression model (\ref{eq:mY}), $\hat{\alpha}_{1z,\text{\tiny RWL}}$ is a regularized calibrated estimator of $\alpha_{1z}$, defined as a minimizer of the Lasso penalized objective function
\begin{eqnarray}
L_{1z,\text{\tiny RWL}}(\alpha_{1z};\hat{\gamma}_{z,\text{\tiny RCAL}}, \hat{\alpha}_{z,\text{\tiny RWL}}) =L_{1z,\text{\tiny WL}}(\alpha_{z1};\hat{\gamma}_{z,\text{\tiny RCAL}}, \hat{\alpha}_{z,\text{\tiny RWL}}) +\lambda \|({\alpha}_{1z})_{1:q_2} \|_1, \label{eq:rcal-loss3a}
\end{eqnarray}
with the loss function
\begin{align}
& L_{1z,\text{\tiny WL}}(\alpha_{1z};\hat{\gamma}_z, \hat{\alpha}_z) =\tilde{E}\left( \one\{Z=z\} w_z(X;\hat{\gamma}_z)[-DY \alpha^\T_{1z}h(X)+ m_z (X; \hat{\alpha}_z)\Psi_{\scriptscriptstyle Y}\{\alpha^\T _{1z}h(X)\}]  \right), \label{eq:rcal-loss3b}
\end{align}
where the weight function $w_z(x;\gamma)$ is the same as above. Interestingly, (\ref{eq:rcal-loss3b}) can be equivalently expressed as
a weighted (quasi-)likelihood loss
\begin{align}
& L_{1z,\text{\tiny WL}}(\alpha_{1z};\hat{\gamma}_z, \hat{\alpha}_z) =\tilde{E}\left(\one\{Z=z\}
w_{1z}(X;\hat{\gamma}_z,\hat{\alpha}_z)[-\tilde{Y}_{1z} \alpha^\T _{1z}h(X)+\Psi_{\scriptscriptstyle Y}\{\alpha^\T _{1z}h(X)\}] \right),\label{eq:rcal-loss3c}
\end{align}
with the pseudo-response $\tilde{Y}_{1z} =Y D/ m_z(X; \hat{\alpha}_z)$ and weight
$w_{1z}(X;\hat{\gamma}_z,\hat{\alpha}_z)=w_z(X;\hat{\gamma}_z) m_z$ $(X; \hat{\alpha}_z)$.
Hence existing software for Lasso penalized weighted estimation such as \texttt{glmnet} (Friedman et al.~2010) and \texttt{RCAL} can be employed to minimize (\ref{eq:rcal-loss3a}) as well as (\ref{eq:rcal-loss2a}).
The loss (\ref{eq:rcal-loss3b}) or (\ref{eq:rcal-loss3c}) differs sharply from that of the likelihood loss (\ref{eq:ML-lossY}), in terms of the residuals implied.

Compared with regularized likelihood estimation in Section~\ref{sec:existing-est}, our method involves using a different set of estimators
$(\hat{\gamma}_{z,\text{\tiny RCAL}}, \hat{\alpha}_{z,\text{\tiny RWL}},\hat{\alpha}_{1z,\text{\tiny RWL}} )$,
which are called regularized calibrated estimators.
Similarly as in Tan (2020b), these estimators are derived to allow model-assisted, asymptotic confidence intervals for $\theta_1$ based on $\hat{\theta}_{1,\text{\tiny RCAL}}$.
See Proposition \ref{pro:RCAL} for a summary and Section~\ref{sec:theory} for further discussion.
We also point out several interesting properties algebraically associated with our estimators.
First, by the Karush--Kuhn--Tucker (KKT) condition for minimizing (\ref{eq:rcal-loss1a}), the fitted instrument propensity score
$\hat{\pi}_{1,\text{\tiny RCAL}} (X)$ satisfies
\begin{align}
& \frac{1}{n} \sum_{i=1}^n \frac{Z_i}{\hat{\pi}_{1,\text{\tiny RCAL}} (X_i)} = 1,  \label{eq:KKT-pi1} \\
& \frac{1}{n} \left| \sum_{i=1}^n \frac{Z_i f_j(X_i)}{\hat{\pi}_{1,\text{\tiny RCAL}} (X_i)} - \sum_{i=1}^n f_j(X_i) \right| \le \lambda, \quad j =1,\ldots, p. \label{eq:KKT-pi2}
\end{align}
where equality holds in (\ref{eq:KKT-pi2}) for any $j$ such that the $j$th estimate $(\hat{\gamma}_{1,\text{\tiny RCAL}})_j$ is nonzero.
These equations also hold with $Z_i$ replaced by $1-Z_i$ and $\hat{\pi}_{1,\text{\tiny RCAL}}$ replaced by $1 - \hat{\pi}_{0,\text{\tiny RCAL}}$.
Eq.~(\ref{eq:KKT-pi1}) shows that
the inverse probability weights, $1/\hat{\pi}_{1,\text{\tiny RCAL}} (X_i)$ with $Z_i=1$, sum to the sample size $n$, whereas
Eq.~(\ref{eq:KKT-pi2}) implies that
the weighted average of each covariate $f_j(X_i)$ over the instrument group $\{Z_i=1\}$ may differ from the overall average of $f_j(X_i)$ by no more than $\lambda$.
Such differences are of interest in showing how a weighted instrument group resembles the overall sample. In contrast,
similar results are not available when using the regularized likelihood estimator $\hat{\gamma}_{\text{\tiny RML}}$.
Moreover, Tan (2020a) studied a comparison between calibrated and maximum likelihood estimation in logistic regression.
Minimization of the calibration loss (\ref{eq:rcal-loss1b}) or (\ref{eq:rcal-loss1c}) achieves a stronger control of
relative errors of fitted propensity scores than that of the likelihood loss (\ref{eq:loss-ps-ml}).

By the KKT condition associated with the intercept in $\alpha_1$ for minimizing (\ref{eq:rcal-loss2a}),
the fitted treatment regression function $\hat{m}_{1,\text{\tiny RWL}}(X)$ satisfies
\begin{align}
& \frac{1}{n} \sum_{i=1}^n Z_i \frac{1-\hat{\pi}_{1,\text{\tiny RCAL}} (X_i) }{\hat{\pi}_{1,\text{\tiny RCAL}} (X_i) }
\left\{ D_i - \hat{m}_{1,\text{\tiny RWL}}(X_i) \right\}= 0 . \label{eq:KKT-mD}
\end{align}
A similar equation holds with $Z_i$ replaced by $1-Z_i$, $\hat{\pi}_{1,\text{\tiny RCAL}}$ by $1 - \hat{\pi}_{0,\text{\tiny RCAL}}$,
and $\hat{m}_{1,\text{\tiny RWL}}$ by $\hat{m}_{0,\text{\tiny RWL}}$.
As a result of (\ref{eq:KKT-mD}), our augmented IPW estimator for $E\{D(1)\}$, defined as $\hat{E}_{\text{\tiny RCAL}} \{D(1)\} =
\tilde{E}\{\varphi_{\scriptscriptstyle D_1}(O;\hat{\pi}_{1,\text{\tiny RCAL}},\hat{m}_{1,\text{\tiny RWL}}) \}$, can be simplified to
\begin{align*}
\tilde E \left[ \hat{m}_{1,\text{\tiny RWL}}(X) + \frac{Z}{\hat{\pi}_{1,\text{\tiny RCAL}}(X) } \left\{ D -  \hat{m}_{1,\text{\tiny RWL}}(X)\right\} \right]
 = \tilde E \left\{ Z D + (1-Z) \hat{m}_{1,\text{\tiny RWL}}(X)\right\}.
\end{align*}
Hence $\hat{E}_{\text{\tiny RCAL}} \{D(1)\}$ always falls within the range of
the binary treatment values $\{D_i: Z_i=1, i=1,\ldots,n\}$ and the predicted values $\{\hat{m}_{1,\text{\tiny RWL}} (X_i): Z_i=0, i=1,\ldots,n\}$,
which are by definition in the interval $[0,1]$.
This boundedness property is not satisfied by the usual estimator $\hat{E}_{\text{\tiny RML}} \{D(1)\} =
\tilde{E}\{\varphi_{\scriptscriptstyle D_1}(O;\hat{\pi}_{\text{\tiny RML}},\hat{m}_{1,\text{\tiny RML}}) \}$,
but is desirable for stabilizing the behavior of augmented IPW estimators, especially when used in the denominator of (\ref{eq:dr-phi}).

By the KKT condition associated with the intercept in $\alpha_{11}$ for minimizing (\ref{eq:rcal-loss3a}),
the fitted functions $\hat{m}_{1,\text{\tiny RWL}}(X)$ and  $\hat{m}_{11,\text{\tiny RWL}}(X)$  jointly satisfy
\begin{align}
& \frac{1}{n} \sum_{i=1}^n Z_i \frac{1-\hat{\pi}_{1,\text{\tiny RCAL}} (X_i) }{\hat{\pi}_{1,\text{\tiny RCAL}} (X_i) }
\left\{ D_i Y_i - \hat{m}_{1,\text{\tiny RWL}}(X_i) \hat{m}_{11,\text{\tiny RWL}}(X_i) \right\}= 0 . \label{eq:KKT-mY}
\end{align}
A similar equation holds with $Z_i$ replaced by $1-Z_i$, $\hat{\pi}_{1,\text{\tiny RCAL}}$ by $1 - \hat{\pi}_{0,\text{\tiny RCAL}}$,
and $(\hat{m}_{1,\text{\tiny RWL}}, \hat{m}_{11,\text{\tiny RWL}})$ by $(\hat{m}_{0,\text{\tiny RWL}},\hat{m}_{10,\text{\tiny RWL}})$.
By (\ref{eq:KKT-mY}), our augmented IPW estimator for $E\{D(1) Y(1)\}$, defined as $\hat{E}_{\text{\tiny RCAL}} \{D(1) $ $\times Y(1)\}
= \tilde{E}\{\varphi_{\scriptscriptstyle D_1Y_{11}}(O;\hat{\pi}_{1,\text{\tiny RCAL}},\hat{m}_{1,\text{\tiny RWL}},\hat{m}_{11,\text{\tiny RWL}}) \}$,
can be simplified to
\begin{align*}
\tilde E \left[(\hat{m}_1 \hat{m}_{11})_{\text{\tiny RWL}} + \frac{Z}{\hat{\pi}_{1,\text{\tiny RCAL}}(X) } \left\{ D Y - (\hat{m}_1 \hat{m}_{11})_{\text{\tiny RWL}} \right\} \right]
 = \tilde E \left\{ Z DY + (1-Z) (\hat{m}_1 \hat{m}_{11})_{\text{\tiny RWL}} \right\} ,
\end{align*}
where $ (\hat{m}_1 \hat{m}_{11})_{\text{\tiny RWL}} = \hat{m}_{1,\text{\tiny RWL}} (X) \hat{m}_{11,\text{\tiny RWL}} (X)$.
As a consequence, $\hat{E}_{\text{\tiny RCAL}} \{D(1) Y(1)\}$ always falls within the range of
the observed values $\{D_i Y_i: Z_i=1, i=1,\ldots,n\}$ and the predicted values $\{\hat{m}_{1,\text{\tiny RWL}} (X_i) \hat{m}_{11,\text{\tiny RWL}} (X_i):
Z_i=0, i=1,\ldots,n\}$.

We present a high-dimensional analysis of the proposed estimator $\hat\theta_{1,\text{\tiny RCAL}}$ in Section \ref{sec:theory}
provided that instrument propensity score model (\ref{eq:pi}) is correctly specified but treatment and outcome models (\ref{eq:mD})--(\ref{eq:mY}) may be misspecified.
Our main result shows that under suitable conditions,
$\hat{\theta}_{1,\text{\tiny RCAL}}$ is consistent for $\theta_1$ and admits the asymptotic expansion
\begin{align}
& \hat{\theta}_{1,\text{\tiny RCAL}} =
\frac{\tilde{E} \{\tau_{\scriptscriptstyle DY_1} (O;  \bar{\pi}_\bullet, \bar{m}_\bullet,\bar{m}_{1\bullet} ) \} }
{\tilde{E} \{ \tau_{\scriptscriptstyle D} (O; \bar{\pi}_\bullet, \bar{m}_\bullet )  \} } + o_p(n^{-1/2}), \label{eq:expansion-RCAL}
\end{align}
where $\bar{\pi}_\bullet =(\bar{\pi}_1,\bar{\pi}_0)$,
$\bar{m}_\bullet = (\bar{m}_1, \bar{m}_0)$,
$\bar{m}_{1\bullet} = (\bar{m}_{11}, \bar{m}_{10})$,
and, for $z\in\{0,1\}$, $\bar{\pi}_z (X) = \pi(X; \bar{\gamma}_z)$,
$\bar{m}_z(X) = m_z(X; \bar{\alpha}_z)$, and
$\bar{m}_{1z}(X) =m_{1z}(X;\bar{\alpha}_{1z})$,
with $(\bar{\gamma}_z, \bar{\alpha}_z, \bar{\alpha}_{1z})$ defined as follows.
The target value $\bar{\gamma}_z$ is defined as a minimizer of the expected loss
\begin{align*}
E\{ L_{z,\text{\tiny CAL}}(\gamma) \}= E \left[ \one\{Z=z\} e^{(1-2z)\gamma^\T  f(X)}+ \one\{Z=1-z\} (2z-1) \gamma^\T  f(X) \right], \quad z \in \{0,1\}.
\end{align*}
Because model (\ref{eq:pi}) is correctly specified, the target values $\bar{\gamma}_1$ and $\bar{\gamma}_0$ are identical to the true value $\gamma^*$, so that
$\bar{\pi}_1 (X) = \bar{\pi}_0(X) = \pi^*(X)$ (Tan 2020a).
With possible misspecification of model (\ref{eq:mD}), the target value $\bar{\alpha}_z$ is defined as a minimizer of the expected loss
\begin{align}
E \{ L_{z,\text{\tiny WL}}(\alpha_z; \bar\gamma_z) \} =
E \left(\one\{Z=z\} w_z(X; \bar\gamma_z)[-D\alpha^\T _zg(X)+\Psi_{\scriptscriptstyle D}\{\alpha^\T _z g(X)\}] \right).
\label{eq:eloss-alpha-Dz}
\end{align}
If model (\ref{eq:mD}) is correctly specified, then $\bar{\alpha}_z$ coincides with the true value $\alpha^*_z$ such that
$ \bar{m}_z(X )= m^*_z(X)$.
Otherwise, $\bar{m}_z(X)$ may differ from $m^*_z(X)$.
Similarly, with possible misspecification of model (\ref{eq:mY}), the target value $\bar{\alpha}_{1z}$ is defined as a minimizer of the expected loss
\begin{align}
E \{ L_{1z,\text{\tiny WL}}(\alpha_{1z}; \bar\gamma_z, \bar{\alpha}_z) \} =
E\left( \one\{Z=z\} w_z(X;\bar\gamma_z)[-DY \alpha^\T_{1z}h(X)+ m_z(X; \bar{\alpha}_z)\Psi_{\scriptscriptstyle Y}\{\alpha^\T _{1z}h(X)\}]  \right). \label{eq:eloss-alpha-Y1z}
\end{align}
If model (\ref{eq:mY}) is correctly specified, then $\bar{\alpha}_{1z}$ coincides with the true value $\alpha^*_{1z}$ such that
$ \bar{m}_{1z}(X) = m^*_{1z}(X)$.
But $\bar{m}_{1z}(X)$ may in general differ from $m^*_{1z}(X)$.
Suppose that the Lasso tuning parameters are specified as $\lambda = A_{0z}^\dag \{\log(\me p)/n\}^{1/2}$
for $\hat\gamma_{z,\text{\tiny RCAL}}$ in (\ref{eq:rcal-loss1a}), $\lambda=A_{1z}^\dag \{\log(\me q_1)/n\}^{1/2}$ for $\hat\alpha_{z,\text{\tiny RCAL}}$ in (\ref{eq:rcal-loss2a}),
and $\lambda= A_{2z}^\dag \{ \log (\me q_2)/n \}^{1/2}$ for $\hat\alpha_{1z,\text{\tiny RCAL}}$ in (\ref{eq:rcal-loss3a}),
where $(A_{0z}^\dag, A_{1z}^\dag, A_{2z}^\dag)$ are sufficiently large constants for $z \in \{0,1\}$.
For a vector $b =(b_0,b_1, \ldots, b_k)^\T$, denote $S_b = \{0\} \cup \{j: b_j \not=0, j=1,\ldots,k\}$
and the size of the set $S_b$ as $|S_b|$.

\begin{pro} \label{pro:RCAL}
Suppose that Assumptions~\ref{ass:gamma1}--\ref{ass:alpha11} hold as in Section~\ref{sec:theory}, their corresponding versions
with $Z$ replaced by $1-Z$ and $(\bar\gamma_1,\bar\alpha_1,\bar\alpha_{11})$ replaced by $(-\bar\gamma_0, \bar\alpha_0, \bar\alpha_{10})$,
and $ \sum_{z=0,1} \{ |S_{\bar\gamma_z}| + |S_{\bar\alpha_z}| + |S_{\bar\alpha_{1z}}| \} \log\{\me (q_1\vee q_2)\} = o(n^{1/2})$,
If model (\ref{eq:pi}) is correctly specified, then $ \hat{\theta}_{1,\text{\tiny RCAL}} $ satisfies the asymptotic expansion (\ref{eq:expansion-RCAL}).
Furthermore, the following results hold.
\begin{itemize} 
\item[(i)]
$n^{1/2} ( \hat{\theta}_{1,\text{\tiny RCAL}} -\theta_1 ) \to_{\mathcal D} \N(0,V_1)$,
where $V_1 = \var\{ \tau_{\scriptscriptstyle DY_1} (O;  \bar{\pi}_\bullet, \bar{m}_\bullet,\bar{m}_{1\bullet} ) -
\theta_1  \tau_{\scriptscriptstyle D} (O; \bar{\pi}_\bullet, \bar{m}_\bullet ) \} /
$ $ E^2 \{\tau_{\scriptscriptstyle D} (O; \bar{\pi}_\bullet, \bar{m}_\bullet ) \} $;

\item[(ii)] a consistent estimator of $V_1$ is
\begin{align*}
\hat V_1 = \tilde E \left[ \left\{\tau_{\scriptscriptstyle DY_1} (O;  \hat{\pi}_\bullet, \hat{m}_\bullet,\hat{m}_{1\bullet} ) -
\hat{\theta}_{1,\text{\tiny RCAL}} \tau_{\scriptscriptstyle D} (O; \hat{\pi}_\bullet, \hat{m}_\bullet )\right\}^2 \right] \Big/
\tilde{E}^2 \{\tau_{\scriptscriptstyle D} (O; \hat{\pi}_\bullet, \hat{m}_\bullet ) \},
\end{align*}
where $(\hat{\pi}_\bullet, \hat{m}_\bullet,\hat{m}_{1\bullet} ) = (\hat{\pi}_{\bullet,\text{\tiny RCAL}}, \hat{m}_{\bullet,\text{\tiny RWL}},\hat{m}_{1\bullet,\text{\tiny RWL}} )$;

\item[(iii)] an asymptotic $(1-c)$ confidence interval for $\theta_1$ is $\hat{\theta}_{1,\text{\tiny RCAL}} \pm z_{c/2} \sqrt{\hat V_1/n}$,
where $z_{c/2}$ is the $(1-c/2)$ quantile of \,$\N(0,1)$.
\end{itemize}
\end{pro}

Finally, we describe how our method can be applied to estimate $\theta_0$
and LATE, denoted as $\theta= \theta_1-\theta_0$.  In addition to models (\ref{eq:pi}) and (\ref{eq:mD})--(\ref{eq:mY}), consider the following outcome regression model
in the untreated population for $z\in\{0,1\}$,
\begin{align}
E(Y|Z=z, D=0, X=x)&=m_{0z}(x;\alpha_{0z})=\psi_{\scriptscriptstyle Y}\{\alpha^\T _{0z} h (x)\}, \label{eq:m0}
\end{align}
where $h (x)=\{1,h_1(x),...,h_{q_2}(x)\}^\T $ is a vector of known functions as in model (\ref{eq:mY}) and $\alpha_{0z}$ is a vector of unknown parameters of dimension $1+q_2$.
For augmented IPW estimation of $E \{ (1-D(0)) Y(0)\}$ and $E \{ (1-D(1)) Y(1)\}$, define
\begin{align*}
& \varphi_{\scriptscriptstyle D_0Y_{00}}(O;\hat{\pi}_0,\hat{m}_0, \hat{m}_{00}) = \frac{1-Z}{1-\hat{\pi}_0(X)}(1-D)Y
-\left\{ \frac{1-Z}{1-\hat{\pi}_0(X)}-1\right\} \{1-\hat{m}_0(X)\} \hat{m}_{00}(X) , \\
& \varphi_{\scriptscriptstyle D_1Y_{01}}(O;\hat{\pi}_1,\hat{m}_1, \hat{m}_{01}) = \frac{Z}{\hat{\pi}_1(X)} (1-D) Y
-\left\{ \frac{Z}{\hat{\pi}_1(X)} -1\right\} \{1-\hat{m}_1(X)\} \hat{m}_{01}(X) ,
\end{align*}
where $\hat{m}_{0z}(X) =m_{0z}(X;\hat{\alpha}_{0z})$ be a fitted regression function.
Then a doubly robust estimator of $\theta_0$, similar to that of $\theta_1$ in (\ref{eq:dr-phi}) is
\begin{align}
\hat\theta_0  (\hat{\pi}_\bullet, \hat{m}_\bullet,\hat{m}_{0\bullet} ) =
\frac{\tilde{E} \{\tau_{\scriptscriptstyle DY_0} (O;  \hat{\pi}_\bullet, \hat{m}_\bullet,\hat{m}_{0\bullet} ) \} }
{\tilde{E} \{ \tau_{\scriptscriptstyle D} (O;  \hat{\pi}_\bullet, \hat{m}_\bullet )  \} }, \label{eq:dr-phi0}
\end{align}
where $\hat{m}_{0\bullet} = (\hat{m}_{01}, \hat{m}_{00})$,
$\tau_{\scriptscriptstyle D} (O;  \hat{\pi}_\bullet, \hat{m}_\bullet ) $ is as in (\ref{eq:dr-phi}), and
\begin{align*}
& \tau_{\scriptscriptstyle DY_0} (O; \hat{\pi}_\bullet, \hat{m}_\bullet,\hat{m}_{0\bullet} ) =
\varphi_{\scriptscriptstyle D_0Y_{00}}(O;\hat{\pi}_0,\hat{m}_0, \hat{m}_{00}) -
\varphi_{\scriptscriptstyle D_1Y_{01}}(O;\hat{\pi}_1,\hat{m}_1, \hat{m}_{01} ) .
\end{align*}
Our point estimator of $\theta_0$ is
$\hat{\theta}_{0,\text{\tiny RCAL}} = \hat\theta_0 (\hat{\pi}_{\bullet,\text{\tiny RCAL}}, \hat{m}_{\bullet,\text{\tiny RWL}},\hat{m}_{0\bullet,\text{\tiny RWL}} )$,
and that of LATE, $\theta = \theta_1-\theta_0$, is $\hat{\theta}_{\text{\tiny RCAL}}=\hat{\theta}_{1,\text{\tiny RCAL}}-\hat{\theta}_{0,\text{\tiny RCAL}}$,
where $\hat{\pi}_{z,\text{\tiny RCAL}}(X)$ and $\hat{m}_{z,\text{\tiny RWL}}(X)$ remain the same as before,
and $\hat{m}_{0z,\text{\tiny RWL}}(X) =m_{0z} (X; \hat{\alpha}_{0z,\text{\tiny RWL}})$ with
$\hat{\alpha}_{0z,\text{\tiny RWL}}$ defined as follows.
For $z\in\{0,1\}$, let $\hat{\alpha}_{0z,\text{\tiny RWL}}$ be a minimizer of the Lasso penalized objective function
\begin{align*}
L_{0z,\text{\tiny RWL}}(\alpha_{0z};\hat{\gamma}_z, \hat{\alpha}_z) =
L_{0z,\text{\tiny WL}}(\alpha_{0z};\hat{\gamma}_z, \hat{\alpha}_z) +\lambda \|({\alpha}_{0z})_{1:r } \|_1, 
\end{align*}
with the weighted (quasi-)likelihood loss
\begin{align*}
& L_{0z,\text{\tiny WL}}(\alpha_{0z};\hat{\gamma}_z, \hat{\alpha}_z) =\tilde{E}\left(\one\{Z=z\}
w_{0z}(X;\hat{\gamma}_z,\hat{\alpha}_z)[-\tilde{Y}_{0z} \alpha^\T _{0z}h (X)+\Psi_{\scriptscriptstyle Y}\{\alpha^\T _{0z}h (X)\}] \right), 
\end{align*}
where $\tilde{Y}_{0z} =Y (1-D)/\{1-m_z(X; \hat{\alpha}_z)\}$ and  $w_{0z}(X;\hat{\gamma}_z,\hat{\alpha}_z)=w_z(X;\hat{\gamma}_z)\{1- m_z(X; \hat{\alpha}_z)\}$.
Under similar conditions as in Proposition~\ref{pro:RCAL}, $\hat{\theta}_{0,\text{\tiny RCAL}}$ admits an asymptotic expansion in the form of (\ref{eq:expansion-RCAL}),
and Wald confidence intervals for $\theta_0$ and LATE can be derived accordingly.
In particular,  an asymptotic $(1-c)$ confidence interval for LATE is $\hat{\theta}_{\text{\tiny RCAL}} \pm z_{c/2} \sqrt{\hat V/n}$,
where
\begin{align*}
\hat V =  \frac{\tilde E \left[ \left\{\tau_{\scriptscriptstyle DY_1} (O;  \hat{\pi}_\bullet, \hat{m}_\bullet,\hat{m}_{1\bullet} ) -
\tau_{\scriptscriptstyle DY_0} (O;  \hat{\pi}_\bullet, \hat{m}_\bullet,\hat{m}_{0\bullet} ) -
\hat{\theta}_{\text{\tiny RCAL}} \tau_{\scriptscriptstyle D} (O; \hat{\pi}_\bullet, \hat{m}_\bullet )\right\}^2 \right] }{
\tilde{E}^2 \{\tau_{\scriptscriptstyle D} (O; \hat{\pi}_\bullet, \hat{m}_\bullet ) \} },
\end{align*}
where $(\hat{\pi}_\bullet, \hat{m}_\bullet,\hat{m}_{1\bullet},\hat{m}_{0\bullet} ) = (\hat{\pi}_{\bullet,\text{\tiny RCAL}}, \hat{m}_{\bullet,\text{\tiny RWL}},\hat{m}_{1\bullet,\text{\tiny RWL}},\hat{m}_{0\bullet,\text{\tiny RWL}}  )$.

\begin{rem}[On completely randomized instruments] \label{rem:randomized-iv}
Our method is directly applicable in the special case where the instrument is assumed to be completely randomized, independently of observed covariates.
The instrument propensity score model (\ref{eq:pi}) with an intercept is valid, because $P(Z=1|X)$ is a constant.
With flexible treatment and outcome models (\ref{eq:mD})--(\ref{eq:mY}), the proposed estimator $\hat{\theta}_{\text{\tiny RCAL}}$ based on augmented IPW estimation
is expected to achieve smaller variances than the simple Wald estimator.
Such efficiency gains are analogous to related results on regression adjustment in completely randomized experiments (with full compliance)
in low- and high-dimensional settings (e.g., Davidian et al. 2005; Bloniarz et al. 2016; Wager et al. 2016).
\end{rem}

\subsection{Theoretical analysis} \label{sec:theory}

We develop theoretical analysis which leads to Proposition~\ref{pro:RCAL} for the proposed estimator $\hat{\theta}_{1,\text{\tiny RCAL}}$ in high-dimensional settings,
provided model (\ref{eq:pi}) is correctly specified but models (\ref{eq:mD})--(\ref{eq:mY}) may be misspecified.
Similar analysis can be obtained for  $\hat{\theta}_{0,\text{\tiny RCAL}}$ and $\hat{\theta}_{\text{\tiny RCAL}}$.
Before formal results are presented, we discuss heuristically how the asymptotic expansion (\ref{eq:expansion-RCAL}) can be achieved, due to use of
the regularized calibrated estimators $(\hat{\gamma}_{z,\text{\tiny RCAL}}, \hat{\alpha}_{z,\text{\tiny RWL}},\hat{\alpha}_{1z,\text{\tiny RWL}} )$ for $z\in\{0,1\}$.
For notational brevity, these estimators are denoted as $(\hat{\gamma}_z, \hat{\alpha}_z,\hat{\alpha}_{1z} )$ unless otherwise noted.

There are two main steps in our analysis.
First,  the estimators $(\hat{\gamma}_z, \hat{\alpha}_z,\hat{\alpha}_{1z} )$
can be shown to converge in probability to $(\bar{\gamma}_z, \bar{\alpha}_z,\bar{\alpha}_{1z} )$ with the $L_1$-norm error bounds:
\begin{align}
& \| \hat{\gamma}_z - \bar{\gamma}_z \|_1 = O_p(1) |S_{\bar\gamma_z}|\{\log(\me p)/n\}^{1/2}, \label{eq:conv-gamma}\\
& \| \hat{\alpha}_z  - \bar{\alpha}_z \|_1 = O_p(1) \{|S_{\bar\gamma_z}|+ |S_{\bar\alpha_z}|\} \{\log(\me q_1)/n\}^{1/2}, \label{eq:conv-alpha-D}\\
& \| \hat{\alpha}_{1z} - \bar{\alpha}_{1z} \|_1 = O_p(1) \{ |S_{\bar\gamma_z}|+ |S_{\bar\alpha_z}| + |S_{\bar\alpha_{1z}}| \} \{\log(\me (q_1\vee q_2))/n\}^{1/2}, \label{eq:conv-alpha-DY}
\end{align}
where $(\bar{\gamma}_z, \bar{\alpha}_z,\bar{\alpha}_{1z} )$
are the target values defined as minimizers of the corresponding expected loss functions in Section~\ref{sec:methods}.
For simplicity, we do not discuss prediction $L_2$-norm error bounds for $(\hat\gamma_z,\hat\alpha_z,\hat\alpha_{1z})$, which are also involved in our rigorous analysis later.
While these results are built on existing high-dimensional, sparse analysis of Lasso penalized M-estimators (Buhlmann \& van de Geer 2011; Huang \& Zhang 2012; Tan 2020a),
additional arguments are needed to carefully handle the dependency of $\hat{\alpha}_z$ on $\hat{\gamma}_z$
and that of $\hat{\alpha}_{1z}$ on $(\hat{\gamma}_z, \hat{\alpha}_z)$,
hence the presence of $|S_{\bar\gamma_z}|$ in the bound for $\hat{\alpha}_z$ and  $|S_{\bar\gamma_z}|$ and $|S_{\bar\alpha_z}|$ in that
for $ \hat{\alpha}_{1z} $.

Second, for $z\in\{0,1\}$, the augmented IPW estimators of $E\{D(z)\}$ and $E\{ D(z) Y(1)\}$ involved in $\hat{\theta}_{1,\text{\tiny RCAL}}$ can be shown to admit
the following asymptotic expansions,
\begin{align}
\tilde{E} \left\{ \varphi_{\scriptscriptstyle D_z}(O;\hat{\pi}_z,\hat{m}_z ) \right\} & =
 \tilde{E} \left\{ \varphi_{\scriptscriptstyle D_z}(O;\bar{\pi}_z,\bar{m}_z ) \right\} + o_p(n^{-1/2}) , \label{eq:expansion-D}\\
\tilde{E} \left\{ \varphi_{\scriptscriptstyle D_z Y_{1z}}(O;\hat{\pi}_z,\hat{m}_z, \hat{m}_{1z}) \right\} & =
 \tilde{E} \left\{ \varphi_{\scriptscriptstyle D_z Y_{1z}}(O;\bar{\pi}_z,\bar{m}_z, \bar{m}_{1z}) \right\}  + o_p(n^{-1/2}) , \label{eq:expansion-DY},
\end{align}
where $(\bar{\pi}_z,\bar{m}_z, \bar{m}_{1z})$ are defined as in Section~\ref{sec:methods}.
From (\ref{eq:expansion-D})--(\ref{eq:expansion-DY}), the expansion (\ref{eq:expansion-RCAL}) for $\hat{\theta}_{1,\text{\tiny RCAL}}$ then follows by the delta method.
To show (\ref{eq:expansion-D}) for $z=1$, consider a Taylor expansion
\begin{align}
& \tilde{E}\{\varphi_{\scriptscriptstyle D_1}(O;\hat{\pi}_1,\hat{m}_1)\}
=\tilde{E}\{\varphi_{\scriptscriptstyle D_1}(O;\bar{\pi}_1,\bar{m}_1)\}
+ (\hat{\gamma}_1-\bar{\gamma}_1)^\T \delta_{\bar{\gamma}_1}+(\hat{\alpha}_1-\bar{\alpha}_1)^\T \delta_{\bar{\alpha}_1}+o_p(n^{-1/2}),
\label{eq:taylor-D}
\end{align}
where the remainder is taken to be $o_p(n^{-1/2})$ under suitable conditions, and
\begin{align*}
& \delta_{\bar{\gamma}_1} =\frac{\partial}{\partial \gamma}\tilde{E}\{\varphi_{\scriptscriptstyle D_1}(O;{\pi},{m}_1)\}\Bigr\rvert_{({\gamma},{\alpha}_1)
= (\bar{\gamma}_1,\bar{\alpha}_1)}
= - \tilde{E}\left\{Z\frac{1-\bar{\pi}_1}{\bar{\pi}_1}(D-\bar{m}_1 )f  \right\}, \\ 
& \delta_{\bar{\alpha}_1}
= \frac{\partial}{\partial {\alpha}_1}\tilde{E}\{\varphi_{\scriptscriptstyle D_1}(O;{\pi},{m}_1)\}\Bigr\rvert_{({\gamma},{\alpha}_1)=(\bar{\gamma}_1,\bar{\alpha}_1)}
= \tilde{E}\left\{\left(1-\frac{Z}{\bar{\pi}_1}\right) \bar{m}^\prime_1 g \right\}. 
\end{align*}
Here $\bar{m}^\prime_1(x) = \psi_{\scriptscriptstyle D}^\prime \{\bar{\alpha}^\T_1 g(x)\}$
and $\psi_{\scriptscriptstyle D}^\prime$ denotes the derivative of $\psi_{\scriptscriptstyle D}$.
A crucial point is that the expectations of the gradients $\delta_{\bar{\gamma}}$ and $\delta_{\bar{\alpha}_1}$ reduce to 0,
\begin{align}
- E \left\{Z\frac{1-\bar{\pi}_1}{\bar{\pi}_1}(D-\bar{m}_1 )f  \right\}  & = 0 , \label{eq:cond-alpha-D} \\
E\left\{\left(1-\frac{Z}{\bar{\pi}_1}\right) \bar{m}^\prime_1 g_1 \right\} & = 0, \label{eq:cond-gamma-D}
\end{align}
provided that model (\ref{eq:pi}) is correctly specified but model (\ref{eq:mD}) may be misspecified.
In fact, under correctly specified model (\ref{eq:pi}), $\bar{\pi}_1$ coincides with $\pi^*$ (Tan 2020a) and hence
condition (\ref{eq:cond-gamma-D}) holds.
Moreover, condition (\ref{eq:cond-alpha-D}) follows from the gradient equation for $\bar{\alpha}_1$
as a minimizer of the expected loss (\ref{eq:eloss-alpha-Dz}) for $z=1$,
because $f$ is a subvector of $g$ and
the gradient of (\ref{eq:eloss-alpha-Dz}) at $\bar{\alpha}_1$
matches the expectation of $\delta_{\bar{\alpha}_1}$ in (\ref{eq:cond-alpha-D}) with $f$ replaced by $g$:
\begin{align*}
& \frac{\partial}{\partial \alpha_1}
E \left(Zw_1(X; \bar\gamma_1)[-D\alpha^\T _1g(X)+\Psi_{\scriptscriptstyle D}\{\alpha^\T _1 g(X)\}] \right)
\Bigr\rvert_{\alpha_1 = \bar{\alpha}_1 } \\
& = - E \left[ Z\frac{1-\bar{\pi}_1}{\bar{\pi}_1} \{D- \psi_{\scriptscriptstyle D} ( \bar{\alpha}^\T_1 g)\} g  \right] . 
\end{align*}
From conditions (\ref{eq:cond-alpha-D})--(\ref{eq:cond-gamma-D}), the sum of the two terms
$(\hat{\gamma}_1-\bar{\gamma}_1)^\T \delta_{\bar{\gamma}_1}$ and
$(\hat{\alpha}_1-\bar{\alpha}_1)^\T \delta_{\bar{\alpha}_1}$
can be shown to be of order $\{|S_{\bar\gamma_1}| + |S_{\bar\alpha_1}| \} \{\log(\me q_1)/n\} $,
which becomes $o_p(n^{-1/2})$ and hence (\ref{eq:taylor-D}) leads to (\ref{eq:expansion-D}) for $z=1$ provided that $\{|S_{\bar\gamma_1}| + |S_{\bar\alpha_1}| \} \log(\me q_1) = o(n^{1/2})$.

Similarly, to show (\ref{eq:expansion-DY}) for $z=1$, consider a Taylor expansion
\begin{align}
\tilde{E}\{\varphi_{\scriptscriptstyle D_1Y_{11}}(O;\hat{\pi}_1,\hat{m}_1, \hat{m}_{11})\}&
=\tilde{E}\{\varphi_{\scriptscriptstyle D_1Y_{11}}(O;\bar{\pi}_1,\bar{m}_1, \bar{m}_{11})\}+(\hat{\gamma}_1-\bar{\gamma}_1)^\T \Delta_{\bar{\gamma}_1} \nonumber \\
&\quad +(\hat{\alpha}_1-\bar{\alpha}_1)^\T \Delta_{\bar{\alpha}_1}+(\hat{\alpha}_{11}-\bar{\alpha}_{11})^\T \Delta_{\bar{\alpha}_{11}}+o_p(n^{-1/2}), \label{eq:taylor-DY}
\end{align}
where the remainder is taken to be $o_p(n^{-1/2})$ under suitable conditions, and
\begin{align*}
& \Delta_{\bar{\gamma}} = - \tilde{E}\left\{Z\frac{1-\bar{\pi}_1}{\bar{\pi}_1}(DY - \bar{m}_1 \bar{m}_{11})f \right\}, \\ 
& \Delta_{\bar{\alpha}_1} =\tilde{E}\left\{\left(1-\frac{Z}{\bar{\pi}_1}\right) \bar{m}^\prime_1 \bar{m}_{11} g \right\}, \quad 
\Delta_{\bar{\alpha}_{11}} = \tilde{E}\left\{\left(1-\frac{Z}{\bar{\pi}_1}\right) \bar{m}_1 \bar{m}^\prime_{11} h \right\}. 
\end{align*}
Here $\bar{m}^\prime_{11}(x) = \psi_{\scriptscriptstyle Y}^\prime \{\bar{\alpha}^\T_{11} h(x)\}$
and $\psi_{\scriptscriptstyle Y}^\prime$ denotes the derivative of $\psi_{\scriptscriptstyle Y}$.
Under correctly specified model (\ref{eq:pi}),  $\bar{\pi}_1$ coincides with $\pi^*$ (Tan 2020a) and hence
the expectations of the gradients $\Delta_{\bar{\gamma}}$ and $\Delta_{\bar{\alpha}_1}$ reduce to 0,
\begin{align}
E \left\{\left(1-\frac{Z}{\bar{\pi}_1}\right) \bar{m}^\prime_1 \bar{m}_{11} g \right\}=0, \quad
E \left\{\left(1-\frac{Z}{\bar{\pi}_1}\right) \bar{m}_1 \bar{m}^\prime_{11} h \right\} =0. \label{eq:cond-alpha-Y}
\end{align}
Moreover, whether model (\ref{eq:mY}) is correctly specified or not, the expectation of the gradient $\Delta_{\bar{\alpha}_{11}}$ reduces to 0,
\begin{align}
E \left\{Z\frac{1-\bar{\pi}_1}{\bar{\pi}_1}(DY - \bar{m}_1 \bar{m}_{11})f \right\} = 0, \label{eq:cond-gamma-Y}
\end{align}
because  $f$ is a subvector of $h$, and
$\bar{\alpha}_{11}$ is defined as a minimizer of the expected loss (\ref{eq:eloss-alpha-Y1z}) for $z=1$ such that the following gradient is 0 at
$\bar{\alpha}_{11}$:
\begin{align*}
& \frac{\partial}{\partial \alpha_{11}}
E\left( Z w_1(X;\bar\gamma_1)[-DY \alpha^\T_{11}h(X)+ m_1(X; \bar{\alpha}_1)
\Psi_{\scriptscriptstyle Y}\{\alpha^\T _{11}h(X)\}]  \right)
\Bigr\rvert_{\alpha_{11} = \bar{\alpha}_{11} } \\
& = - E \left[ Z\frac{1-\bar{\pi}_1}{\bar{\pi}_1} \{DY - \bar{m}_1 \psi_{\scriptscriptstyle Y} (\bar{\alpha}^\T _{11} h) \} \right]. 
\end{align*}
From these mean-zero conditions on the gradients, the sum of three terms
$(\hat{\gamma}-\bar{\gamma})^\T \Delta_{\bar{\gamma}}$.
$(\hat{\alpha}_1-\bar{\alpha}_1)^\T \Delta_{\bar{\alpha}_1}$,
and $(\hat{\alpha}_{11}-\bar{\alpha}_{11})^\T \Delta_{\bar{\alpha}_{11}}$
can be shown to be of order $\{|S_{\bar\gamma_1}| + |S_{\bar\alpha_1}| + |S_{\bar\alpha_{11}}| \} \{\log(\me(q_1 \vee q_2))/n\} $,
which becomes $o_p(n^{-1/2})$ and hence (\ref{eq:taylor-DY}) leads to (\ref{eq:expansion-DY}) for $z=1$ provided that
$\{|S_{\bar\gamma_1}| + |S_{\bar\alpha_1}| + |S_{\bar\alpha_{11}}| \} \log\{ \me (q_1 \vee q_2 )\} = o(n^{1/2})$,
where $q_1 \vee q_2 = \max(q_1,q_2)$.

\begin{rem}[On likelihood vs calibrated estimation] \label{rem:ortho-cond}
We compare calibrated estimation with usual likelihood-based estimation.
From the preceding discussion, the mean-zero conditions (\ref{eq:cond-alpha-D})--(\ref{eq:cond-gamma-D}) and (\ref{eq:cond-alpha-Y})--(\ref{eq:cond-gamma-Y}) are
crucial for the desired expansions (\ref{eq:expansion-D})--(\ref{eq:expansion-DY}) to hold.
For example,
if the expectations of $\delta_{\bar{\gamma}_1}$ and $\delta_{\bar{\alpha}_1}$
were nonzero, then the two terms $(\hat{\gamma}_1-\bar{\gamma}_1)^\T \delta_{\bar{\gamma}_1}$ and
$(\hat{\alpha}_1-\bar{\alpha}_1)^\T \delta_{\bar{\alpha}_1}$ in (\ref{eq:taylor-D})
would be of order $\{\log(p)/n\}^{1/2}$ and $\{\log(q_1)/n\}^{1/2}$ in high-dimensional settings, as seen from (\ref{eq:conv-gamma})--(\ref{eq:conv-alpha-D}).
These mean-zero conditions can be satisfied in different manners.
If models (\ref{eq:pi}) and (\ref{eq:mD})--(\ref{eq:mY}) are correctly specified,
then (\ref{eq:cond-alpha-D})--(\ref{eq:cond-gamma-D}) and (\ref{eq:cond-alpha-Y})--(\ref{eq:cond-gamma-Y}) directly hold,
with the target values $(\bar{\gamma}_1, \bar{\alpha}_1, \bar{\alpha}_{11})$ identical to
the true values $(\gamma^*, \alpha^*_1, \alpha^*_{11})$.
This reasoning is applicable with $(\hat{\gamma}_z, \hat{\alpha}_z,\hat{\alpha}_{1z} )$
replaced by the regularized likelihood estimators
$(\hat{\gamma}_{\text{\tiny RML}}, \hat{\alpha}_{z,\text{\tiny RML}},\hat{\alpha}_{1z,\text{\tiny RML}} )$,
and would lead to asymptotic expansion (\ref{eq:expansion-RML}) for  $\hat{\theta}_{1,\text{\tiny RML}}$, as studied in Chernozhukov et al.~(2018).
In contrast, for our method,
while conditions (\ref{eq:cond-gamma-D}) and (\ref{eq:cond-gamma-Y}) are satisfied by relying on model (\ref{eq:pi}) being correctly specified,
conditions (\ref{eq:cond-alpha-D}) and (\ref{eq:cond-alpha-Y}) are achieved
with possible misspecification of models (\ref{eq:mD})--(\ref{eq:mY}), by carefully choosing (``calibrating'') the loss functions
for the estimators $\hat{\alpha}_{z,\text{\tiny RWL}}$ and $\hat{\alpha}_{1z,\text{\tiny RWL}}$.
Effectively, the loss function (\ref{eq:rcal-loss2b}) and (\ref{eq:rcal-loss3b}) for $\hat{\alpha}_{z,\text{\tiny RWL}}$ and $\hat{\alpha}_{1z,\text{\tiny RWL}}$
are derived by integrating with respect to $\alpha_z$ and $\alpha_{1z}$ respectively the gradients of
$\tilde{E} \left\{ \varphi_{\scriptscriptstyle D_z}(O;\pi(\cdot;\gamma), m_z(\cdot; \alpha_z )) \right\}$
and $ \tilde{E} \left\{ \varphi_{\scriptscriptstyle D_z Y_{1z}}(O;\pi(\cdot;\gamma), m_z(\cdot; \alpha_z) ,
m_{1z}(\cdot;\alpha_{1z})) \right\}$ in $\gamma$, in the case of $f=g=h$.
See Tan (2020b, Section 3.2)  for a related discussion.
\end{rem}

\begin{rem}[On low- vs high-dimensional estimation] \label{rem:low-high-dim}
Our preceding discussion is mainly concerned with high-dimensional settings where the numbers of regressors $p$, $q_1$, and $q_2$ are close to or larger than the sample size $n$.
The asymptotic expansions (\ref{eq:expansion-D})--(\ref{eq:expansion-DY}) from calibrated estimation are desirable in facilitating construction of confidence intervals,
because the first-order terms such as
$(\hat{\gamma}_1-\bar{\gamma}_1)^\T \delta_{\bar{\gamma}_1}$ and $(\hat{\alpha}_1-\bar{\alpha}_1)^\T \delta_{\bar{\alpha}_1}$ are made to be negligible up to order $n^{-1/2}$.
Otherwise, these first-order terms would be at least of order $\{\log(p)/n\}^{1/2}$ and difficult to quantify.
For completeness, it should also be noted that for previous methods studied in low-dimensional settings (Tan 2006a, Ogburn et al.~2015), valid confidence intervals can be obtained from
the more general asymptotic expansions (\ref{eq:taylor-D}) and (\ref{eq:taylor-DY}) along with usual influence functions
for likelihood-based or similar estimators $(\hat\gamma_1,\hat\alpha_1, \hat\alpha_{11}$), where the first-order terms are of order $n^{-1/2}$.
\end{rem}

\begin{rem}[On choices of model-assisted inference] \label{rem:model-assisted}
Our method allows model-assisted inference, relying on correct specification of instrument propensity score model (\ref{eq:pi}) but not
treatment and outcome regression models (\ref{eq:mD})--(\ref{eq:mY}).
Similar ideas can be employed to develop model-assisted inference relying on correct specification of models (\ref{eq:mD})--(\ref{eq:mY}) but not model (\ref{eq:pi}),
or doubly robust inference relying on correct specification of either model (\ref{eq:pi}) or models (\ref{eq:mD})--(\ref{eq:mY}).
In each case, additional modifications would be needed with increasing analytical and computational complexity.
For example, model (\ref{eq:pi}) needs to be properly expanded
such that regularized estimation of $\gamma$ could be designed to satisfy all three mean-zero conditions in (\ref{eq:cond-alpha-D}) and (\ref{eq:cond-alpha-Y}).
In contrast, by handling model-assisted inference based on model (\ref{eq:pi}), our method is developed in a practically convenient manner, involving sequential estimation
in the three models (\ref{eq:pi}), (\ref{eq:mD}), and (\ref{eq:mY}).
\end{rem}

\begin{rem}[On asymmetry between propensity score and outcome regression] \label{rem:asymmetry}
Our choice of inference based on instrument propensity score model (\ref{eq:pi}) is also related to a fundamental asymmetry
between propensity score and outcome regression approaches (Tan 2007).
As reflected by the estimator (\ref{eq:OR}), fitted treatment and outcome regression functions $\hat{m}_z(X)$ and $\hat{m}_{1z}(X)$
are desired to be valid for all observed covariates $X$, but such functions for those values of $X$ with $P(Z=z|X)$ close to 0
are effectively determined by extrapolation according to models (\ref{eq:mD})--(\ref{eq:mY}), which can be built and checked from only the truncated data $\{(D_i, Y_i, X_i): Z_i=z, i=1,\ldots,n\}$.
In contrast, model (\ref{eq:pi}) can be readily learned from $\{(Z_i,X_i): i=1,\ldots,n\}$ without data truncation.
\end{rem}

\begin{rem}[On calibrated estimation for propensity scores]
By a careful examination of the outline above, our theoretical analysis  also remains valid with $\hat{\gamma}_z$ replaced by
the regularized likelihood estimator $\hat{\gamma}_{\text{\tiny RML}}$.
In particular, the mean-zero conditions (\ref{eq:cond-gamma-D}) and (\ref{eq:cond-gamma-Y}) would still be satisfied
under correctly specified model (\ref{eq:pi}).
Nevertheless, we prefer the regularized likelihood estimator $\hat{\gamma}_{z,\text{\tiny RCAL}}$
for two additional reasons. One is the informative form of the KKT condition (\ref{eq:KKT-pi2}). The other is
an advantage of calibrated estimation in controlling relative errors of propensity scores for inverse probability weighting,
regardless of outcome regression, as studied in Tan (2020a).
\end{rem}

In the remainder of this section, we present formal results underlying Proposition~\ref{pro:RCAL}.
While convergence of the regularized calibrated estimators $(\hat\gamma_z, \hat\alpha_z)$ and the AIPW estimator
for $E\{D(z)\}$ can be obtained directly from Tan (2020b),
our analysis needs to carefully tackle convergence of $\hat\alpha_{1z}$ and the AIPW estimator for $E\{D(z) Y(1)\}$.
The situation is more complicated than in Tan (2020b), as well as the earlier literature on Lasso penalized $M$-estimation (Buhlmann \& van de Geer 2011; Huang \& Zhang 2012),
mainly because the loss function for $\hat \alpha_{1z}$, i.e.,
$L_{1z,\text{\tiny WL}}(\alpha_{1z};\hat{\gamma}_z, \hat{\alpha}_z)$ from (\ref{eq:rcal-loss3b}),
involves not only data-dependent weight $w_z(X;\hat\gamma_z)$ but also data-dependent mean function $\hat m_z(X ; \hat\alpha_z)$.
We extend a technical strategy in Tan (2020b) to control such dependency and establish
the desired convergence of $\hat\alpha_{1z}$ under similar conditions as required in unweighted Lasso penalized $M$-estimation in
high-dimensional settings. The error bounds obtained, however, depend on the sparsity sizes of the target values
$\bar\gamma_z$ and $\bar\alpha_z$, in addition to that of $\bar\alpha_{1z}$.

First, we summarize the results which can be deduced directly from Tan (2020b) about $(\hat\gamma_1, \hat\alpha_1)$ and
the AIPW estimator $ \tilde E \{  \varphi_{\scriptscriptstyle D_1}(O;\hat{\pi}_1,\hat{m}_1 ) \}$ for $E\{D(1)\}$.
Suppose that the Lasso tuning parameters are specified as $\lambda = A_0 \lambda_0$ for $\hat\gamma_1$ in (\ref{eq:rcal-loss1a}) and $\lambda=A_1 \lambda_1$ for $\hat\alpha_1$ in (\ref{eq:rcal-loss2a}),
where $\lambda_0 = \{ \log( (1+p)/\epsilon)/n\}^{1/2}$ and $\lambda_1 = \{ \log(  (1+q_1)/\epsilon)/n\}^{1/2} \, (\ge \lambda_0)$.
For a matrix $\Sigma$ with row indices $\{0,1,\ldots,k\}$,
a compatibility condition (Buhlmann \& van de Geer 2011) is said to hold with a subset $S \in \{0,1,\ldots,k\}$ and constants $\nu >0$ and $\mu>1$ if
$\nu^2  (\sum_{j\in S} |b_j|)^2 \le |S| ( b^\T \Sigma b )$
for any vector $b=(b_0,b_1,\ldots,b_k)\in \bbR^{1+k}$ satisfying $\sum_{j\not\in S} |b_j| \le \mu \sum_{j\in S} |b_j|$.

\begin{ass} \label{ass:gamma1}
Suppose that the following conditions are satisfied. \vspace{-.1in}
\begin{itemize} \addtolength{\itemsep}{-.1in}
\item[(i)] $\max_{j=0,1,\ldots,p} |f_j(X)| \le C_{f0}$ almost surely for a constant $C_{f0} \ge 1$.

\item[(ii)] $\bar\gamma_1^\T f(X)$ is bounded below by a constant $C_{f1}$ almost surely.

\item[(iii)] A compatibility condition holds for $\Sigma_f$ with the subset $S_{\bar\gamma_1} = \{0\} \cup \{j: (\bar\gamma_1)_j\not= 0, j=1,\ldots,p\}$
and some constants $\nu_0 >0$ and $\mu_0 >1$, where $\Sigma_f = E\{ Z w_1 (X;\bar\gamma_1) f(X) f^\T (X)\}$.

\item[(iv)] $| S_{\bar\gamma_1} | \lambda_0 $ is sufficiently small.
\end{itemize}
\end{ass}

\begin{ass} \label{ass:alpha1}
Suppose that the following conditions are satisfied. \vspace{-.1in}
\begin{itemize} \addtolength{\itemsep}{-.1in}
\item[(i)] $\max_{j=0,1,\ldots,p} |g_j(X)| \le C_{g0}$ almost surely for a constant $C_{g0} \ge 1$.

\item[(ii)] $\bar\alpha_1^\T g(X)$ is bounded in absolute values by $C_{g1} >0$ almost surely.

\item[(iii)] $\psi_{\scriptscriptstyle D}^\prime (u) \le \psi_{\scriptscriptstyle D}^\prime (\tilde u) \me^{C_{g2} |u-\tilde u|}$ for any $(u,\tilde u)$, where $C_{g2} >0$ is a constant.

\item[(iv)]  A compatibility condition holds for $\Sigma_g$ with the subset $S_{\bar\alpha_1} = \{0\} \cup \{j: (\bar\alpha_1)_j\not= 0, j=1,\ldots,p\}$,
and some constants $\nu_1 >0$ and $\mu_1 >1$, where $\Sigma_g = E\{ Z w_1 (X;\bar\gamma_1) g(X) g^\T (X)\}$.

\item[(v)] $| S_{\bar\gamma_1} | \lambda_0 + | S_{\bar\alpha_1} | \lambda_1$ is sufficiently small.
\end{itemize}
\end{ass}

\begin{thm}[Tan 2020b] \label{thm:gamma1-alpha1}
Suppose that Assumptions~\ref{ass:gamma1}--\ref{ass:alpha1} hold and $\lambda_0 \le 1$.
For sufficiently large constants $A_0$ and $A_1$, we have probability $1- c_0 \epsilon$,
\begin{align}
& \| \hat\gamma_1 - \bar\gamma_1 \|_1 \le M_0 |S_{\bar\gamma_1}| \lambda_0 , \quad
(\hat\gamma_1 - \bar\gamma_1)^\T \tilde \Sigma_f (\hat\gamma_1 - \bar\gamma_1) \le M_0 |S_{\bar\gamma_1}| \lambda_0^2,  \label{eq:bound-gamma1} \\
& \| \hat\alpha_1 - \bar\alpha_1 \|_1 \le M_1 (| S_{\bar\gamma_1} | \lambda_0 + | S_{\bar\alpha_1} | \lambda_1) , \quad \
(\hat\alpha_1 - \bar\alpha_1)^\T \tilde \Sigma_g (\hat\alpha_1 - \bar\alpha_1) \le M_1 (| S_{\bar\gamma_1} | \lambda_0^2 + | S_{\bar\alpha_1}| \lambda_1^2), \label{eq:bound-alpha1}
\end{align}
where $c_0$, $M_0$, and $M_1$ are positive constants and $\tilde \Sigma_f$ and $\tilde \Sigma_g$ are sample versions of
$\Sigma_f$ and $\Sigma_g$. i.e., $\tilde \Sigma_f = \tilde E\{ Z w_1 (X;\bar\gamma_1) f(X) f^\T (X)\}$ and $\tilde \Sigma_g = \tilde E\{ Z w_1 (X;\bar\gamma_1) g(X) g^\T (X)\}$.
Moreover, if model (\ref{eq:pi}) is correctly specified, then we also have with probability $1- c_0 \epsilon$,
\begin{align}
& \left| \tilde E \left\{ \varphi_{\scriptscriptstyle D_1}(O;\hat{\pi}_1,\hat{m}_1 ) \right\} -
 \tilde E \left\{ \varphi_{\scriptscriptstyle D_1}(O;\bar{\pi}_1,\bar{m}_1 ) \right\} \right|
\le M_{10} (|S_{\bar\gamma_1}| \lambda_0 + |S_{\bar\alpha_1}| \lambda_1 ) \lambda_1 ,  \label{eq:expansion-bound-D}\\
& \left|\tilde E\left[ \left\{  \varphi_{\scriptscriptstyle D_1}(O;\hat{\pi}_1,\hat{m}_1 )  -  \varphi_{\scriptscriptstyle D_1}(O;\bar{\pi}_1,\bar{m}_1 ) \right\}^2 \right] \right|
 \le M_{20} (|S_{\bar\gamma_1}| \lambda_0 + |S_{\bar\alpha_1}| \lambda_1 ) , \label{eq:var-est-D}
\end{align}
where $M_{10}$ and $M_{20}$ are positive constants.
\end{thm}

Inequalities (\ref{eq:bound-gamma1})--(\ref{eq:bound-alpha1}) lead directly to the desired convergence (\ref{eq:conv-gamma})--(\ref{eq:conv-alpha-D}) for $(\hat\gamma_1,\hat\alpha_1)$.
Moreover, inequality (\ref{eq:expansion-bound-D}) yields the asymptotic expansion (\ref{eq:expansion-D}) for the AIPW estimator $\hat E \{D(1)\} =
\tilde E \{ \varphi_{\scriptscriptstyle D_1}(O;\hat{\pi}_1,\hat{m}_1 ) \}$ provided
$\{ |S_{\bar\gamma_1}| + |S_{\bar\alpha_1}| \}  \log(\me q_1) = o(n^{1/2})$.
Inequality (\ref{eq:var-est-D}) can be used to show that the sample variance $\tilde E ( [\varphi_{\scriptscriptstyle D_1}(O;\hat{\pi}_1,\hat{m}_1) - \hat E\{D(1)\} ]^2 )$
is a consistent estimator for $\var\{ \varphi_{\scriptscriptstyle D_1}(O;\bar{\pi}_1,\bar{m}_1) \}$, provided $\{ |S_{\bar\gamma_1}| + |S_{\bar\alpha_1}| \}  \log^{1/2}(\me q_1) = o(n^{1/2})$,
which is satisfied under $\{ |S_{\bar\gamma_1}| + |S_{\bar\alpha_1}| \}  \log(\me q_1) = o(n^{1/2})$.
While consistent variance estimation is sufficient for justifying Wald confidence intervals for $E \{ D(1) \}$ by the Slutsky theorem,
the convergence rate in (\ref{eq:var-est-D}) can be improved under additional conditions. See Tan (2020b, Theorem 4).

Next, we discuss theoretical analysis of $\hat\alpha_{11}$, with the Lasso tuning parameter $\lambda = A_2 \lambda_2$ in (\ref{eq:rcal-loss3a}), where
$\lambda_2 = \{ \log((1+q_2) /\epsilon)/n\}^{1/2} \, (\ge \lambda_0)$.
As the loss $L_{11,\text{\tiny WL}}(\alpha_{11}; \hat\gamma_z, \hat{\alpha}_z)$ is convex in $\alpha_{11}$, the corresponding Bregman divergence is defined as
\begin{align*}
& D _{1z,\text{\tiny WL}}(\alpha^\prime_{1z}, \alpha_{1z}; \hat\gamma_z, \hat{\alpha}_z)  \\
& =L _{1z,\text{\tiny WL}}(\alpha^\prime_{1z}; \hat\gamma_z, \hat{\alpha}_z) -
L _{1z,\text{\tiny WL}}(\alpha_{1z}; \hat\gamma_z, \hat{\alpha}_z) -
(\alpha^\prime_{1z} - \alpha_{1z})^\T  \frac{\partial L _{1z,\text{\tiny WL}}}{\partial\alpha_{1z} }
(\alpha_{1z}; \hat\gamma_z, \hat{\alpha}_z) .
\end{align*}
The symmetrized Bregman divergence is easily shown to be
\begin{align}
& D ^\dag_{1z,\text{\tiny WL}}(\alpha^\prime_{1z}, \alpha_{1z}; \hat\gamma_z, \hat{\alpha}_z)  =
D _{1z,\text{\tiny WL}}(\alpha^\prime_{1z}, \alpha_{1z}; \hat\gamma_z, \hat{\alpha}_z) +
D _{1z,\text{\tiny WL}}(\alpha_{1z}, \alpha^\prime_{1z}; \hat\gamma_z, \hat{\alpha}_z) \nonumber \\
& = (\alpha^\prime_{1z} - \alpha_{1z})^\T
\tilde E\left[ \one\{Z=z\} w_z(X;\hat\gamma_z) m_z(X; \hat{\alpha}_z)
\{ \psi_{\scriptscriptstyle Y} (\alpha^{\prime\T} _{1z} h ) - \psi_{\scriptscriptstyle Y} (\alpha^\T _{1z} h) \} \right]. \label{eq:sym-bregman}
\end{align}
After statement of the assumptions required, Theorem~\ref{thm:alpha11} establishes the convergence of $\hat\gamma_2$ to the target value $\bar\gamma_2$ in the
both $L_1$ norm $\|\hat\alpha_{11}-\bar\alpha_{11}\|_1$ and the symmetrized Bregman divergence.

\begin{ass} \label{ass:alpha11}
Suppose that the following conditions are satisfied. \vspace{-.1in}
\begin{itemize} \addtolength{\itemsep}{-.1in}
\item[(i)] $\max_{j=0,1,\ldots,p} |h_j(X)| \le C_{h0}$ almost surely for a constant $C_{h0} \ge 1$.

\item[(ii)] $ DY - m_1(X; \bar{\alpha}_1) \psi_{\scriptscriptstyle Y} (\bar\alpha^\T _{11} h)$ is uniformly sub-gaussian given $X$ with parameters $(\sigma_0,\sigma_1)$.

\item[(iii)] $\bar\alpha_{11}^\T h(X)$ is bounded in absolute values by $C_{h1} >0$ almost surely.

\item[(iv)] $\psi_{\scriptscriptstyle Y}^\prime (u) \le \psi_{\scriptscriptstyle Y}^\prime (\tilde u) \me^{C_{h2} |u-\tilde u|}$ for any $(u,\tilde u)$, where $C_{h2} >0$ is a constant.

\item[(v)] A compatibility condition holds for $\Sigma_h$ with the subset $S_{\bar\alpha_{11}} = \{0\} \cup \{j: (\bar\alpha_{11})_j\not= 0, j=1,\ldots,p\}$,
and some constants $\nu_2 >0$ and $\mu_2 >1$,  where $\Sigma_h = E\{ Z w_1 (X;\bar \gamma_1) h(X) h^\T (X)\}$.

\item[(vi)] $| S_{\bar\gamma_1} | \lambda_0 + | S_{\bar\alpha_1} | \lambda_1 + | S_{\bar\alpha_{11}} | \lambda_2 $ is sufficiently small:
$| S_{\bar\gamma_1} | \lambda_0  \le \varrho_0$, $| S_{\bar\gamma_1} | \lambda_0 + | S_{\bar\alpha_1} | \lambda_1  \le \varrho_1$,
and $| S_{\bar\alpha_{11}} | \lambda_2  \le \varrho_2$, such that
$\varrho_3=\nu_2^{-2} B_{h1} (1+\mu_2)^2 \varrho_2 < 1$,
$\varrho_4 = M_1 \varrho_1 C_{g0} C_{g2} \me^{C_{g2} C_{g0} M_1 \varrho_1} <1$
$\varrho_5 = C_{h2} C_{h0} (A_2-B_1)^{-1} \mu_{22}^2 \nu_{21}^{-2} C_{g3}^{-1} C_{h3}^{-1} \varrho_2 < 1$, and
$\varrho_6 = C_{h2} C_{h0} (A_2-B_1)^{-1}\mu_{21}^{-2} C_{g3}^{-1} C_{h3}^{-1} $ $\times (M_{01} \varrho_0 + M_{11}\varrho_1) < 1$,
where
$\mu_{21} = 1-2A_2 /\{ (\mu_2+1)(A_2-B_1)\} \in (0,1]$, $\mu_{22}= (\mu_2+1)(A_2-B_1)$,
$\nu_{21} = \nu_2 (1-\varrho_3)^{1/2}$,
$(B_1, B_{h1})$ are from Lemmas~\ref{lem:prob-grad}--\ref{lem:prob-mat-h},
$(M_{01}, M_{11})$ are from Lemma~\ref{lem:remove-hat},
and $(C_{g3}, C_{h3})$ are from Lemma~\ref{lem:local-quad} in the Supplement.
\end{itemize}
\end{ass}

\begin{thm} \label{thm:alpha11}
In the setting of Theorem~\ref{thm:gamma1-alpha1}, suppose that Assumption \ref{ass:alpha11} also holds.
Then for $\lambda_2 = A_2 \lambda_2$ and $A_2 > B_1 (\mu_2+1)/(\mu_2 -1)$, we have with probability $1- (c_0+8)\epsilon$,
\begin{align}
& D^\dag_{11,\text{\tiny WL}} (  \hat \alpha_{11} , \bar \alpha_{11};  \bar \gamma_1, \bar\alpha_1) + (A_2 - B_1) \lambda_2 \|\hat\alpha_{11} - \bar\alpha_{11} \|_1 \nonumber \\
& \le C_{g3}^{-1} C_{h3}^{-1} \left\{ 2 \mu_{21}^{-2} (M_{01}+M_{11}) |S_{\bar\gamma_1}| \lambda_0^2 + 2 \mu_{21}^{-2} M_{11} |S_{\bar\alpha_1}| \lambda_1^2 +
\mu_{22}^2 \nu_{21}^{-2}  |S_{\bar\alpha_{11}}| \lambda_2^2 \right\} , \label{eq:bound-alpha11}
\end{align}
where $(\mu_{21}, \mu_{22}, \nu_{21}, B_1, M_{01}, M_{11}, C_{g3}, C_{h3})$ are defined in Assumption~\ref{ass:alpha11}.
\end{thm}

From the proof, an upper bound can also be obtained with probability $1- (8+c_0) \epsilon$ on the weighted prediction $L_2$ norm (in the scale of linear predictors),
\begin{align}
& \tilde E \{ Z w_1(X;\bar\gamma_1) (\hat\alpha_{11}^\T h - \bar\alpha_{11}^\T h)^2 \} = (\hat\alpha_{11} - \bar\alpha_{11})^\T \tilde \Sigma_h (\hat\alpha_{11} - \bar\alpha_{11}) \nonumber \\
& \le C_{g3}^{-1} C_{h3}^{-1} (1-\varrho_4 \vee \varrho_5)^{-1} D^\dag_{11,\text{\tiny WL}} (  \hat \alpha_{11} , \bar \alpha_{11};  \bar \gamma_1, \bar\alpha_1) , \label{eq:bound-alpha11-quad}
\end{align}
where $\tilde \Sigma_h$ is the sample version of $\Sigma_h$, i.e., $\tilde \Sigma_h = \tilde E\{ Z w_1 (X;\bar \gamma_1) h(X) h^\T (X)\}$.
For notational simplicity of subsequent discussion, let $M_2$ be a constant such that the right-hand side of (\ref{eq:bound-alpha11}) is upper bounded by
$\min\{ C_{g3} C_{h3} (1-\varrho_4 \vee \varrho_5) , (A_2 - B_1)^{-1} \} \times
M_2 (| S_{\bar\gamma_1} | \lambda_0^2 + | S_{\bar\alpha_1} | \lambda_1^2 + | S_{\bar\alpha_{11}} | \lambda_2^2 )$. Then
we have with probability $1 - (8+c_0)\epsilon$,
\begin{align*}
& \| \hat\alpha_{11} - \bar\alpha_{11} \|_{11} \le M_2 (| S_{\bar\gamma_1} | \lambda_0 + | S_{\bar\alpha_1} | \lambda_1 + | S_{\bar\alpha_{11}} | \lambda_2 ) ,\\
& (\hat\alpha_{11} - \bar\alpha_{11})^\T \tilde \Sigma_h (\hat\alpha_{11} - \bar\alpha_{11}) \le M_2 (| S_{\bar\gamma_1} | \lambda_0^2 + | S_{\bar\alpha_1} | \lambda_1^2 + | S_{\bar\alpha_{11}} | \lambda_2^2 ),
\end{align*}
which lead to the desired convergence (\ref{eq:conv-alpha-DY}).
With the preceding results for $(\hat\gamma_1,\hat\alpha_1,\hat\alpha_{11})$, Theorem~\ref{thm:expansion-DY} provides
an error bound for the AIPW estimator for $E \{ D(1) Y(1)\}$.

\begin{thm} \label{thm:expansion-DY}
In the setting of Theorem~\ref{thm:alpha11}, if model (\ref{eq:pi}) is correctly specified, then we have with probability $1-(14+c_0)\epsilon$, \vspace{-.1in}
\begin{align}
& \left| \tilde E \left\{ \varphi_{\scriptscriptstyle D_1Y_{11}}(O;\hat{\pi}_1,\hat{m}_1, \hat{m}_{11} ) \right\} -
\tilde E \left\{ \varphi_{\scriptscriptstyle D_1Y_{11}}(O;\bar{\pi}_1,\bar{m}_1, \bar{m}_{11} ) \right\} \right| \nonumber \\
& \le M_{30} |S_{\bar\gamma_1}| \lambda_0(\lambda_1\vee\lambda_2) + M_{31} |S_{\bar\alpha_1}| \lambda_1 (\lambda_1\vee\lambda_2) + M_{32} |S_{\bar\alpha_{11}}| \lambda_2^2 , \label{eq:expansion-bound-DY}
\end{align}
where $(M_{30},M_{31},M_{32})$ are positive constants from Lemma~\ref{lem:expansion-DY}.
In addition, we have with probability $1-(8+c_0)\epsilon$, \vspace{-.1in}
\begin{align}
& \left| \tilde E \left[ \left\{ \varphi_{\scriptscriptstyle D_1Y_{11}}(O;\hat{\pi}_1,\hat{m}_1, \hat{m}_{11} ) -
 \varphi_{\scriptscriptstyle D_1Y_{11}}(O;\bar{\pi}_1,\bar{m}_1, \bar{m}_{11} ) \right\}^2 \right] \right| \nonumber \\
& \le M_{40} (|S_{\bar\gamma_1}| \lambda_0^2 + |S_{\bar\alpha_1}| \lambda_1^2 + |S_{\bar\alpha_{11}}| \lambda_2^2) +
M_{41} (|S_{\bar\gamma_1}| \lambda_0 + |S_{\bar\alpha_1}| \lambda_1 + |S_{\bar\alpha_{11}}| \lambda_2)^2 , \label{eq:var-est-DY}
\end{align}
where $(M_{40}, M_{41})$ are positive constants from Lemma~\ref{lem:var-est-DY}.
\end{thm}

Inequality (\ref{eq:expansion-bound-DY}) yields the asymptotic expansion (\ref{eq:expansion-DY}) for the AIPW estimator $\hat E \{D(1)Y(1)\} =
\tilde E \{\varphi_{\scriptscriptstyle D_1Y_{11}}(O;\hat{\pi}_1,\hat{m}_1, \hat{m}_{11} ) \}$ provided
$\{ |S_{\bar\gamma_1}| + |S_{\bar\alpha_1}| + |S_{\bar\alpha_{11}}| \}  \log\{\me( q_1\vee q_2)\} = o(n^{1/2})$.
Inequality (\ref{eq:var-est-DY}) can be used to show that the sample variance $\tilde E ( [\varphi_{\scriptscriptstyle D_1Y_{11}}(O;\hat{\pi}_1,\hat{m}_1, \hat{m}_{11} ) - \hat E\{D(1)Y(1)\} ]^2 )$
is a consistent estimator for $\var\{ \varphi_{\scriptscriptstyle D_1Y_{11}}(O;\bar{\pi}_1,\bar{m}_1, \bar{m}_{11} )\}$,
provided $\{ |S_{\bar\gamma_1}| + |S_{\bar\alpha_1}| + |S_{\bar\alpha_{11}}| \}  \log^{1/2}\{\me( q_1\vee q_2)\} = o(n^{1/2})$,
which is satisfied under $\{ |S_{\bar\gamma_1}| + |S_{\bar\alpha_1}| + |S_{\bar\alpha_{11}}| \}  \log\{\me( q_1\vee q_2)\} = o(n^{1/2})$.

Our theoretical analysis above deals with convergence of $(\hat\gamma_1,\hat\alpha_1,\hat\alpha_{11})$ and
the asymptotic expansions of the AIPW estimators for $E\{D(1)\}$ and $E\{ D(1)Y(1)\}$, i.e., (\ref{eq:expansion-D}) and (\ref{eq:expansion-DY}) with $z=1$.
Similar results can be obtained as Theorems~\ref{thm:gamma1-alpha1}--\ref{thm:expansion-DY} with $Z$ replaced by $1-Z$ and
$(\hat\gamma_1,\hat\alpha_1,\hat\alpha_{11})$ replaced by $(-\hat\gamma_0, \hat\alpha_0,\hat\alpha_{10})$,
provided that Assumptions~\ref{ass:gamma1}--\ref{ass:alpha11} are modified accordingly.
Combining both results for $z=1$ and $z=0$ leads to Proposition~\ref{pro:RCAL} by standard arguments.

\section{Simulation studies} \label{sec:simulation}

We present simulation studies to compare pointwise properties of $\hat\theta_{1,\text{\tiny RML}}$ based on regularized likelihood estimation without or with post-Lasso refitting
and $\hat\theta_{1,\text{\tiny RCAL}}$ based on regularized calibrated estimation
and coverage properties of the associated confidence intervals.
In addition, motivated by Remark~\ref{rem:randomized-iv}, we compare these methods with the IPW (i.e.~Wald) estimator $\hat{\theta}_{1,\text{\tiny IPW}}(\hat{\pi}_0)$, with $\hat{\pi}_0=\tilde{E}(Z)$,
in the setting where the instrument is assumed to be completely randomized.

\subsection{Implementation details}

Both the regularized likelihood and calibrated methods are implemented using the R package \texttt{RCAL}  (Tan 2020a). The penalized versions of loss functions (\ref{eq:loss-ps-ml}), (\ref{eq:ML-lossD}) and (\ref{eq:ML-lossY}) for computing $\hat{\gamma}_{\text{\tiny RML}}$, $\hat{\alpha}_{z,\text{\tiny RML}}$ and $\hat{\alpha}_{ {1z},\text{\tiny RML}}$, or penalized loss functions (\ref{eq:rcal-loss1a}), (\ref{eq:rcal-loss2a}) and (\ref{eq:rcal-loss3a})  for computing $\hat{\gamma}_{z,\text{\tiny RCAL}}$, $\hat{\alpha}_{z,\text{\tiny RWL}}$ and $\hat{\alpha}_{ {1z},\text{\tiny RWL}}$, are minimized for fixed tuning parameters $\lambda$ using algorithms similar to those in Friedman et al.~(2010),
but with the coordinate descent method replaced by an active set method
as in Osborne et al.~(2000) for solving each Lasso penalized least squares
problem.
All variables in $f({X})$, $g({X})$ and $h({X})$ are standardized to have sample means 0 and variances 1.

We determine the value of the Lasso tuning parameter $\lambda$ using 5-fold cross validation based on the corresponding loss function. Let $(\mathcal{I}_k)^5_{k=1}$ be a $5$-fold random partition of
the observation indices $\{1, 2, . . . , n\}$. For a loss function $L(\gamma)$, either the average negative log-likelihood $L_{ \text{\tiny ML}}(\gamma)$ in (\ref{eq:loss-ps-ml}), or the calibration loss $L_{z,\text{\tiny CAL}}(\gamma)$ in (\ref{eq:rcal-loss1b})--(\ref{eq:rcal-loss1c}) for $z=0,1$, denote by $L(\gamma;\mathcal{I})$ the loss function obtained when the sample average $\tilde{E}()$ is computed over only the subsample indexed by $\mathcal{I}$. The 5-fold cross-validation criterion is defined as $\text{CV}_5(\lambda)=(1/5)\sum^5_{k=1} L(\hat{\gamma}_{\lambda}^{(k)}; \mathcal{I}_k)$, where $\hat{\gamma}_{\lambda}^{(k)}$ is a minimizer of the penalized loss $L(\gamma; \mathcal{I}^c_k)+\lambda\|\gamma_{1:p}\|_1$ over the subsample of size $4n/5$ indexed by $\mathcal{I}^c_k$, the complement of $\mathcal{I}_k$. Then $\lambda$ is selected by minimizing $\text{CV}_5(\lambda)$ over the discrete set $\{\lambda^{\ast}/2^j: j=0,1,...,10\}$, where for $\hat{\pi}_0=\tilde{E}(Z)$, the value $\lambda^{\ast}$ is computed as either $\lambda^{\ast}=\max_{j=1,...,p} |\tilde{E}\{(T-\hat{\pi}_0)f_j({X})\}|$ when the likelihood loss (\ref{eq:loss-ps-ml}) is used, or $\lambda^{\ast}=\max_{j=1,...,p} |\tilde{E}\{[(1-Z)/(1-\hat{\pi}_0)-1]f_j({X})\}|$ or $\max_{j=1,...,p} |\tilde{E}\{(Z/\hat{\pi}_0-1)f_j({X})\}|$ when calibration loss (\ref{eq:rcal-loss1b}) or (\ref{eq:rcal-loss1c}) is used respectively. It can be shown that in each case, the penalized loss $L(\gamma)+\lambda\|\gamma_{1:p}\|_1$ over the original sample of size $n$ has a minimum at $\gamma_{1:p}=0$ for all $\lambda\geq \lambda^{\ast}$.

The computation of $(\hat{\alpha}_{z, \text{\tiny RML}},\hat{\alpha}_{{1z},\text{\tiny RML}})$ or $(\hat{\alpha}_{z,\text{\tiny RWL}},\hat{\alpha}_{{1z}, \text{\tiny RWL}})$ proceeds similarly as above. In the latter case, cross-validation based on $L_{z,\text{\tiny WL}}(\alpha_{z};\hat{\gamma}_{z})$ is performed with $\hat{\gamma}_{z}$ held at the fixed value $\hat{\gamma}_{z,\text{\tiny RCAL}}$ obtained in the prior step, and cross-validation based on $L_{1z, \text{\tiny WL}}(\alpha_{{1z}};\hat{\gamma}_{z},\hat{\alpha}_{z})$ is performed with $(\hat{\gamma}_{z},\hat{\alpha}_{z})$ held at the fixed values $(\hat{\gamma}_{z,\text{\tiny RCAL}},\hat{\alpha}_{z,\text{\tiny RWL}})$ in the prior steps.

\subsection{Conditionally randomized instrument}
Let ${X}=(X_1,...,X_p)$ be independent variables where each $X_j$ is $N(0,1)$ truncated to the interval $(-2.5,2.5)$, and then standardized to have mean 0 and variance 1. Consider the transformed variables $W_1=\exp(0.5X_1)$, $W_2=10+\{1+\exp(X_1)\}^{-1}X_2$, $W_3=(0.04X_1X_3+0.6)^3$ and $W_4=(X_2+X_4+20)^{2}$. Let ${X}^{\dag}=(X^{\dag}_1,...,X^{\dag}_p)$, where $X^{\dag}_j=\{W_j-E(W_j)\}/\sqrt{\mbox{Var}(W_j)}$ for $j=1,...,4$, and $X^{\dag}_j=X_j$ for $5 \leq j\leq p$. This setup follows that in the preprint of Tan (2020b) and ensures strict one-to-one mapping between $X$ and $X^{\dag}$. Figure \ref{fig:scatterplot} in
 the Supplement shows the scatter plots  from a simulated data sample of the variables $(X^{\dag}_1, X^{\dag}_2,X^{\dag}_3,X^{\dag}_4)$, which are correlated with each other as would be found in real data. Consider the following data-generating configurations:
\begin{itemize}\addtolength{\itemsep}{-.1in}
\item[(C1)] Generate $Z$ given ${X}$ from a Bernoulli distribution with $P(Z=1|X)=\{1+\exp(-X^{\dag}_1+0.5X^{\dag}_2-0.25X^{\dag}_3-0.1X^{\dag}_4)\}^{-1}$.
Then, independently, generate $U$ from a standard Logistic distribution,  $D=\one(1-2.5Z+0.25X^{\dag}_1+X^{\dag}_2+0.5X^{\dag}_3-1.5X^{\dag}_4\geq U)$ and $Y(1)$ from a Normal distribution with variance 1 and
mean $E\{Y(1)|Z,X,U\}=0.5X^{\dag}_1+X^{\dag}_2+X^{\dag}_3+X^{\dag}_4+2U$.
\item[(C2)] Generate $(Z,U)$ as in (C1), but generate $D=\one(1-2.5Z+0.25X_1+X_2+0.5X_3-1.5X_4\geq U)$ and $Y(1)$ from a Normal distribution with variance 1 and
mean $E\{Y(1)|Z,X,U\}=0.5X_1+X_2+X_3+X_4+2U$.
\item[(C3)] Generate $Z$ given ${X}$ from a Bernoulli distribution with $P(Z=1|X)=\{1+\exp(-X_1+0.5X_2-0.25X_3-0.1X_4)\}^{-1}$, then generate $(U,D,Y(1))$ as in (C1).
\end{itemize}
Set $Y=Y(1)$ if $D=1$. The observed data consist of independent and identically distributed copies $\{(Y_iD_i, D_i,Z_i,X_i):i=1,...,n\}$. Consider the following model specifications:
\begin{itemize} \addtolength{\itemsep}{-.1in}
\item[(M1)]  Logistic instrument propensity score model (6), logistic $D$-outcome model (9) and linear $Y$-outcome model (10) with $f_j(X)=g_j(X)=h_j(X)=X_j^{\dag}$ for $j=1,...,p$.

\item[(M2)]  Logistic instrument propensity score model (6) and logistic $D$-outcome model (9)  with $f_j(X)=g_j(X)=X_j^{\dag}$ for $j=1,...,p$, and linear $Y$-outcome model (10) with
$h_j(X)=X_j^{\dag}$, $j=1,...,p$,
$h_{p+1}(X)=\hat{m}_{\scriptscriptstyle D_0}(X)$, $ h_{p+1+k}(X)=\{\hat{m}_{\scriptscriptstyle D_0}(X)-\xi_{0k}\}_{+}$,
$h_{p+4}(X)=\hat{m}_{\scriptscriptstyle D_1}(X)$, and $h_{p+4+k}(X)=\{\hat{m}_{\scriptscriptstyle D_1}(X)-\xi_{1k}\}_{+}$, $1\leq k \leq 3$,
where $\xi_{zk}$ is the $k^{\text{th}}$ quartile of the fitted values $\hat{m}_{\scriptscriptstyle D_z}(X)$ and $c_{+}=\max(0,c)$.
\end{itemize}
For (M2),  linear spline bases $(\hat{m}_{\scriptscriptstyle D_1},\hat{m}_{\scriptscriptstyle D_0})$ are included as additional functions in $h(X)$.
As discussed in Remark~\ref{rem:model-Y}, the dependence of $m^{\ast}_{\scriptscriptstyle Y_{dz}}(X)$ on $m^{\ast}_{\scriptscriptstyle D_{z}}(X)$ is in general unknown.
A simple strategy is then to incorporate splines bases in $(\hat{m}_{\scriptscriptstyle D_1},\hat{m}_{\scriptscriptstyle D_0})$ to enlarge $Y$-outcome model.

\begin{table}
\caption{Summary of results for estimation of $\theta_1$ with a conditionally randomized instrument.} \label{table1}  \vspace{-.1in}
\footnotesize
\begin{center}
\begin{tabular*}{1\textwidth}{@{\extracolsep\fill} c c ll c ll c ll} \hline
			\toprule
              &\multicolumn{3}{c}{(C1) cor IPS, more cor OR} & \multicolumn{3}{c}{(C2) cor IPS, less cor OR} & \multicolumn{3}{c}{(C3) mis IPS, more cor OR}\\
              \cmidrule(lr){2-4}\cmidrule(lr){5-7}\cmidrule(lr){8-10}
	    & RCAL & RML & RML2&RCAL & RML & RML2&RCAL & RML & RML2\\
	     \hline \rule{0pt}{3ex}
&  \multicolumn{9}{c}{ (M1) $n=800,p=400$}\\
Bias & $-.146$ & $-.195$ & $-.007$& $-.054$ & $-.208$ & $-.118$ & $.043$ & $-.022$ & $.119$\\
$\sqrt{\text{Var}}$ & .433 & .435 & .896 & .518 & .521 & 1.231  & .429 & .434 & .804  \\
$\sqrt{\text{EVar}}$ &.418 & .400 & 2.533 & .510 & .486 & 11.410 & .418 & .422 & .957\\
Cov90 & .854 & .811 & .868 & .889 & .830 & .848& .886 & .876 & .885  \\
Cov95 & .908 & .889 & .930 & .935 & .897 & .898& .932 & .933 & .939\\ \rule{0pt}{3ex}

&  \multicolumn{9}{c}{  (M2) $n=800,p=400$}\\
Bias & $-.155$ & $-.201$ & $-.026$& $-.056$ & $-.210$ & $-.143$& $.025$ & $-.026$ & $.120$ \\
$\sqrt{\text{Var}}$ & .432 & .438 & .752 & .522 & .521 & 1.301& .427 & .433 & .784  \\
$\sqrt{\text{EVar}}$ &.415 & .401 & 1.290 &.509 & .486 & 11.804 & .415 & .422 & .790\\
Cov90 & .848 & .803 & .872 & .884 & .832 & .851 & .884 & .877 & .883 \\
Cov95 & .909 & .882 & .928 & .933 & .895 & .894& .937 & .929 & .941\\\rule{0pt}{3ex}
&  \multicolumn{9}{c}{ (M1) $n=800,p=1000$}\\
Bias & $-.198$ & $-.239$ & $-.145$& $-.087$ & $-.227$ & $-.204$ & $.047$ & $-.028$ & $.092$\\
$\sqrt{\text{Var}}$ & .428 & .424 & .632 & .518 & .521 & .749  & .438 & .451 & .961  \\
$\sqrt{\text{EVar}}$ &.411 & .393 & .600 & .493 & .477 & .742 & .411 & .413 & .816\\
Cov90 & .837 & .808 & .832 & .879 & .831 & .833& .882 & .864 & .857  \\
Cov95 & .900 & .869 & .901 & .933 & .888 & .899& .945 & .924 & .920\\ \rule{0pt}{3ex}

&  \multicolumn{9}{c}{  (M2) $n=800,p=1000$}\\
Bias & $-.223$ & $-.242$ & $-.146$& $-.099$ & $-.227$ & $-.205$& $.011$ & $-.030$ & $.068$ \\
$\sqrt{\text{Var}}$ & .430 & .424 & .632 & .515 & .520 & .759& .441 & .450 & .714  \\
$\sqrt{\text{EVar}}$ &.407 & .393 & .605 & .491 & .476 & .744 & .407 & .412 & .677\\
Cov90 & .820 & .803 & .831 & .874 & .824 & .828 & .875 & .862 & .858 \\
Cov95 & .880 & .868 & .900 & .929 & .888 & .894& .939 & .928 & .917\\\hline
\end{tabular*}\\[.1in]
\parbox{1\textwidth}{\small Note: RCAL denotes $\hat\theta_{1,\text{\tiny RCAL}}$, RML denotes $\hat\theta_{1,\text{\tiny RML}}$ and RML2 denotes the variant where the nuisance parameters are estimated by refitting models with only the variables selected from the corresponding Lasso estimation.
Bias and $\sqrt{\text{Var}}$ are the Monte Carlo bias and standard deviation of the points estimates, $\sqrt{\text{EVar}}$ is the
square root of the mean of the variance estimates, and Cov90 or Cov95 is the coverage proportion of
the 90\% or 95\% confidence intervals, based on 1000 repeated simulations. The true values of $\theta_1$ under (C1)--(C3) are calculated
using Monte Carlo integration with 100 repeated samples each of size $10^7$.}
\end{center}  \vspace{-.1in}
\end{table}

The instrument propensity score (IPS) model is correct in configurations (C1) and (C2) but misspecified in (C3).
The outcome regression (OR) $D$-model is correct in configurations (C1) and (C3), but misspecified in (C2),
while the $Y$-model in either (M1) or (M2) is misspecified in all configurations (C1)--(C3), but
it can be regarded as being ``closer" to the truth in (C1) and (C3) than in (C2) due to using $X^{\dag}$ instead of $X$ as regressors.
Therefore the models in both (M1) and (M2) can be classified as follows in configurations (C1)--(C3):
\begin{itemize} \addtolength{\itemsep}{-.1in}
\item[(C1)] IPS model correctly specified,  OR models ``more correctly'' specified;
\item[(C2)] IPS model correctly specified,  OR models ``less correctly'' specified;
\item[(C3)] IPS model misspecified,  OR models ``more correctly'' specified.
\end{itemize}

Similarly as in Kang \& Schafer (2007) for $p=4$, the OR $D$- and $Y$-models in case (C2) and IPS model in (C3) appear adequate by standard diagnosis techniques.
See Figures \ref{fig:scatter-c1}--\ref{fig:scatter-c3} in the Supplement for  scatterplots of $Y$ against $X^{\dag}_j$ within $\{D = 1\}$,
boxplots of $X^{\dag}_j$ within $\{D = 0\}$ and $\{D = 1\}$ as well as boxplots of $X^{\dag}_j$ within $\{Z = 0\}$ and $\{Z = 1\}$ for $j =1,...,4$.

For $n = 800$ and $p = 400$ or $1000$, Table \ref{table1} summarizes the results  based on 1000 repeated simulations. The methods RCAL and RML perform similarly to each other in terms of absolute bias, variance and coverage in (C1) and (C3), but RCAL yields noticeably smaller absolute  biases and better coverage than {RML} and RML2 in (C2).
The post-Lasso refitting method {RML2} appears to achieve coverages closer to the nominal probabilities in (C1), but yield substantially higher variances in all three cases (C1)--(C3).
These properties can also be seen from the QQ-plots of the estimates and t-statistics in Figures \ref{fig:qq-n800-p400-m1}--\ref{fig:qq-n800-p1000-m2} in the Supplement.
The performances of each of the three methods are similar with models (M1) or (M2) specified. Hence in the settings studied, there is little benefit in adding the spline terms in the outcome $Y$-model.

\subsection{Completely randomized instrument}
We generate data under the following configurations  with a completely randomized instrument:
\begin{itemize} \addtolength{\itemsep}{-.1in}
\item[(C4)] Generate $Z$ from a Bernoulli distribution with $P(Z=1)=0.5$, and, independently, generate $U$ from a standard Logistic distribution. Then, generate $D=\one(1-2.5Z+0.25X^{\dag}_1+X^{\dag}_2+0.5X^{\dag}_3-1.5X^{\dag}_4\geq U)$ and $Y(1)$ from a Normal distribution with variance 1 and
mean $E\{Y(1)|Z,X,U\}=0.5X^{\dag}_1+X^{\dag}_2+X^{\dag}_3+X^{\dag}_4+2U$.
\item[(C5)] Generate $(Z,U)$ as in (C4), but generate $D=\one(1-2.5Z+0.25X_1+X_2+0.5X_3-1.5X_4\geq U)$ and $Y(1)$ from a Normal distribution with variance 1 and
mean $E\{Y(1)|Z,X,U\}=0.5X_1+X_2+X_3+X_4+2U$.
\end{itemize}

For $n = 800$ and $p=1000$, Table \ref{table2}  summarizes the results  based on 1000 repeated simulations. See the Supplement for $p=400$ results. The methods RCAL, RML and IPW yield small bias and adequate coverage proportions in (C4) and (C5). The refitting method {RML2} also yields small bias, but coverage proportions noticeably below the nominal probabilities.
The $L_2$ average interval lengths, $\sqrt{\text{EVar}}$, from RCAL and RML are comparable,
and are $\approx 9\%$ shorter than those of IPW in (C4), and $\approx 6\%$ shorter in (C5). Such efficiency gains are comparable to those reported in previous simulation studies dealing with the average treatment effect, e.g. the interval lengths of Lasso-adjusted methods are $\approx 10\%$ shorter than those of the unadjusted difference-of-means estimator in Bloniarz et al.~(2016).
In instrumental variable analysis, such variance reduction is particularly helpful because the Wald estimator usually suffers from large standard errors.

\begin{table}[b!]
\caption{Summary of results for estimation of $\theta_1$ with a completely randomized instrument.} \label{table2}  \vspace{-.1in}
\footnotesize
\begin{center}
\def\arraystretch{.9}
\begin{tabular*}{0.8\textwidth}{@{\extracolsep\fill}c c ccl c ccl} \hline
			\toprule
              &\multicolumn{4}{c}{(C4) cor IPS, more cor OR} & \multicolumn{4}{c}{(C5) cor IPS, less cor OR}\\
              	     \cmidrule(lr){2-5}\cmidrule(lr){6-9}
	  & IPW & RCAL & RML  & RML2 & IPW & RCAL & RML  & RML2\\
	     \hline\rule{0pt}{3ex}
&  \multicolumn{8}{c}{  (M1) $n=800,p=1000$}\\\rule{0pt}{3ex}
Bias &.019 & $-.008$ & $-.002$   & $-.029$ &.046& .015 & .020 &   $-.003$ \\
$\sqrt{\text{Var}}$ &.439 &.409 & .406      & .424 &.544 & .521 & .522 &   .545 \\
$\sqrt{\text{EVar}}$&.444 &.410  & .401      & .378 &.529& .501 & .498 &  .473\\
Cov90 & .903&.904 & .902       & .843 &890& .889 & .886    & .829   \\
Cov95 & .952&.956 & .951       & .925 &937& .942 & .937    & .908  \\\rule{0pt}{3ex}
&  \multicolumn{8}{c}{ (M2) $n=800,p=1000$ }\\
Bias & &$-.019$ & $.000$  & $-.030$ &&$.010$ & .021 & $-.004$ \\
$\sqrt{\text{Var}}$ && .411 & .405   & .424 &&.520 & .521   & .545 \\
$\sqrt{\text{EVar}}$ &---&.407 & .401   & .379 &---& .500 & .498  & .474 \\
Cov90 && .903 & .907  &.843 && .889 & .890   & .836  \\
Cov95 && .949 & .948  &.922 && .942 &.939  & .901\\ \hline
\end{tabular*}\\[.1in]
\parbox{0.8\textwidth}{\small Note: See the footnote of Table \ref{table1}. IPW denotes $\hat{\theta}_{1,\text{\tiny IPW}}(\hat{\pi}_0)$,
with asymptotic variance estimated by accounting for variation of $\hat{\pi}_0=\tilde{E}(Z)$.
}
\end{center}  \vspace{-.1in}
\end{table}

\section{Effect of education on earnings}

The causal relationship between education and earnings has been of considerable interest in economics.
Card (1995) proposed proximity to college as an instrument for completed education. The argument is that
proximity to college could be taken as being randomized conditionally on observed covariates,
and its influence on earnings could be only through that on schooling decision.
Consider the analytic sample  in Card (1995) from National Longitudinal Survey (NLS) of Young men, which comprises  3,010 men with  valid education and wage responses in the 1976 interview.
Similarly as in Tan (2006a), we define the treatment as education after high school, i.e. $D=\one(\text{years of schooling}>12)$, the instrument $Z$ a binary indicator for proximity to a 4-year college, and the outcome $Y$ a surrogate outcome constructed  for the log of hourly earnings
at age 30. The raw vector of covariates $X$  include a race indicator, indicators for nine regions of residence and for residence in SMSA in 1966, mother's and father's  years of schooling  (\texttt{momed} and \texttt{daded} respectively) and indicators for missing values, indicators for living with both natural parents, with one natural parent and one step parent, and with mother only at age 14, and the Knowledge of World of Work score (\texttt{kww}) in 1966 and a missing
indicator. We use mean imputation for the missing values, and standardize all continuous variables with sample means 0 and variances 1.

We reanalyze the NLS data to estimate LATE of education beyond high school on log hourly earnings, using more flexible, higher dimensional models than previously allowed.
We apply $(\hat\theta_{0,\text{\tiny RCAL}},\hat\theta_{1,\text{\tiny RCAL}})$ based on regularized calibrated estimation (RCAL) and $(\hat\theta_{0,\text{\tiny RML}},\hat\theta_{1,\text{\tiny RML}})$
based on regularized likelihood estimation (RML) as well as the post-Lasso variant (RML2). The specification for $f(X)=g(X)$ consists of all the indicator variables mentioned above, \texttt{momed}, \texttt{daded},
linear spline bases in \texttt{kww} as well as interactions between the spline terms with all the indicator variables. The vector $h(X)$ augments $f(X)$ and $g(X)$ by adding linear spline terms for each fitted treatment regression $\hat{m}_{z}$, $z\in\{0,1\}$. We vary the model complexity by considering the number of knots in the set $k\in \{3,9,15\}$, with knots at the $i/(k+1)$-quantiles for $i=1,...,k$. The tuning parameter $\lambda$ is determined using $5$-fold cross validation based on the corresponding penalized loss functions, as described in Section~\ref{sec:simulation}. As an anchor specification, we also consider main-effect models with $f(X)=g(X)=(1,X^\T)^\T$ and $h(X)=(1,X^\T,\hat{m}_0,\hat{m}_1)^\T$, whereby the nuisance parameters are estimated using non-penalized likelihood or calibration estimation.

\begin{table}[t!]
\caption{Estimates of the effect of education beyond high school on log earnings.} \label{table3}  \vspace{-.1in}
\footnotesize
\begin{center}
\def\arraystretch{.9}
\begin{tabular*}{0.8\textwidth}{@{\extracolsep\fill}c ccc ccc} \hline
		\toprule
	    & RCAL & RML & RML2 \\
	    \midrule
	    	                 \multicolumn{4}{c}{Non-penalized main effects $(p=q_1=19, q_2=21)$ }\\
$\theta_0$ & $6.565\pm 0.236$ & $6.582\pm 0.269$ &  \\
$\theta_1$ & $6.402\pm 0.297$ & $6.302\pm 0.348$ & --- \\
LATE       & $0.164\pm 0.365$ & $0.280\pm 0.436$ & \\ 	
	                   \multicolumn{4}{c}{Linear spline with 3 knots $(p=q_1=114, q_2=122)$ }\\
$\theta_0$ & $6.667\pm 0.334$ & $6.635\pm 0.323$ & $6.586\pm 0.357$ \\
$\theta_1$ & $6.269\pm 0.382$ & $6.261\pm 0.365$ & $6.260\pm 0.409$ \\
LATE       & $0.405\pm 0.522$ & $0.374\pm 0.493$ & $0.326\pm 0.537$ \\
	                   \multicolumn{4}{c}{Linear spline with 9 knots $(p=q_1=213, q_2=233)$}\\
$\theta_0$ & $6.683\pm 0.310$ & $6.625\pm 0.315$ & $6.585\pm 0.350$ \\
$\theta_1$ & $6.197\pm 0.359$ & $6.230\pm 0.359$ & $6.196\pm 0.399$ \\
LATE       & $0.485\pm 0.495$ & $0.395\pm 0.480$ & $0.389\pm 0.530$ \\ 	
	                   \multicolumn{4}{c}{Linear spline with 15 knots $(p=q_1=312, q_2=344)$}\\
$\theta_0$ & $6.684\pm 0.310$ & $6.623\pm 0.310$ & $6.583\pm 0.340$ \\
$\theta_1$ & $6.180\pm 0.360$ & $6.226\pm 0.359$ & $6.208\pm 0.393$  \\
LATE       & $0.504\pm 0.498$ & $0.398\pm 0.477$ & $0.375\pm 0.518$ \\ \hline
\end{tabular*}\\[.1in]
\parbox{0.8\textwidth}{\small Note: Estimate $\pm$ $1.96\times$standard error.
As defined in Section~\ref{sec:existing-est}, $p$, $q_1$, or $q_2$ is the number of regressors in IPS, outcome $D$-model, or outcome $Y$-model.}
\end{center}  \vspace{-.1in}
\end{table}

Table \ref{table3} shows the estimates of $(\theta_0, \theta_1)$ and LATE of education beyond high school on log hourly earnings.
Regularized estimation from RCAL, RML and RML2 yield similar point estimates; the differences are small compared with the standard errors.
The RCAL and RML estimates have noticeably smaller standard errors than RML2.
Interestingly, for splines with 15 knots,
the LATE is estimated from RCAL with 95\% confidence interval $0.504\pm 0.498$, which excludes 0,
whereas those from RML and RML2 include 0.

While the validity of confidence intervals is difficult to assess using real data,
Figure \ref{fig:qq-application} in the Supplement shows that the standardized sample influence functions for estimation of LATE.
The curves from RCAL appear to be more normally distributed than RML or RML2, especially in the tails.
In addition,  Figures \ref{fig:sd_3knots}--\ref{fig:sd_15knots} in the Supplement present the standardized calibration differences for all the variables $f_j({X})$, $j=1,...,p$, similarly as in Tan (2020a).
Compared with RML and RML2, our method RCAL consistently yields smaller maximum absolute standardized differences and involves fewer nonzero estimates of $\gamma_j$ in IPS models.

\section{Conclusion}

We develop a computationally tractable method and appropriate theoretical analysis, to obtain model-assisted confidence intervals for population LATEs
in high-dimensional settings. There are various interesting topics which warrant further investigation.
Both the instrument and treatment are assumed to be binary here.
It is desirable to extend our method to handle multi-valued instruments and treatments
and to estimate treatment effects under other identification assumptions.
Another methodological question is whether doubly robust confidence intervals can be derived
in a computationally and theoretically satisfactory manner for practical use.



\vspace{.3in}
\centerline{\bf\Large References}

\begin{description}\addtolength{\itemsep}{-.1in}

\vspace{-.05in} \item Abadie, A. (2003) Semiparametric instrumental variable estimation of treatment response models, {\em Journal of Econometrics}, 113, 231--263.

\vspace{-.05in} \item Angrist, J.D., Imbens, G.W. and Rubin, D.B. (1996) Identification of causal effects using instrumental variables, {\em Journal of the American Statistical Association}, 91, 444--455.

\vspace{-.05in} \item Avagyan, V. and Vansteelandt, S. (2017) Honest data-adaptive inference for the average treatment effect under model misspecification using penalised bias-reduced double-robust estimation, {\em arXiv preprint}, arXiv:1708.03787.

\vspace{-.05in} \item Bloniarz, A., Liu, H., Zhang, C., Sekhon, J.S. and Yu, B. (2016) Lasso adjustments of treatment effect estimates in randomized experiments, {\em Proceedings of the National Academy of Sciences}, 113, 7383--7390.

\vspace{-.05in} \item  Baiocchi, M., Cheng, J., and Small, D.S. (2014). Instrumental variable methods for causal inference, {\em Statistics in Medicine}, 33, 2297--2340.

\vspace{-.05in} \item Belloni, A., Chernozhukov, V. and Hansen, C. (2014) Inference on treatment effects after selection among high-dimensional controls, {\em The Review of Economic Studies}, 81, 608--650.

\vspace{-.05in} \item Bradic, J., Wager, S. and Zhu, Y. (2019) Sparsity double robust inference of average treatment effects, {\em arXiv preprint}, arXiv:1905.00744.

\vspace{-.05in} \item Buhlmann, P. and van de Geer, S. (2011) {\em Statistics for High-Dimensional Data: Methods, Theory and Applications}, New York: Springer.

\vspace{-.05in} \item Card, D. (1995) Using geographic variation in college proximity to estimate the return to schooling, in {\em Aspects of Labour Market Behaviour: Essays in Honour of John Vanderkamp}, 201--222, Toronto: University of Toronto Press.

\vspace{-.05in} \item Chernozhukov, V., Chetverikov, D., Demirer, M., Duflo, E., Hansen, C., Newey, W. and Robins, J.M. (2018) Double/debiased machine learning for treatment and structural parameters,  {\em The Econometrics Journal}, 21, C1--C68.

\vspace{-.05in} \item Davidian, M., Tsiatis, A.A. and Leon, S. (2005) Semiparametric estimation of treatment effect in a pretest--posttest study with missing data,  {\em Statistical Science}, 20, 261--301.

\vspace{-.05in} \item Farrell, M. (2015) Robust inference on average treatment effects with possibly more covariates than observations, {\em Journal of Econometrics}, 189, 1--23.

\vspace{-.05in} \item  Friedman, J., Hastie, T. and Tibshirani, R. (2010). Regularization paths for generalized linear models via coordinate descent, {\em Journal of Statistical Software}, 33, 1--22.

\vspace{-.05in} \item Fr{\"o}lich, M. (2007) Nonparametric IV estimation of local average treatment effects with covariates, {\em Journal of Econometrics}, 139, 35--75.



\vspace{-.05in} \item Hirano, K., Imbens, G.W., Rubin, D.B. and Zhou, X.-H. (2000), Assessing the effect of an influenza vaccine in an encouragement design, {\em Biostatistics}, 1, 69--88.

\vspace{-.05in}  \item Huang, J. and Zhang, C.-H. (2012) Estimation and selection via absolute penalized convex minimization and its multistage adaptive applications,
{\em Journal of Machine Learning Research}, 13, 1839–-1864.

\vspace{-.05in} \item Imbens, G.W. (2014) Instrumental variables: An econometrician's perspective, {\em Statistical Science}, 29, 323--358.


\vspace{-.05in} \item Kang, J.D.Y. and  Schafer, J.L. (2007) Demystifying double robustness: A comparison of alternative strategies for estimating a population mean from incomplete data, {\em Statistical Science}, 4, 523--539.

\vspace{-.05in} \item  Kim, J.K. and Haziza, D. (2014) Doubly robust inference with missing data in survey sampling, {\em Statistica Sinica}, 24, 375--394.

\vspace{-.05in} \item Little, R.J.A. and Yau, L. (1998). Statistical techniques for analyzing data from prevention trials: Treatment of no-shows using Rubin's causal model, {\em Psychological Methods}, 3, 147--159.

\vspace{-.05in} \item  McCullagh, P. and Nelder, J. (1989) {\em Generalized Linear Models} (2nd edition), New York: Chapman \& Hall.


\vspace{-.05in} \item Neyman, J. (1990) On the application of probability theory to agricultural experiments. Essay on principles. Section 9, {\em Statistical Science}, 5, 465--472.

\vspace{-.05in} \item Ning, Y., Peng, S. and Imai, K. (2020) Robust estimation of causal effects via high-dimensional covariate balancing propensity score, {\em Biometrika}, to appear.

\vspace{-.05in} \item Ogburn, E.L., Rotnitzky, A. and Robins, J.M. (2015) Doubly robust estimation of the local average treatment effect curve,  {\em Journal of the Royal Statistical Society}, Ser.~B, 77, 373--396.

\vspace{-.05in} \item Okui, R., Small, D.S., Tan, Z. and Robins, J.M. (2012) Doubly robust instrumental variable regression, {\em Statistica Sinica}, 22, 173--205.

\vspace{-.05in} \item Osborne, M., Presnell, B., and Turlach, B. (2000) A new approach to variable selection in least squares problems,  {\em IMA Journal of Numerical Analysis}, 20, 389-–404.

\vspace{-.05in} \item Robins, J.M. (1994) Correcting for non-compliance in randomized trials using structural nested mean models, {\em Communications in Statistics-Theory and methods}, 23, 2379--2412.

\vspace{-.05in} \item Robins, J.M., Rotnitzky, A. and Zhao, L.P. (1994) Estimation of regression coefficients when some regressors are not always observed, {\em Journal of the American Statistical Association}, 89, 846--866.

\vspace{-.05in} \item Rubin, D.B. (1974) Estimating causal effects of treatments in randomized and nonrandomized studies, {\em Journal of educational Psychology}, 66, 688--701.

\vspace{-.05in} \item S{\"a}rndal, C.E., Swensson, B. and Wretman, J. (2003) {\em Model Assisted Survey Sampling}, {Springer Science \& Business Media}.

\vspace{-.05in} \item  Smucler, E., Rotnitzky, A. and Robins, James M. (2019) A unifying approach for doubly-robust $\ell_1$ regularized estimation of causal contrasts, {\em arXiv preprint}, arXiv:1904.03737.

\vspace{-.05in} \item Tan, Z. (2006a) Regression and weighting methods for causal inference using instrumental variables, {\em Journal of the American Statistical Association}, 101, 1607--1618.

\vspace{-.05in} \item Tan, Z. (2006b) A distributional approach for causal inference using propensity scores, {\em Journal of the American Statistical Association}, 101, 1619--1637.

\vspace{-.05in} \item Tan, Z. (2007) Comment: Understanding {OR}, {PS} and {DR}, {\em Statistical Science}, 22, 560--568.

\vspace{-.05in} \item Tan, Z. (2010a) Marginal and nested structural models using instrumental variables, {\em Journal of the American Statistical Association}, 105, 157--169.

\vspace{-.05in} \item Tan, Z. (2010b) Bounded, efficient, and doubly robust estimation with inverse weighting, {\em Biometrika}, 97, 661--682.

\vspace{-.05in} \item Tan, Z. (2020a) Regularized calibrated estimation of propensity scores with model misspecification and high-dimensional data, {\em Biometrika}, 107, 137--158.

\vspace{-.05in} \item Tan, Z. (2020b) Model-assisted inference for treatment effects using regularized calibrated estimation with high-dimensional data, {\em Annals of Statistics}, 48, 811--837.

\vspace{-.05in} \item Tibshirani, R. (1996) Regression shrinkage and selection via the lasso,  {\em Journal of the Royal Statistical Society}, Ser.~B, 58, 267--288.

\vspace{-.05in} \item Uysal, S.D. (2011) Doubly robust IV estimation of the local average treatment effects, Unpublished manuscript.

\vspace{-.05in} \item van de Geer, S., Buhlmann, P., Ritov, Y. and Dezeure, R. (2014) On asymptotically optimal confidence regions and tests for high-dimensional models, {\em Annals of Statistics}, 42, 1166--1202.

\vspace{-.05in} \item Vansteelandt, S.and Goetghebeur, E. (2003) Causal inference with generalized structural mean models, {\em Journal of the Royal Statistical Society}, Ser.~B, 65, 817--835.

\vspace{-.05in} \item Vermeulen, K. and Vansteelandt, S. (2015) Bias-reduced doubly robust estimation, {\em Journal of the American Statistical Association}, 110,1024--1036.

\vspace{-.05in} \item Vytlacil, E. (2002) Independence, monotonicity, and latent index models: An equivalence result, {\em Econometrica}, 70, 331--341.

\vspace{-.05in} \item Wager, S., Du, W., Taylor, J. and Tibshirani, R.J. (2016) High-dimensional regression adjustments in randomized experiments, {\em Proceedings of the National Academy of Sciences}, 113, 12673--12678.

\vspace{-.05in} \item Wang, L., and Tchetgen Tchetgen, E. (2019) Bounded, efficient and multiply robust estimation of average treatment effects using instrumental variables, {\em Journal of the Royal Statistical Society}, Ser.~B, 80, 531--550.

\vspace{-.05in} \item Wright, P.G. (1928) {\em Tariff on Animal and Vegetable Oils}, {Macmillan Company, New York}.

\vspace{-.05in} \item Zhang, C.-H. and Zhang, S.S. (2014) Confidence intervals for low-dimensional parameters with high-dimensional data, {\em Journal of the Royal Statistical Society}, Ser.~B, 76, 217--242.

\end{description}



\clearpage

\setcounter{page}{1}

\setcounter{section}{0}
\setcounter{equation}{0}

\setcounter{figure}{0}
\setcounter{table}{0}

\renewcommand{\theequation}{S\arabic{equation}}
\renewcommand{\thesection}{\Roman{section}}

\renewcommand\thefigure{S\arabic{figure}}
\renewcommand\thetable{S\arabic{table}}

\setcounter{lem}{0}

\begin{center}
{\Large Supplementary Material for}

{\Large ``High-dimensional Model-assisted Inference for Local Average Treatment Effects with Instrumental Variables}

\vspace{.1in} {\large Baoluo Sun \& Zhiqiang Tan}

\end{center}

 \section{Additional results in simulation studies}

\begin{figure}[!htb]
\caption{\small Scatter plots of $(X^\dag_1,X^\dag_2, X^\dag_3,X^\dag_4)$ with marginal histograms from a simulated sample of size $n=800$.}
\label{fig:scatterplot} \vspace{.15in}
\begin{tabular}{c}
\includegraphics[width=6.2in, height=6in]{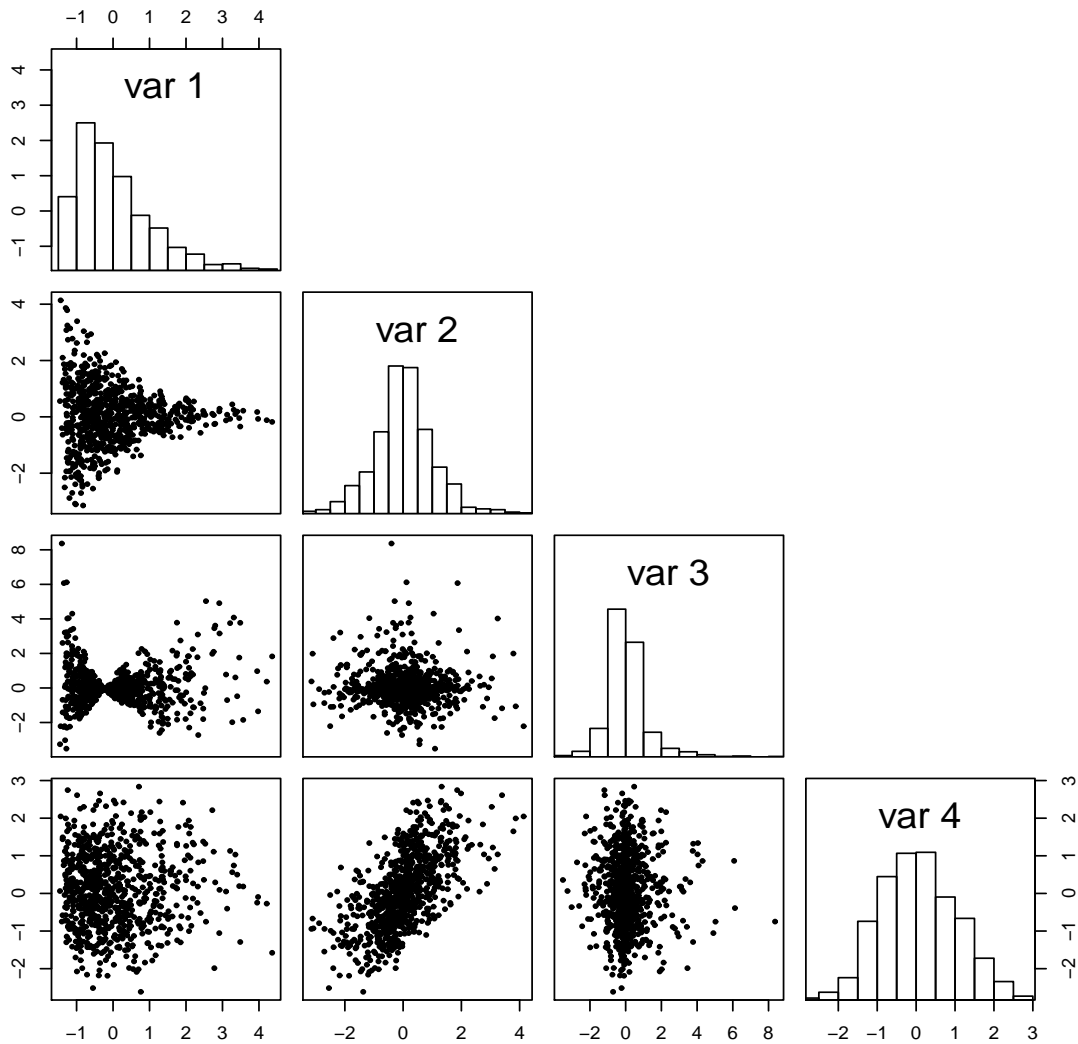} \vspace{-.25in}
\end{tabular}
\end{figure}

\begin{sidewaysfigure}
\caption{\small  Scatterplots of $Y$ against $(X^{\dag}_1, X^{\dag}_2,X^{\dag}_3,X^{\dag}_4)$ within $\{D = 1\}$, boxplots of $(X^{\dag}_1, X^{\dag}_2,X^{\dag}_3,X^{\dag}_4)$ within $\{D = 0\}$ and  $\{D = 1\}$ as well as boxplots of $(X^{\dag}_1, X^{\dag}_2,X^{\dag}_3,X^{\dag}_4)$ within $\{Z = 0\}$ and  $\{Z = 1\}$ from a sample of size $n = 800$ in case (C1).}
\label{fig:scatter-c1} \vspace{.15in}
\centering
\includegraphics[angle=360, totalheight=6.2in]{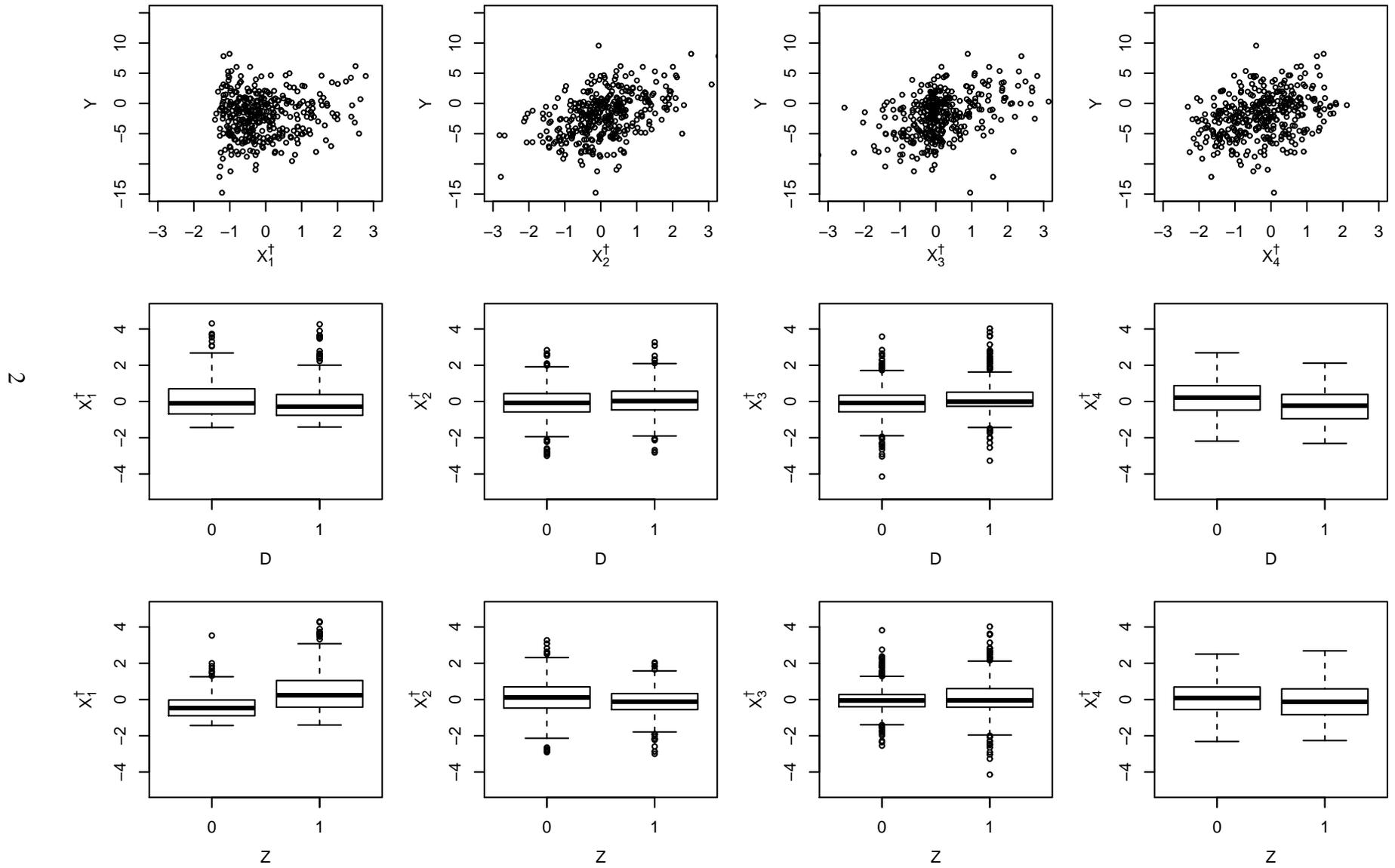} \vspace{-.25in}
\end{sidewaysfigure}

\begin{sidewaysfigure}
\caption{\small  Scatterplots of $Y$ against $(X^{\dag}_1, X^{\dag}_2,X^{\dag}_3,X^{\dag}_4)$ within $\{D = 1\}$, boxplots of $(X^{\dag}_1, X^{\dag}_2,X^{\dag}_3,X^{\dag}_4)$ within $\{D = 0\}$ and  $\{D = 1\}$ as well as boxplots of $(X^{\dag}_1, X^{\dag}_2,X^{\dag}_3,X^{\dag}_4)$ within $\{Z = 0\}$ and  $\{Z = 1\}$ from a sample of size $n = 800$ in case (C2).}
\label{fig:scatter-c2} \vspace{.15in}
\centering
\includegraphics[angle=360, totalheight=6.2in]{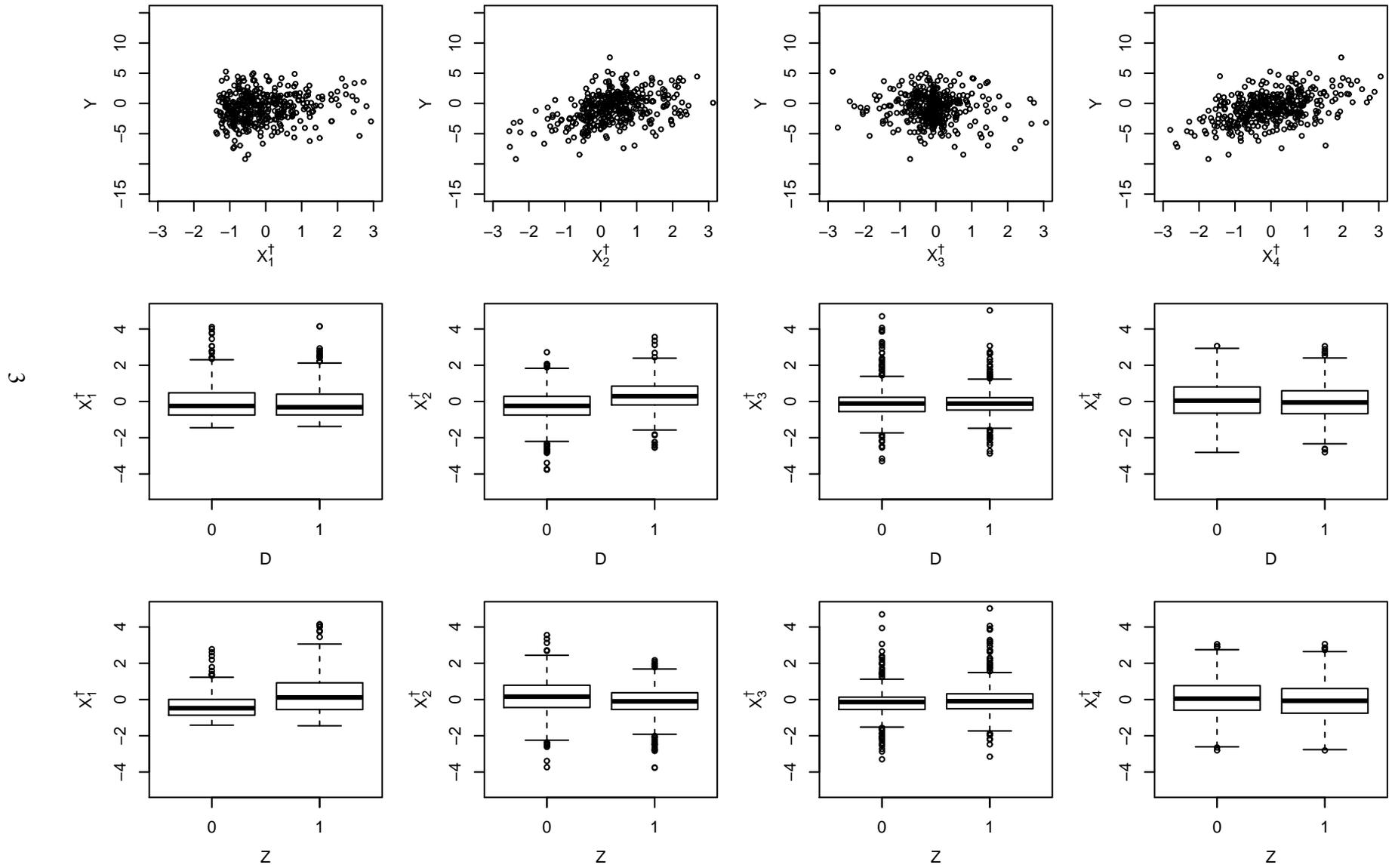} \vspace{-.25in}
\end{sidewaysfigure}

\begin{sidewaysfigure}
\caption{\small  Scatterplots of $Y$ against $(X^{\dag}_1, X^{\dag}_2,X^{\dag}_3,X^{\dag}_4)$ within $\{D = 1\}$, boxplots of $(X^{\dag}_1, X^{\dag}_2,X^{\dag}_3,X^{\dag}_4)$ within $\{D = 0\}$ and  $\{D = 1\}$ as well as boxplots of $(X^{\dag}_1, X^{\dag}_2,X^{\dag}_3,X^{\dag}_4)$ within $\{Z = 0\}$ and  $\{Z = 1\}$ from a sample of size $n = 800$ in case (C3).}
\label{fig:scatter-c3} \vspace{.15in}
\centering
\includegraphics[angle=360, totalheight=6.2in]{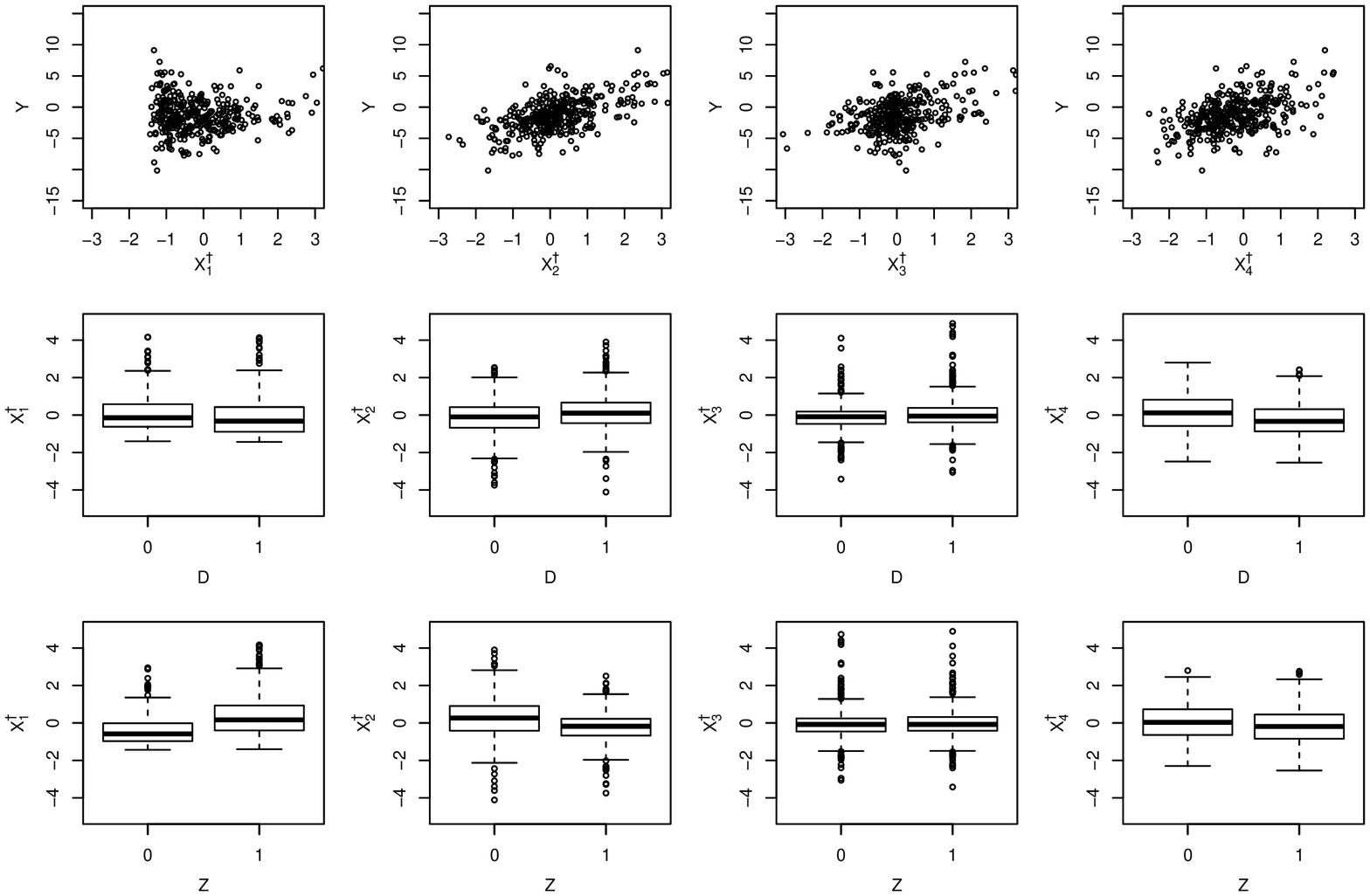} \vspace{-.25in}
\end{sidewaysfigure}

\begin{sidewaysfigure}
\caption{\small QQ-plots of the estimates (first row) and $t$-statistics (second row) against standard normal ($n=800$, $p=400$),
based on $\hat\theta_{1,\text{\tiny RML}}$ ($\circ$), the post-Lasso variant (\textcolor{blue}{$\triangle$}) and $\hat\theta_{1,\text{\tiny RCAL}}$ (\textcolor{red}{$\times$}) under specification (M1). For readability, only a subset of 101 order statistics are shown as points on the QQ lines.}
\label{fig:qq-n800-p400-m1} \vspace{.15in}
\centering
\includegraphics[angle=360, totalheight=6.2in]{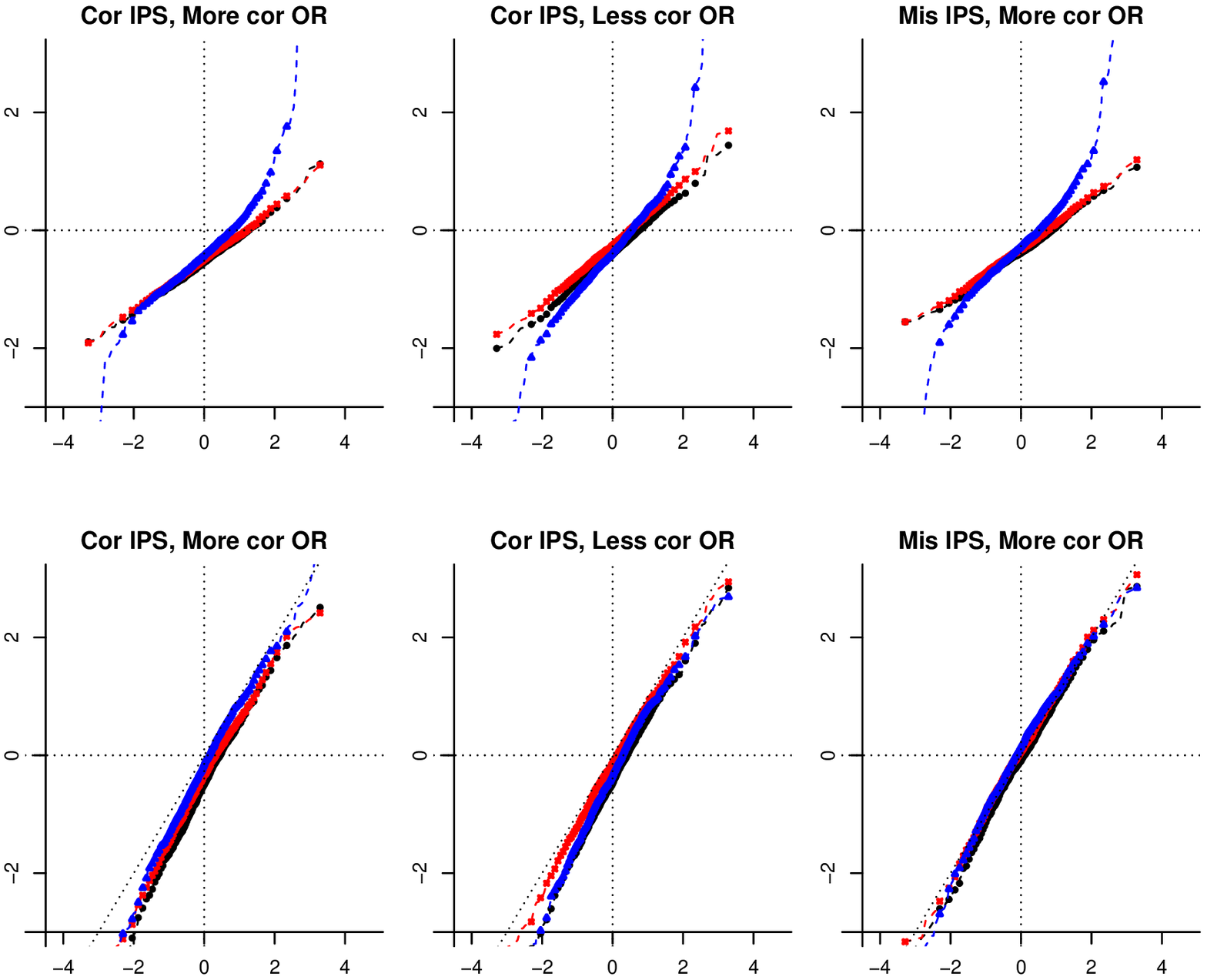} \vspace{-.25in}
\end{sidewaysfigure}
\clearpage

\begin{sidewaysfigure}
\caption{\small QQ-plots of the estimates (first row) and $t$-statistics (second row) against standard normal ($n=800$, $p=1000$),
based on $\hat\theta_{1,\text{\tiny RML}}$ ($\circ$), the post-Lasso variant (\textcolor{blue}{$\triangle$}) and $\hat\theta_{1,\text{\tiny RCAL}}$ (\textcolor{red}{$\times$}) under specification (M1). For readability, only a subset of 101 order statistics are shown as points on the QQ lines.}
\label{fig:qq-n800-p1000-m1} \vspace{.15in}
\centering
\includegraphics[angle=360, totalheight=6.2in]{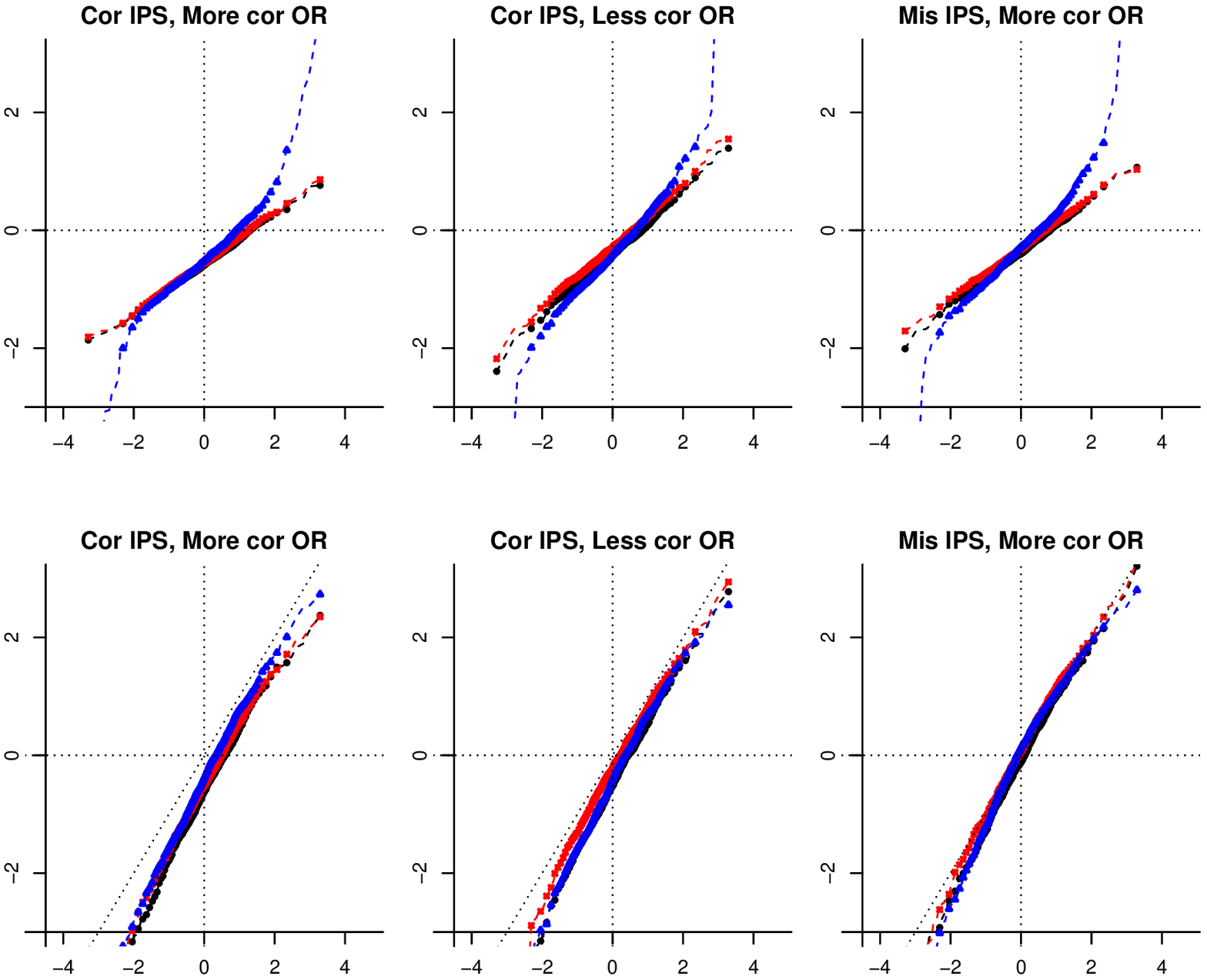} \vspace{-.25in}
\end{sidewaysfigure}
\clearpage

\begin{sidewaysfigure}
\caption{\small QQ-plots of the estimates (first row) and $t$-statistics (second row) against standard normal ($n=800$, $p=400$),
based on $\hat\theta_{1,\text{\tiny RML}}$ ($\circ$), the post-Lasso variant (\textcolor{blue}{$\triangle$}) and $\hat\theta_{1,\text{\tiny RCAL}}$ (\textcolor{red}{$\times$}) under specification (M2). For readability, only a subset of 101 order statistics are shown as points on the QQ lines.}
\label{fig:qq-n800-p400-m2} \vspace{.15in}
\centering
\includegraphics[angle=360, totalheight=6.2in]{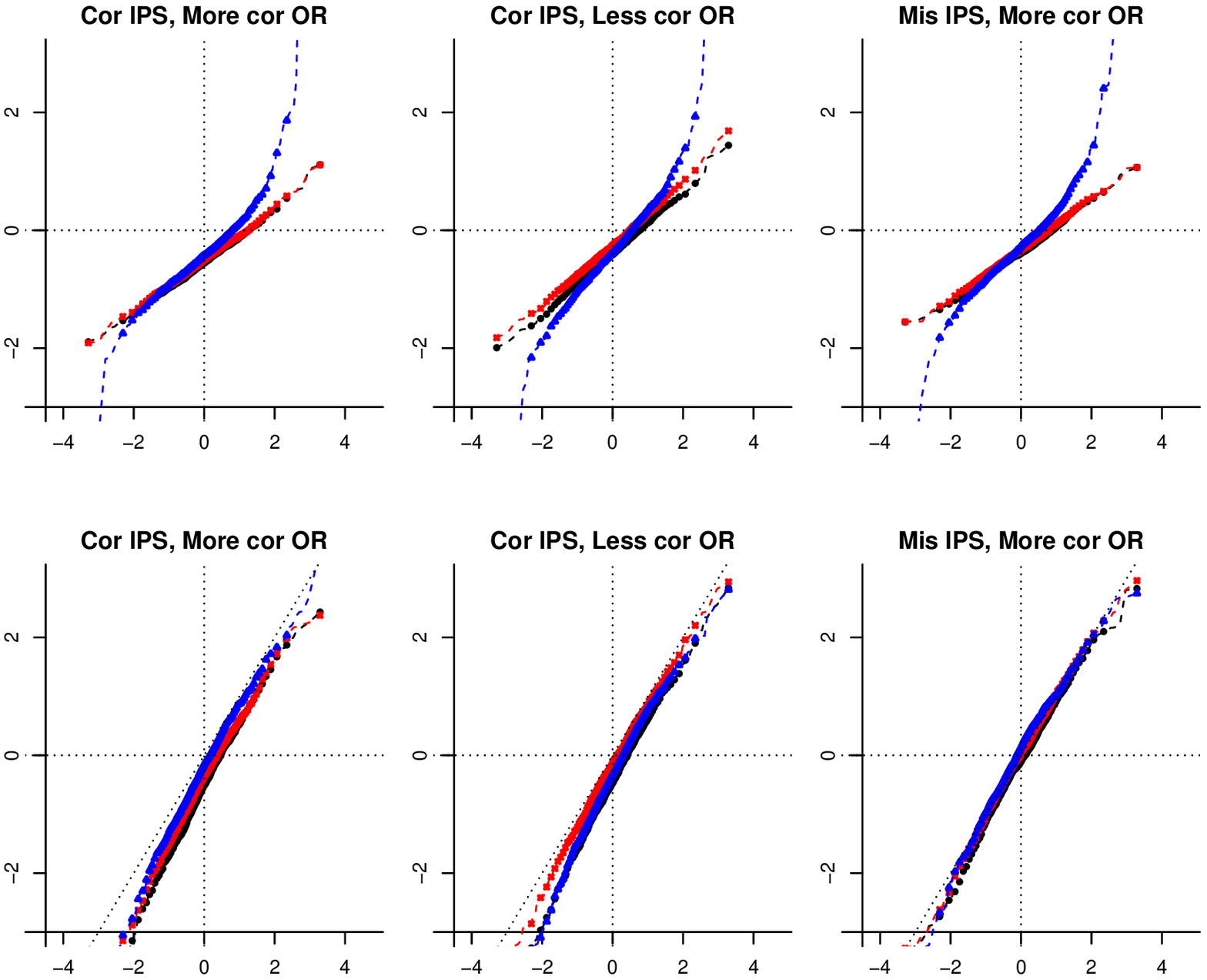} \vspace{-.25in}
\end{sidewaysfigure}

  \begin{sidewaysfigure}
\caption{\small QQ-plots of the estimates (first row) and $t$-statistics (second row) against standard normal ($n=800$, $p=1000$),
based on $\hat\theta_{1,\text{\tiny RML}}$ ($\circ$), the post-Lasso variant (\textcolor{blue}{$\triangle$}) and $\hat\theta_{1,\text{\tiny RCAL}}$ (\textcolor{red}{$\times$}) under specification (M2). For readability, only a subset of 101 order statistics are shown as points on the QQ lines.}
\label{fig:qq-n800-p1000-m2} \vspace{.15in}
\centering
\includegraphics[angle=360, totalheight=6.2in]{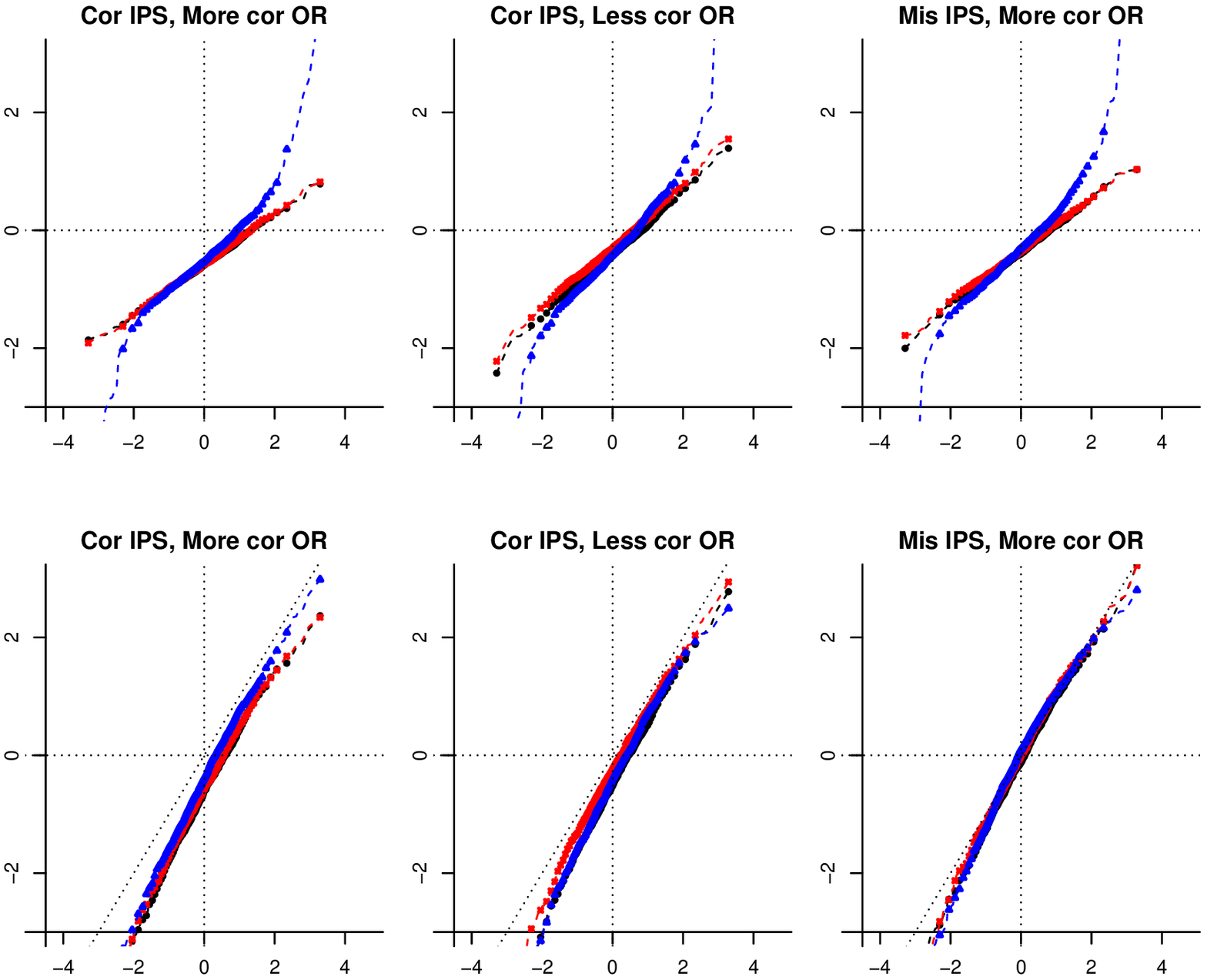} \vspace{-.25in}
\end{sidewaysfigure}

\begin{table}
\caption{Summary of results for estimation of $\theta_1$ with a completely randomized instrument.} \label{table2-supp}  \vspace{-.1in}
\footnotesize
\begin{center}
\begin{tabular*}{0.7\textwidth}{@{\extracolsep\fill}c c ccl c ccl} \hline
			\toprule
              &\multicolumn{4}{c}{(C4) cor IPS, more cor OR} & \multicolumn{4}{c}{(C5) cor IPS, less cor OR}\\
              	     \cmidrule(lr){2-5}\cmidrule(lr){6-9}
	  & IPW & RCAL & RML  & RML2 & IPW & RCAL & RML  & RML2\\
	     \hline\rule{0pt}{3ex}
&  \multicolumn{8}{c}{  (M1) $n=800,p=400$}\\
Bias & .019 & $-.012$ & $-.005$  & $-.019$ & .046 & .023 & .029  & .040 \\
$\sqrt{\text{Var}}$ & .439 & .424 & .417 & .435  & .544 & .508 & .508  & .526 \\
$\sqrt{\text{EVar}}$ & .444 &.408 & .400  & .378 & .529 & .504 & .500  & .529\\
Cov90 & .903 & .891 & .885  & .841  & .890 & .903 & .900  & .889   \\
Cov95 & .952 & .946 & .942  & .906  & .937 & .943 & .945 & .936  \\\rule{0pt}{3ex}
&  \multicolumn{8}{c}{  (M2) $n=800,p=400$}\\
Bias & &$-.016$ & $-.002$   & $-.023$ & &.023 & .030 &  .004 \\
$\sqrt{\text{Var}}$ & &.423 & .416 &   .432 & &.506 & .508 &  .514 \\
$\sqrt{\text{EVar}}$&--- &.407 & .400 &  .378 &--- &.504 & .500 & .476\\
Cov90 & &.889 & .891 & .844 & &.903 & .903   & .879   \\
Cov95 & &.942 & .940 & .911 & &.946 & .947   & .931  \\
&  \multicolumn{8}{c}{  (M1) $n=800,p=1000$}\\\rule{0pt}{3ex}
Bias & & $-.008$ & $-.002$   & $-.029$ && .015 & .020 &   $-.003$ \\
$\sqrt{\text{Var}}$ & &.409 & .406      & .424 && .521 & .522 &   .545 \\
$\sqrt{\text{EVar}}$&--- &.410  & .401      & .378 &---& .501 & .498 &  .473\\
Cov90 & &.904 & .902       & .843 && .889 & .886    & .829   \\
Cov95 & &.956 & .951       & .925 && .942 & .937    & .908  \\\rule{0pt}{3ex}
&  \multicolumn{8}{c}{ (M2) $n=800,p=1000$ }\\
Bias & &$-.019$ & $.000$  & $-.030$ &&$.010$ & .021 & $-.004$ \\
$\sqrt{\text{Var}}$ && .411 & .405   & .424 &&.520 & .521   & .545 \\
$\sqrt{\text{EVar}}$ &---&.407 & .401   & .379 &---& .500 & .498  & .474 \\
Cov90 && .903 & .907  &.843 && .889 & .890   & .836  \\
Cov95 && .949 & .948  &.922 && .942 &.939  & .901\\ \hline
\end{tabular*}\\[.1in]
\parbox{0.7\textwidth}{\small Note: See the footnote of Table \ref{table2}. For completeness, the results with $n=800$ and $p=1000$ are also included.}
\end{center}  \vspace{-.1in}
\end{table}

\clearpage
 \section{Additional results in empirical application}

\begin{figure}[!b]
 \centering
\caption{\small Boxplots of the weights $1/(1-\hat{\pi})$ or $1/\hat{\pi}$ within the $Z=0$ and $Z=1$ groups (1st and 2nd columns respectively), each normalized to sum to the sample size $n=3010$, as well as QQ-plots with a 45-degree line of the standardized sample influence functions for non-penalized main effects model (top row) and linear spline specification with 3, 9 and 15 knots (2nd to 4th rows respectively).}
\label{fig:qq-application} \vspace{.15in}
\begin{tabular}{c}
\includegraphics[width=5in, height=7in]{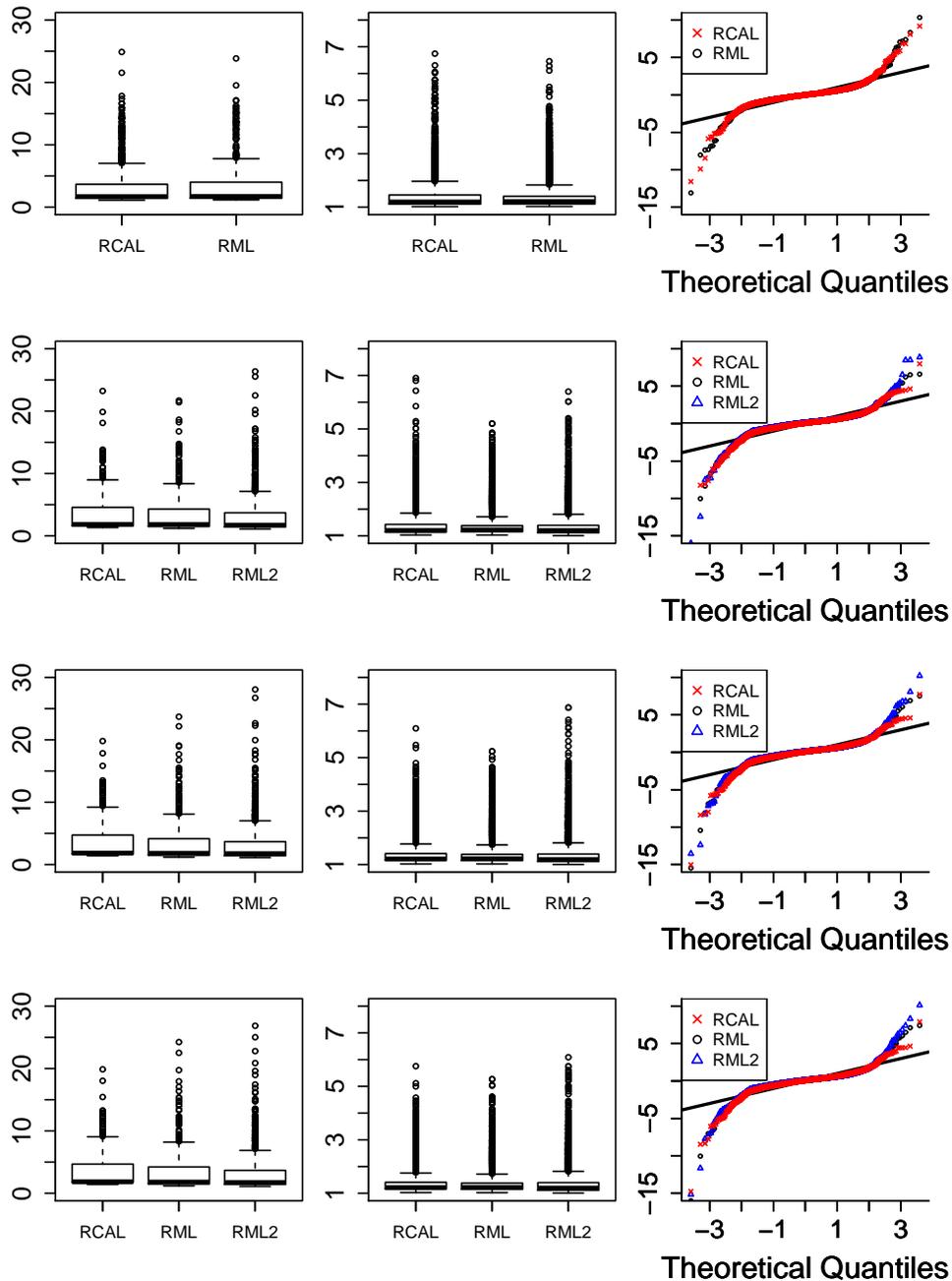} \vspace{-.25in}
\end{tabular}
\end{figure}

\clearpage
The standardized calibration difference in the $Z=1$ group for a function $h(X)$ using an estimated $\hat\pi(X)$ is
$\text{\small CAL}^1(\hat{\pi},h)= \frac{\tilde{E} [ \{Z/\hat{\pi}({X})-1\}h(X) ] /\tilde{E}\{Z/\hat{\pi}({X})\}}{ \tilde V^{1/2} (h(X))}$,
where $\tilde V()$ denotes the sample variance.

\begin{figure}[!b]
 \centering
\caption{\small Standardized calibration differences $\text{\small CAL}^1(\hat{\pi};f_j)$ over index $j\in\{1,...,114\}$ under linear spline specification with 3 knots for the estimators $\hat{\pi}=\tilde{E}(Z)$, $\hat{\pi}_{\text{\tiny 1,RCAL}}$, $\hat{\pi}_{\text{\tiny RML}}$ and the post-Lasso variant $\hat{\pi}_{\text{\tiny RML2}}$ respectively (top 2 rows), with $\lambda$ selected from cross validation. Two horizontal lines are placed at the maximum absolute standardized differences. Marks (x)
are plotted at the indices $j$ corresponding to 30 nonzero estimates of $\gamma_j$ for $\hat{\pi}_{\text{\tiny 1,RCAL}}$ and 31 nonzero estimates of $\gamma_j$ for $\hat{\pi}_{\text{\tiny RML}}$ and $\hat{\pi}_{\text{\tiny RML2}}$. Scatter plots of the fitted propensity scores $\{\hat{\pi}_{\text{\tiny 1,RCAL}}(X_i),\hat{\pi}_{\text{\tiny RML}}({X}_i)\}$ and $\{\hat{\pi}_{\text{\tiny 1,RCAL}}(X_i),\hat{\pi}_{\text{\tiny RML2}}({X}_i)\}$ respectively (bottom row), in the group of individuals who stay near a 4-year college, i.e. $\{Z_i=1, i=1,...,n\}$.}
\label{fig:sd_3knots} \vspace{.15in}
\begin{tabular}{c}
\includegraphics[width=5in, height=6in]{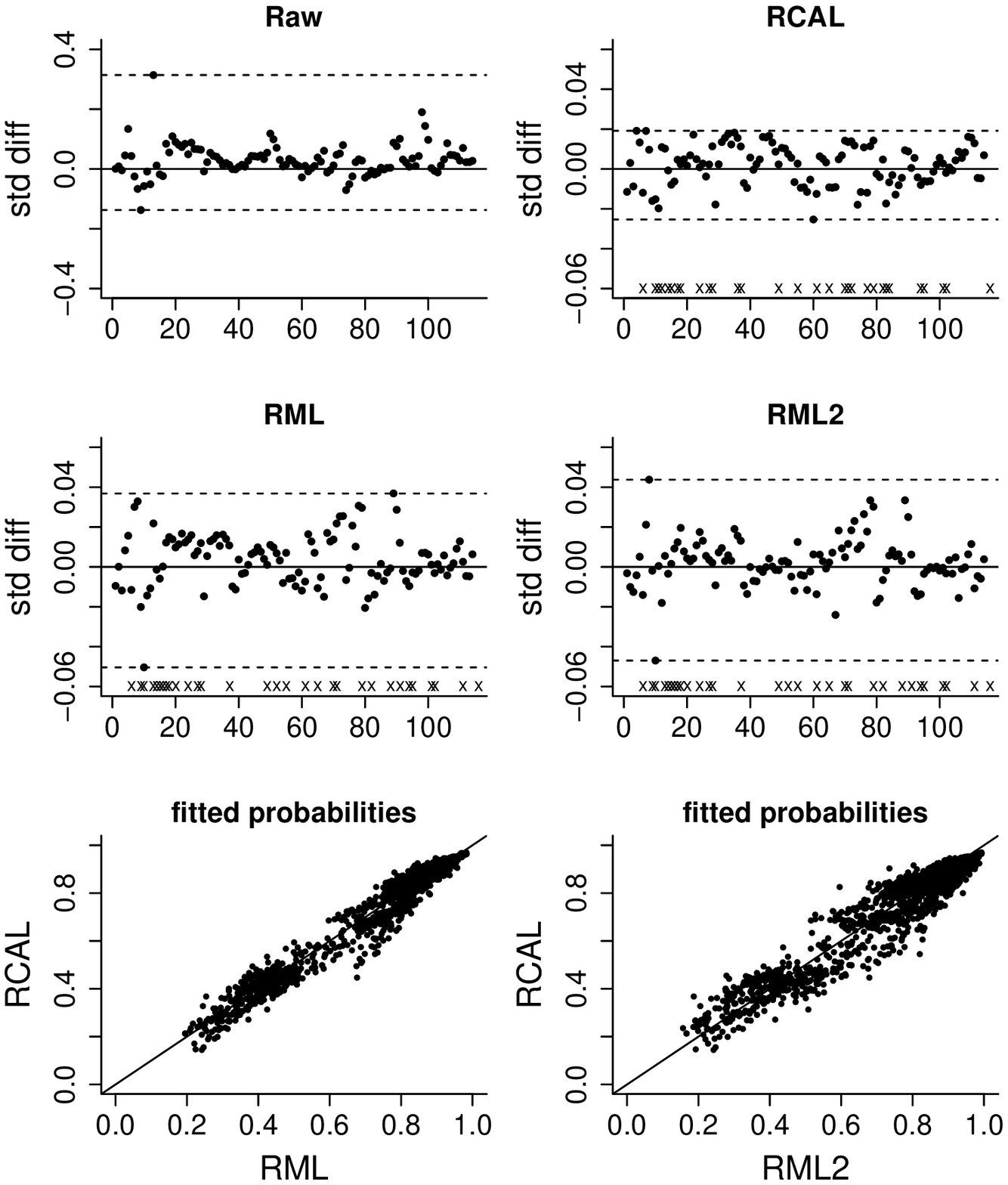} \vspace{-.25in}
\end{tabular}
\end{figure}

 \begin{figure}
 \centering
\caption{\small Standardized calibration differences $\text{\small CAL}^1(\hat{\pi};f_j)$ over index $j\in\{1,...,213\}$ under linear spline specification with 9 knots for the estimators $\hat{\pi}=\tilde{E}(Z)$, $\hat{\pi}_{\text{\tiny 1,RCAL}}$, $\hat{\pi}_{\text{\tiny RML}}$ and the post-Lasso variant $\hat{\pi}_{\text{\tiny RML2}}$ respectively (top 2 rows), with $\lambda$ selected from cross validation. Two horizontal lines are placed at the maximum absolute standardized differences. Marks (x)
are plotted at the indices $j$ corresponding to 21 nonzero estimates of $\gamma_j$ for $\hat{\pi}_{\text{\tiny 1,RCAL}}$ and 37 nonzero estimates of $\gamma_j$ for $\hat{\pi}_{\text{\tiny RML}}$ and $\hat{\pi}_{\text{\tiny RML2}}$. Scatter plots of the fitted propensity scores $\{\hat{\pi}_{\text{\tiny 1,RCAL}}(X_i),\hat{\pi}_{\text{\tiny RML}}({X}_i)\}$ and $\{\hat{\pi}_{\text{\tiny 1,RCAL}}(X_i),\hat{\pi}_{\text{\tiny RML2}}({X}_i)\}$ respectively (bottom row), in the group of individuals who stay near a 4-year college, i.e. $\{Z_i=1, i=1,...,n\}$.}
\label{fig:sd_9knots} \vspace{.15in}
\begin{tabular}{c}
\includegraphics[width=5in, height=6in]{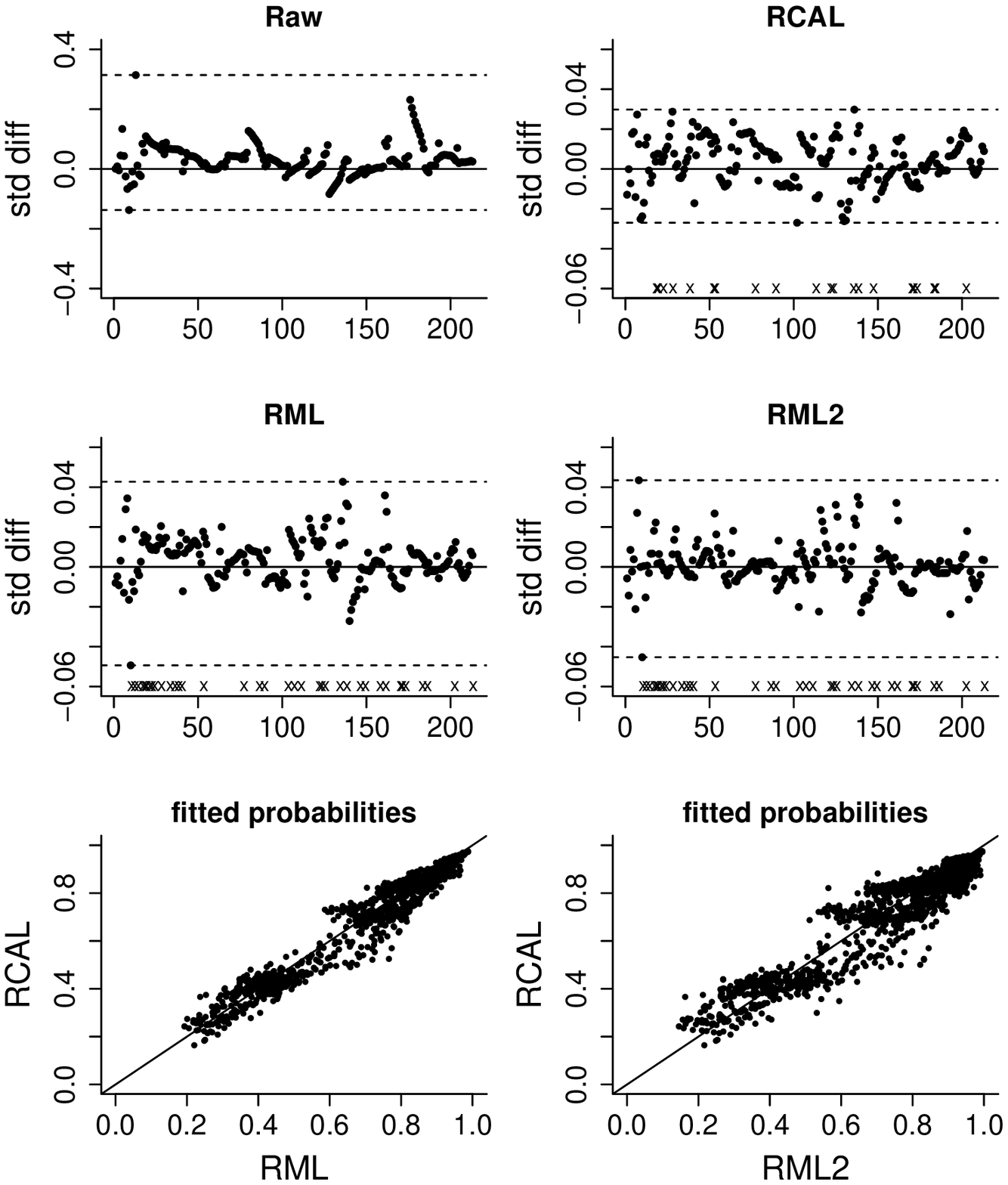} \vspace{-.25in}
\end{tabular}
\end{figure}

 \begin{figure}
 \centering
\caption{\small Standardized calibration differences $\text{\small CAL}^1(\hat{\pi};f_j)$ over index $j\in\{1,...,312\}$ under linear spline specification with 15 knots for the estimators $\hat{\pi}=\tilde{E}(Z)$, $\hat{\pi}_{\text{\tiny 1,RCAL}}$, $\hat{\pi}_{\text{\tiny RML}}$ and the post-Lasso variant $\hat{\pi}_{\text{\tiny RML2}}$ respectively (top 2 rows), with $\lambda$ selected from cross validation. Two horizontal lines are placed at the maximum absolute standardized differences. Marks (x)
are plotted at the indices $j$ corresponding to 20 nonzero estimates of $\gamma_j$ for $\hat{\pi}_{\text{\tiny 1,RCAL}}$ and 34 nonzero estimates of $\gamma_j$ for $\hat{\pi}_{\text{\tiny RML}}$ and $\hat{\pi}_{\text{\tiny RML2}}$. Scatter plots of the fitted propensity scores $\{\hat{\pi}_{\text{\tiny 1,RCAL}}(X_i),\hat{\pi}_{\text{\tiny RML}}({X}_i)\}$ and $\{\hat{\pi}_{\text{\tiny 1,RCAL}}(X_i),\hat{\pi}_{\text{\tiny RML2}}({X}_i)\}$ respectively (bottom row), in the group of individuals who stay near a 4-year college, i.e. $\{Z_i=1, i=1,...,n\}$.}
\label{fig:sd_15knots} \vspace{.15in}
\begin{tabular}{c}
\includegraphics[width=5in, height=6in]{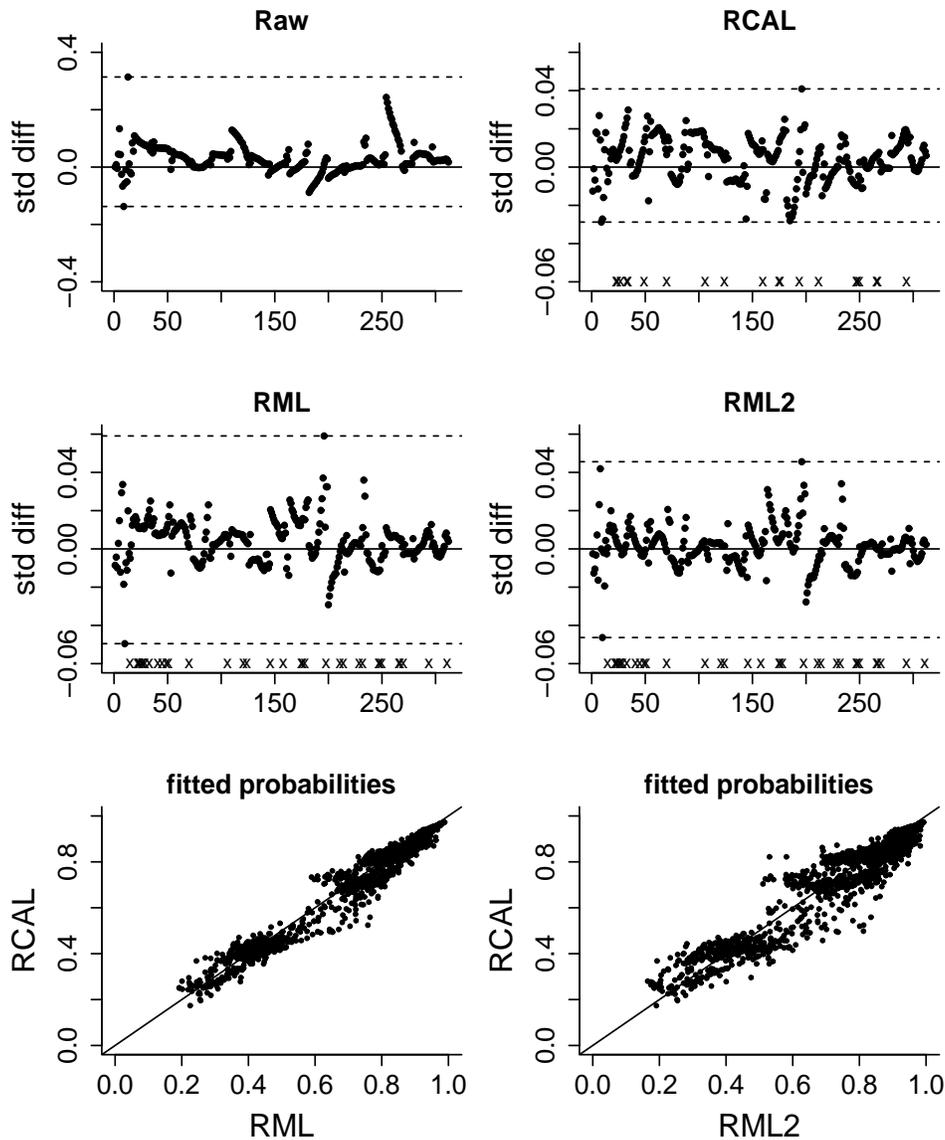} \vspace{-.25in}
\end{tabular}
\end{figure}

\clearpage
\section{Technical details}

\subsection{Probability lemmas} \label{sec:prob-lem}

Denote by $\Omega_0$ the event that (\ref{eq:bound-gamma1})--(\ref{eq:bound-alpha1}) hold.
Then $P(\Omega_0) \ge 1 - c_0 \epsilon$ under Theorem~\ref{thm:gamma1-alpha1}.
The following Lemmas~\ref{lem:prob-grad}--\ref{lem:prob-mat-f2} are used in the proof of Theorem~\ref{thm:alpha11} (Section~\ref{sec:prf-thm-alpha11}).

\begin{lem} \label{lem:prob-grad}
Under Assumptions~\ref{ass:gamma1}(ii) and \ref{ass:alpha11}(i)--(ii), there exists a positive constant $B_1$,
depending on $(C_{f1}, C_{h0}, \sigma_0, \sigma_1)$, such that
$P(\Omega_1 ) \ge 1- 2\epsilon$, where $\Omega_1$ denotes the event
\begin{align*}
\sup_{j=0,1,\ldots,r} \left| \tilde E \big[  Z  w_1(X; \bar\gamma_1)
\{ DY - m_1(X; \bar{\alpha}_1) \psi_{\scriptscriptstyle Y} (\bar\alpha^\T _{11} h) \} h_j(X) \big] \right| \le B_1 \lambda_2 .
\end{align*}
\end{lem}

\begin{prf}
This can be shown similarly as Lemma~2 in the Supplement of Tan (2020b).
\end{prf}

Recall
\begin{align*}
& \Sigma_{h} = E  \left[ Z  w_1(X;\bar\gamma_1) h(X) h^\T(X) \right]  , \quad
\tilde \Sigma_{h} = \tilde  E  \left[ Z  w_1(X;\bar\gamma_1) h(X) h^\T(X) \right].
\end{align*}

\begin{lem} \label{lem:prob-mat-h}
Under Assumptions~\ref{ass:gamma1}(ii) and \ref{ass:alpha11}(i), there exists a positive constant $B_{h1}$,
depending on $(C_{f1}, C_{h0})$, such that
$P(\Omega_{h1} ) \ge 1- 2\epsilon$, where $\Omega_{h1}$ denotes the event
\begin{align*}
\sup_{j,k=0,1,\ldots, r} \left| ( \tilde \Sigma_{h})_{jk} - (\Sigma_{h})_{jk} \right| \le B_{h1} \lambda_2 .
\end{align*}
\end{lem}

\begin{prf}
This can be shown similarly as Lemma~1(ii) in the Supplement of Tan (2020b).
\end{prf}

Recall
\begin{align*}
& \Sigma_f = E  \left[ Z  w_1(X;\bar\gamma_1) f(X) f^\T(X) \right]  , \quad
 \tilde \Sigma_f = \tilde  E  \left[ Z  w_1(X;\bar\gamma_1) f(X) f^\T(X) \right].
\end{align*}

\begin{lem} \label{lem:prob-mat-f}
Under Assumptions~\ref{ass:gamma1}(i)--(ii), there exists a positive constant $B_{f1}$,
depending on $(C_{f0}, C_{f1})$, such that
$P(\Omega_{f1} ) \ge 1- 2\epsilon$, where $\Omega_{f1}$ denotes the event
\begin{align*}
\sup_{j,k=0,1,\ldots,p} \left| ( \tilde \Sigma_f)_{jk} - (\Sigma_f)_{jk} \right| \le B_{f1} \lambda_0 .
\end{align*}
\end{lem}

\begin{prf}
This is Lemma~1(ii) in the Supplement of Tan (2020b).
\end{prf}

Denote
\begin{align*}
& \Sigma_{f2} = E  \left[ Z  w_1(X;\bar\gamma_1)  \{ DY - m_1(X; \bar{\alpha}_1) \psi_{\scriptscriptstyle Y} (\bar\alpha^\T _{11} h) \}^2 f(X) f^\T(X) \right]  , \\
& \tilde \Sigma_{f2} = \tilde  E  \left[ Z  w_1(X;\bar\gamma_1)  \{ DY - m_1(X; \bar{\alpha}_1) \psi_{\scriptscriptstyle Y} (\bar\alpha^\T _{11} h) \}^2 f(X) f^\T(X) \right].
\end{align*}

\begin{lem} \label{lem:prob-mat-f2}
Under Assumptions~\ref{ass:gamma1}(i)--(ii) and \ref{ass:alpha11}(ii), there exists a positive constant $B_{f2}$,
depending on $(C_{f0}, C_{f1}, \sigma_0, \sigma_1)$, such that if $\lambda_0 \le 1$, then
$P(\Omega_{f2} ) \ge 1- 2\epsilon$, where $\Omega_{f2}$ denotes the event
\begin{align*}
\sup_{j,k=0,1,\ldots,p} \left| ( \tilde \Sigma_{f2})_{jk} - (\Sigma_{f2})_{jk} \right| \le B_{f2} \lambda_0 .
\end{align*}
\end{lem}

\begin{prf}
This can be shown similarly as Lemma~3 in the Supplement of Tan (2020b).
\end{prf}

In the event $\Omega_{f1} \cap \Omega_{f2}$, we have for any vector $b \in \bbR^{1+p}$,
\begin{align*}
& \left| (\tilde E - E) \left\{  Z  w_1(X;\bar\gamma_1) (b^\T f)^2 \right\} \right| \le B_{f1} \lambda_0 \| b\|_1^2 , \\
& \left| (\tilde E - E) \left\{  Z  w_1(X;\bar\gamma_1) \{ DY - m_1(X; \bar{\alpha}_1) \psi_{\scriptscriptstyle Y} (\bar\alpha^\T _{11} h) \}^2 (b^\T f)^2 \right\} \right| \le B_{f2} \lambda_0 \| b\|_1^2 .
\end{align*}
By Assumption~\ref{ass:alpha11}(ii), $E [ \{ DY - m_1(X; \bar{\alpha}_1) \psi_{\scriptscriptstyle Y} (\bar\alpha^\T _{11} h) \}^2  |X ] \le \sigma_0^2 + \sigma_1^2$
and hence
\begin{align*}
& E  \left[ Z  w_1(X;\bar\gamma_1)  \left\{ DY - m_1(X; \bar{\alpha}_1) \psi_{\scriptscriptstyle Y} (\bar\alpha^\T _{11} h) \right\}^2 (b^\T f)^2 \right] \\
& \le (\sigma_0^2 + \sigma_1^2) E  \left\{ Z  w_1(X;\bar\gamma_1)  f(X) (b^\T f)^2 \right\}.
\end{align*}
Combining the preceding three inequalities shows that in the event $\Omega_{f1} \cap \Omega_{f2}$,
\begin{align}
& \tilde E  \left\{  Z  w_1(X;\bar\gamma_1) \{ DY - m_1(X; \bar{\alpha}_1) \psi_{\scriptscriptstyle Y} (\bar\alpha^\T _{11} h) \}^2 (b^\T f)^2 \right\}  \nonumber \\
& \le (\sigma_0^2 + \sigma_1^2) \tilde E  \left\{ Z  w_1(X;\bar\gamma_1) (b^\T f)^2 \right\} + \{B_{f2} + (\sigma_0^2 + \sigma_1^2) B_{f1}\} \lambda_0 \|b\|_1^2 .  \label{eq:ineq-mat-f2}
\end{align}

The following Lemmas~\ref{lem:prob-mat-f3}--\ref{lem:prob-grad-pi-alpha11} are used in the proof of Theorem~\ref{thm:expansion-DY} (Section~\ref{sec:prf-thm-expansion-DY}).
Denote
\begin{align*}
& \Sigma_{f3} = E  \left\{ Z  w_1(X;\bar\gamma_1)  |DY - m_1(X; \bar{\alpha}_1) \psi_{\scriptscriptstyle Y} (\bar\alpha^\T _{11} h) | f(X) f^\T(X) \right\}  , \\
& \tilde \Sigma_{f3} = \tilde  E  \left\{ Z  w_1(X;\bar\gamma_1)  |DY - m_1(X; \bar{\alpha}_1) \psi_{\scriptscriptstyle Y} (\bar\alpha^\T _{11} h) | f(X) f^\T(X) \right\} .
\end{align*}

\begin{lem} \label{lem:prob-mat-f3}
Under Assumptions~\ref{ass:gamma1}(i)--(ii) and \ref{ass:alpha11}(ii), there exists a positive constant $B_{f3}$,
depending on $(C_{f0}, C_{f1}, \sigma_0, \sigma_1)$, such that
$P(\Omega_{f3} ) \ge 1- 2\epsilon$, where $\Omega_{f2}$ denotes the event
\begin{align*}
\sup_{j,k=0,1,\ldots,p} \left| ( \tilde \Sigma_{f3})_{jk} - (\Sigma_{f3})_{jk} \right| \le B_{f3} \lambda_0 .
\end{align*}
\end{lem}

\begin{prf}
This can be shown similarly as Lemma~4 in the Supplement of Tan (2020b).
\end{prf}

Similarly as (\ref{eq:ineq-mat-f2}), we have in the event $\Omega_{f1} \cap \Omega_{f3}$, for any vector $b \in \bbR^{1+p}$,
\begin{align}
& \tilde E  \left\{  Z  w_1(X;\bar\gamma_1) | DY - m_1(X; \bar{\alpha}_1) \psi_{\scriptscriptstyle Y} (\bar\alpha^\T _{11} h) | (b^\T f)^2 \right\}  \nonumber \\
& \le (\sigma_0^2 + \sigma_1^2)^{1/2} \tilde E  \left\{ Z  w_1(X;\bar\gamma_1) (b^\T f)^2 \right\} + \{B_{f3} + (\sigma_0^2 + \sigma_1^2)^{1/2} B_{f1}\} \lambda_0 \|b\|_1^2 .  \label{eq:ineq-mat-f3}
\end{align}

\begin{lem} \label{lem:prob-grad-pi-alpha1}
Suppose that Assumptions~\ref{ass:gamma1}(ii), \ref{ass:alpha1}(i)-(iii), and \ref{ass:alpha11}(iii) hold. Then
$P(\Omega_{2,r_1} ) \ge 1- 2 \epsilon$ for any $r_1 \ge 0$, where $\Omega_{2,r_1}$ denotes the event
\begin{align*}
& \sup_{ \| \alpha_1 - \bar\alpha_1\| \le r_1}
\left| \tilde E \left[\left\{1- \frac{Z}{\pi^* (X)} \right\} \bar m_{11} (X) \{ \psi_{\scriptscriptstyle D}(\alpha_1^\T g) - \psi_{\scriptscriptstyle D} (\bar \alpha_1^\T g ) \} \right] \right| \\
& \le 4 (\me^{-C_{f1}}+1) C_{g0} \tilde C_{h1} \psi_{\scriptscriptstyle D}^\prime (0) \me^{C_{g2} (C_{g1} + C_{g0} r_1)} r_1 \lambda_1 .
\end{align*}
Here $\tilde C_{h1} = \max\{ |\psi_{\scriptscriptstyle Y}(-C_{h1})|, |\psi_{\scriptscriptstyle Y}(C_{h1})|\}$.
\end{lem}

\begin{prf}
Using (\ref{eq:remove-hat-alpha1-b}), this can be shown similarly as Lemma~13 in the Supplement of Tan (2020b).
\end{prf}

\begin{lem} \label{lem:prob-grad-pi-alpha11}
Suppose that Assumptions~\ref{ass:gamma1}(ii) and \ref{ass:alpha11}(i),(iii),(iv) hold. Then
$P(\Omega_{3,r_2} ) \ge 1- 2 \epsilon$ for any $r_2 \ge 0$, where $\Omega_{3,r_2}$ denotes the event
\begin{align*}
& \sup_{ \alpha_1\in \bbR^{1+q_1}, \| \alpha_{11} - \bar\alpha_{11} \| \le r_2}
\left| \tilde E \left[\left\{1- \frac{Z}{\pi^* (X)} \right\} \psi_{\scriptscriptstyle D} (\alpha_1^\T g) \{ \psi_{\scriptscriptstyle Y}(\alpha_{11}^\T h) - \psi_{\scriptscriptstyle Y} (\bar \alpha_{11}^\T h ) \} \right] \right| \\
& \le 4 (\me^{-C_{f1}}+1) C_{h0} \psi_{\scriptscriptstyle Y}^\prime (0) \me^{C_{h2} (C_{h1} + C_{h0} r_2)} r_2 \lambda_2 .
\end{align*}
\end{lem}

\begin{prf}
Using (\ref{eq:remove-hat-alpha11}), this can be shown similarly as Lemma~13 in the Supplement of Tan (2020b).
\end{prf}

\subsection{Proof of Theorem~\ref{thm:alpha11}} \label{sec:prf-thm-alpha11}

We split the proof of Theorem \ref{thm:alpha11} into a series of lemmas.
The first one is usually called a basic inequality for $\hat\alpha_{11}$, but depending on the first-step estimators $(\hat\gamma_1,\hat\alpha_1)$.

\begin{lem} \label{lem:basic-ineq-hat}
For any coefficient vector $\alpha_{11}$, we have
\begin{align}
& D ^\dag_{11,\text{\tiny WL}}(\hat\alpha_{11}, \alpha_{11}; \hat\gamma_1, \hat{\alpha}_1)
+ \lambda \| (\hat\alpha_{11})_{1:q_2} \|_1 \nonumber \\
& \le
(\hat\alpha_{11} - \alpha_{11})^\T
\tilde E\left[  Z  w_1(X;\hat\gamma_1)
\{ DY - m_1(X; \hat{\alpha}_1) \psi_{\scriptscriptstyle Y} (\alpha^\T _{11} h) \} h \right]
+ \lambda \| (\alpha_{11})_{1:q_2} \|_1 . \label{eq:basic-ineq-hat}
\end{align}
\end{lem}

\begin{prf}
This can be shown similarly as Lemma~6 in the Supplement of Tan (2020b).
\end{prf}

The second lemma deals with the dependency on $(\hat\gamma_1,\hat\alpha_1)$ in the upper bound from the basic inequality (\ref{eq:basic-ineq-hat}).
Denote
\begin{align*}
Q(\hat\alpha_{11},\bar\alpha_{11}; \bar\gamma_1)  =
\tilde E \left[  Z  w_1(X;\bar\gamma_1)  (\hat\alpha_{11}^\T h - \bar\alpha_{11}^\T h )^2  \right].
\end{align*}

\begin{lem} \label{lem:remove-hat}
In the event $\Omega_0 \cap \Omega_{f1} \cap \Omega_{f2}$, we have
\begin{align*}
& (\hat\alpha_{11} - \bar\alpha_{11})^\T
\tilde E\left[  Z  w_1(X;\hat\gamma_1)
\{ DY - m_1(X; \hat{\alpha}_1) \psi_{\scriptscriptstyle Y} (\bar\alpha^\T _{11} h) \} h \right] \\
& \le (\hat\alpha_{11} - \bar\alpha_{11})^\T
\tilde E\left[  Z  w_1(X;\bar\gamma_1)
\{ DY - m_1(X; \bar{\alpha}_1) \psi_{\scriptscriptstyle Y} (\bar\alpha^\T _{11} h) \} h \right] \\
& \quad + \left\{ (M_{01} |S_{\bar\gamma_1}| \lambda_0^2)^{1/2} + M^{1/2}_{11} (|S_{\bar\gamma_1}| \lambda_0^2 + |S_{\bar\alpha_1}| \lambda_1^2)^{1/2} \right\}
\left\{ Q(\hat\alpha_{11},\bar\alpha_{11}; \bar\gamma_1) \right\}^{1/2},
\end{align*}
where $M_{01} = \me^{ 2 C_{f0} M_0 \varrho_0 } [ (\sigma_0^2 + \sigma_1^2) M_0 +  \{B_{f2} + (\sigma_0^2 + \sigma_1^2) B_{f1}\} M_0^2 \varrho_0]$ and,
with $\tilde C_{h1} = \max\{ |\psi_{\scriptscriptstyle Y}(-C_{h1})|, $ $|\psi_{\scriptscriptstyle Y} (C_{h1})|\}$,
$M_{11} = \psi_{\scriptscriptstyle D}^{\prime 2} (0) \tilde C^2_{h1} \me^{2C_{f0} M_0 \varrho_0 +2 C_{g2}(C_{g1}+ C_{g0} M_1 \varrho_1) } M_1$
\end{lem}

\begin{prf}
Consider the following decomposition
\begin{align*}
& (\hat\alpha_{11} - \bar\alpha_{11})^\T
\tilde E\left[  Z  w_1(X;\hat\gamma_1)
\{ DY - m_1(X; \hat{\alpha}_1) \psi_{\scriptscriptstyle Y} (\bar\alpha^\T _{11} h) \} h \right] \\
& = (\hat\alpha_{11} - \bar\alpha_{11})^\T
\tilde E\left[  Z  w_1(X;\bar\gamma_1)
\{ DY - m_1(X; \bar{\alpha}_1) \psi_{\scriptscriptstyle Y} (\bar\alpha^\T _{11} h) \} h \right] + \Delta_1 + \Delta_2,
\end{align*}
where
\begin{align*}
& \Delta_1 = (\hat\alpha_{11} - \bar\alpha_{11})^\T
\tilde E\left[  Z  \{ w_1(X;\hat\gamma_1)- w_1(X;\bar\gamma_1) \}
\{ DY - m_1(X; \bar{\alpha}_1) \psi_{\scriptscriptstyle Y} (\bar\alpha^\T _{11} h) \} h \right] , \\
& \Delta_2 = (\hat\alpha_{11} - \bar\alpha_{11})^\T
\tilde E\left[  Z  w_1(X; \hat\gamma_1) \{ m_1(X; \hat{\alpha}_1) - m_1(X; \bar{\alpha}_1) \}
  \psi_{\scriptscriptstyle Y} (\bar\alpha^\T _{11} h) h \right].
\end{align*}
To handle $\Delta_1$, we have by the mean value theorem and Assumption~\ref{ass:gamma1}(i),
\begin{align}
& |  w_1(X;\hat\gamma_1)- w_1(X;\bar\gamma_1) |  = \left| \me^{ -\hat\gamma_1^\T f } - \me^{ -\bar\gamma_1^\T f} \right| \nonumber \\
& \le \me^{ -\bar\gamma_1^\T f} \me^{ | \hat\gamma_1^\T f  - \bar\gamma_1^\T f | } |\hat\gamma_1^\T f - \bar\gamma_1^\T f|
\le \me^{ C_{f0} \| \hat\gamma_1- \bar\gamma_1\|_1 } w_1(X;\bar\gamma_1) |\hat\gamma_1^\T f - \bar\gamma_1^\T f| . \label{eq:remove-hat-gamma}
\end{align}
By the Cauchy--Schwartz inequality, we have in the event $\Omega_0 \cap \Omega_{f1} \cap \Omega_{f2}$,
\begin{align}
& | \Delta_1 |
 \le \me^{ C_{f0} \| \hat\gamma_1- \bar\gamma_1\|_1 }  \tilde E^{1/2} \left[ Z  w_1(X;\bar\gamma_1)
 \{ DY - m_1(X; \bar{\alpha}_1) \psi_{\scriptscriptstyle Y} (\bar\alpha^\T _{11} h) \}^2
 (\hat\gamma_1^\T f - \bar\gamma_1^\T f)^2 \right] \nonumber \\
& \quad \times \tilde E^{1/2} \left[  Z   w_1(X;\bar\gamma_1)  (\hat\alpha_{11}^\T h - \bar\alpha_{11}^\T h )^2 \right] \nonumber \\
& \le (M_{01} |S_{\bar\gamma_1}| \lambda_0^2 )^{1/2}
 \tilde E^{1/2} \left[  Z   w_1(X;\bar\gamma_1)  (\hat\alpha_{11}^\T h - \bar\alpha_{11}^\T h )^2 \right], \label{eq:remove-hat-prf1}
\end{align}
where $M_{01} = \me^{ 2 C_{f0} M_0 \varrho_0 } [ (\sigma_0^2 + \sigma_1^2) M_0 +  \{B_{f2} + (\sigma_0^2 + \sigma_1^2) B_{f1}\} M_0^2 \varrho_0]$. The second step follows because
\begin{align*}
& \tilde E \left[ Z  w_1(X;\bar\gamma_1) \{ DY - m_1(X; \bar{\alpha}_1) \psi_{\scriptscriptstyle Y} (\bar\alpha^\T _{11} h) \}^2
 (\hat\gamma_1^\T f - \bar\gamma_1^\T f)^2 \right]  \\
& \le (\sigma_0^2 + \sigma_1^2) M_0 |S_{\bar\gamma_1}| \lambda_0^2 +  \{B_{f2} + (\sigma_0^2 + \sigma_1^2) B_{f1}\} M_0^2 \varrho_0 |S_{\bar\gamma_1}| \lambda_0^2
\end{align*}
in the event $\Omega_0 \cap \Omega_{f1} \cap \Omega_{f1}$ by (\ref{eq:bound-gamma1}), (\ref{eq:ineq-mat-f2}) with $b=\hat\gamma_1-\bar\gamma_1$, and Assumption~\ref{ass:alpha11}(vi).
To handle $\Delta_2$, we have $ w_1(X;\hat\gamma_1)  \le \me^{C_{f0} \| \hat\gamma_1 - \bar\gamma_1\|_1}  w_1(X;\bar\gamma_1) $
by the mean value theorem and Assumption~\ref{ass:gamma1}(i).
Moreover, by Assumptions~\ref{ass:alpha1}(i)--(iii),
\begin{align}
& | m_1(X; \hat{\alpha}_1) - m_1(X; \bar{\alpha}_1)| =
| \psi_{\scriptscriptstyle D} (\hat{\alpha}_1^\T g )  - \psi_{\scriptscriptstyle D} (\bar{\alpha}_1^\T g ) | \nonumber \\
& =  \left| (\hat{\alpha}_1-\bar{\alpha}_1 )^\T g  \right|  \int_0^1 \psi_{\scriptscriptstyle D}^\prime
(\bar \alpha_1^\T g + u (\hat \alpha_1 - \bar \alpha_1 )^\T g ) \,\dif u \nonumber \\
& \le \left| (\hat{\alpha}_1-\bar{\alpha}_1 )^\T g  \right|  \psi_{\scriptscriptstyle D}^\prime
 (\bar \alpha_1^\T g ) \int_0^1 \me^{C_{g2} u |(\hat{\alpha}_1-\bar{\alpha}_1 )^\T g | } \,\dif u \label{eq:remove-hat-alpha1} \\
& \le \left| (\hat{\alpha}_1-\bar{\alpha}_1 )^\T g  \right|  \psi_{\scriptscriptstyle D}^\prime (0) \me^{C_{g2} (C_{g1}+ C_{g0} \|\hat{\alpha}_1-\bar{\alpha}_1\|_1) }. \label{eq:remove-hat-alpha1-b}
\end{align}
By the Cauchy--Schwartz inequality and Assumption~\ref{ass:alpha11}(ii),
we have in the event $\Omega_0$,
\begin{align}
& |\Delta_2 | \le  \psi_{\scriptscriptstyle D}^\prime (0)\psi_{\scriptscriptstyle Y} (C_{h1}) \tilde C_{h1}  \me^{C_{g2} (C_{g1}+ C_{g0} \|\hat{\alpha}_1-\bar{\alpha}_1\|_1) } \nonumber \\
& \quad \times \tilde E^{1/2} \left[  Z  w_1(X;\bar\gamma_1)  (\hat{\alpha}_1^\T g - \bar{\alpha}_1^\T g)^2 \right]
\tilde E^{1/2}  \left[  Z  w_1(X;\bar\gamma_1)  (\hat\alpha_{11}^\T h - \bar\alpha_{11}^\T h )^2  \right] \nonumber \\
& \le  M_{11}^{1/2} ( |S_{\bar\gamma_1}| \lambda_0^2 + |S_{\bar\alpha_1}| \lambda_1^2 )^{1/2} \tilde E^{1/2}  \left[  Z  w_1(X;\bar\gamma_1)  (\hat\alpha_{11}^\T h - \bar\alpha_{11}^\T h )^2  \right] , \label{eq:remove-hat-prf2}
\end{align}
where $\tilde C_{h1} = \max\{ |\psi_{\scriptscriptstyle Y}(-C_{h1})|, |\psi_{\scriptscriptstyle Y} (C_{h1})|\}$
and  $M_{11} = \psi_{\scriptscriptstyle D}^{\prime 2} (0 ) \tilde C^2_{h1} \me^{ 2C_{f0} M_0 \varrho_0 +2C_{g2}  (C_{g1}+C_{g0} M_1 \varrho_1 ) } M_1$.
The second step follows by (\ref{eq:bound-alpha1}) and Assumption~\ref{ass:alpha11}(vi).
Combining (\ref{eq:remove-hat-prf1})--(\ref{eq:remove-hat-prf2}) yields the desired inequality.
\end{prf}

The third lemma derives an implication of the basic inequality (\ref{eq:basic-ineq-hat}) using the triangle inequality for the $L_1$ norm,
while incorporating the bound from Lemma~\ref{lem:remove-hat}.

\begin{lem} \label{lem:basic-ineq-bar}
Denote $b =\hat\alpha_{11} - \bar\alpha_{11}$. In the event $\Omega_0 \cap\Omega_1 \cap \Omega_{f1} \cap \Omega_{f2}$, we have
\begin{align}
& D ^\dag_{11,\text{\tiny WL}}(\hat\alpha_{11}, \bar\alpha_{11}; \hat\gamma_1, \hat{\alpha}_1)
+ (A_2- B_1) \lambda_2  \| b \|_1 \nonumber \\
& \le \left\{ (M_{01} |S_{\bar\gamma_1}| \lambda_0^2)^{1/2} + M^{1/2}_{11} (|S_{\bar\gamma_1}| \lambda_0^2 + |S_{\bar\alpha_1}| \lambda_1^2)^{1/2} \right\}
\left\{ Q(\hat\alpha_{11},\bar\alpha_{11}; \bar\gamma_1) \right\}^{1/2}
+ 2 A_2 \lambda_2 \sum_{j\in S_{\bar \alpha_{11}}} |b_j| . \label{eq:basic-ineq-bar}
\end{align}
\end{lem}

\begin{prf}
In the event $\Omega_1$, we have
\begin{align*}
b^\T \tilde E\left[  Z  w_1(X;\bar\gamma_1)
\{ DY - m_1(X; \bar{\alpha}_1) \psi_{\scriptscriptstyle Y} (\bar\alpha^\T _{11} h) \} h \right] \le B_1 \lambda_2 \|b\|_1 .
\end{align*}
From this bound and  Lemmas~\ref{lem:basic-ineq-hat}--\ref{lem:remove-hat}, we have in the event $\Omega_0 \cap\Omega_1 \cap \Omega_{f1} \cap \Omega_{f2}$,
\begin{align*}
& D ^\dag_{11,\text{\tiny WL}}(\hat\alpha_{11}, \alpha_{11}; \hat\gamma_1, \hat{\alpha}_1)
+ A_2 \lambda_2  \| (\hat\alpha_{11})_{1:q_2} \|_1 \\
& \le  B_1 \lambda_2\|b\|_1 + A_2 \lambda_2 \| (\bar \alpha_{11})_{1:q_2} \|_1 \\
& \quad + \left\{ (M_{01} |S_{\bar\gamma_1}| \lambda_0^2)^{1/2} + M^{1/2}_{11} (|S_{\bar\gamma_1}| \lambda_0^2 + |S_{\bar\alpha_1}| \lambda_1^2)^{1/2} \right\}
\left\{ Q(\hat\alpha_{11},\bar\alpha_{11}; \bar\gamma_1) \right\}^{1/2}.
\end{align*}
Using  the identity $ | (\hat\alpha_{11}) _j| = | (\hat\alpha_{11}-\bar\alpha_{11})_j|$ for $j \not\in S_{\bar\alpha_{11}}$
and the triangle inequality $ | (\hat\alpha_{11}) _j| \ge |(\bar\alpha_{11}) _j| - | (\hat\alpha_{11}-\bar\alpha_{11})_j|$ for $j \in S_{\bar\alpha_{11}} \setminus \{0\}$
and rearranging the result yields (\ref{eq:basic-ineq-bar}).
\end{prf}

The following lemma provides a desired bound relating the Bregman divergence $D^\dag_{11,\text{\tiny WL}} (\alpha_{11}, \bar\alpha_{11};$ $ \hat\gamma_1, \hat\alpha_1)$
with the quadratic function $(\alpha - \bar\alpha_{11})^\T \tilde \Sigma_h(\alpha - \bar\alpha_{11})$.

\begin{lem} \label{lem:local-quad}
In the event $\Omega_0$, we have for any vector $b \in \bbR^{1+q_2}$,
\begin{align*}
& D ^\dag_{11,\text{\tiny WL}}( \alpha_{11}, \bar\alpha_{11}; \hat\gamma_1, \hat{\alpha}_1)
\ge C_{g3} C_{h3} \frac{1- \me^{-C_{h2}C_{h0} \|b\|_1}}{C_{h2}C_{h0}\|b\|_1}
(b^\T \tilde \Sigma_h b),
\end{align*}
where $b = \alpha_{11}-\bar\alpha_{11}$, $C_{g3} =  \me^{- C_{f0} M_0 \varrho_0}\psi_{\scriptscriptstyle D} (-C_{g1}) (1-\varrho_4) $, and $C_{h3} = \psi^\prime _{\scriptscriptstyle Y} (0) \me^{-C_{h2} C_{h1}}$.
Throughout, set $(1-\me^{-c})/c=1$ for $c=0$.
\end{lem}

\begin{prf}
Direct calculation yields
\begin{align*}
& D ^\dag_{11,\text{\tiny WL}}(\alpha_{11}, \bar\alpha_{11}; \hat\gamma_1, \hat{\alpha}_1) \\
& = (\alpha_{11} - \bar \alpha_{11})^\T
\tilde E\left[  Z  w_1(X;\hat\gamma_1) m_1(X; \hat{\alpha}_1)
\{ \psi_{\scriptscriptstyle Y} (\alpha^\T _{11} h ) - \psi_{\scriptscriptstyle Y} (\bar \alpha^\T _{11} h) \} \right] \\
& = \tilde E \left[  Z  w_1(X;\hat\gamma_1) m_1(X; \hat{\alpha}_1) \left\{\int_0^1  \psi^\prime _{\scriptscriptstyle Y} (\bar \alpha_{11}^\T h + u (\alpha_{11} - \bar \alpha_{11} )^\T h )  \,\dif u \right\}
(\alpha_{11}^\T h -\bar\alpha_{11}^\T h)^2 \right] .
\end{align*}
By the mean value theorem and Assumption~\ref{ass:gamma1}(i), $ w_1(X;\hat\gamma_1)  \ge \me^{- C_{f0} \| \hat\gamma_1 - \bar\gamma_1\|_1}  w_1(X;\bar\gamma_1) $.
By inequality (\ref{eq:remove-hat-alpha1}) and Assumptions~\ref{ass:alpha1}(i)--(iii), we have
\begin{align*}
& m_1(X; \hat{\alpha}_1)  \ge  m_1(X; \bar {\alpha}_1)\left\{ 1 -  C_{g0} \|\hat{\alpha}_1-\bar{\alpha}_1\|_1
 C_{g2} \me^{ C_{g2} C_{g0} \|\hat{\alpha}_1-\bar{\alpha}_1\|_1 }  \right\} \\
& \ge \psi_{\scriptscriptstyle D} (-C_{g1}) \left\{ 1 -  C_{g0} \|\hat{\alpha}_1-\bar{\alpha}_1\|_1
 C_{g2} \me^{ C_{g2} C_{g0} \|\hat{\alpha}_1-\bar{\alpha}_1\|_1 } \right\} .
\end{align*}
The first step follows because, with $\psi_{\scriptscriptstyle D}(-\infty)=0$,
$ \psi_{\scriptscriptstyle D} (\eta) = \int_{-\infty}^\eta \psi_{\scriptscriptstyle D}^\prime (t) \,\dif t \ge \psi_{\scriptscriptstyle D}^\prime(\eta) \int_{-\infty}^\eta \me^{-C_{g2} (\eta - t)}\,\dif t $ $= \psi_{\scriptscriptstyle D}^\prime(\eta) / C_{g2}$
and hence $ \psi_{\scriptscriptstyle D}^\prime(\eta)  \le C_{g2} \psi_{\scriptscriptstyle D} (\eta)$ for any $\eta \in \bbR$.
Then by (\ref{eq:bound-gamma1})--(\ref{eq:bound-alpha1}) and Assumption~\ref{ass:alpha11}(vi), we obtain in the event $\Omega_0$,
\begin{align*}
& D ^\dag_{11,\text{\tiny WL}}(\alpha^\prime_{11}, \alpha_{11}; \hat\gamma_1, \hat{\alpha}_1) \\
& \ge C_{g3} \tilde E \left[  Z  w_1(X;\bar\gamma_1)  \left\{\int_0^1  \psi^\prime _{\scriptscriptstyle Y} (\bar \alpha_{11}^\T h + u (\hat \alpha_{11} - \bar \alpha_{11} )^\T h )  \,\dif u \right\}
(\alpha_{11}^\T h -\bar\alpha_{11}^\T h)^2 \right] ,
\end{align*}
where $C_{g3} = \me^{- C_{f0} M_0 \varrho_0}  \psi_{\scriptscriptstyle D} (-C_{g1}) (1-\varrho_4)  $,
with $\varrho_4 = C_{g0} M_1 \varrho_1 C_{g2} \me^{ C_{g2} C_{g0} M_1 \varrho_1} <1$.
Furthermore, by Assumption~\ref{ass:alpha11}(iii)--(iv),  we have
\begin{align*}
& D ^\dag_{11,\text{\tiny WL}}(\alpha^\prime_{11}, \alpha_{11}; \hat\gamma_1, \hat{\alpha}_1) \\
& \ge C_{g3} \tilde E \left[  Z  w_1(X;\bar\gamma_1) \psi^\prime _{\scriptscriptstyle Y} (\bar \alpha_{11}^\T h ) \left\{\int_0^1 \me^{-C_{h2} u |(\hat \alpha_{11} - \bar \alpha_{11} )^\T h  |} \,\dif u \right\}
(\alpha_{11}^\T h -\bar\alpha_{11}^\T h)^2 \right] \\
& \ge C_{g3} C_{h3} \left\{\int_0^1 \me^{-C_{h2} C_{h0} \|\hat \alpha_{11} - \bar \alpha_{11}\|_1 } \,\dif u \right\}
\tilde E \left\{  Z  w_1(X;\bar\gamma_1) (\alpha_{11}^\T h -\bar\alpha_{11}^\T h)^2 \right\} ,
\end{align*}
where $C_{h3} = \psi^\prime _{\scriptscriptstyle Y} (0) \me^{-C_{h2} C_{h1}}$.
The desired result follows because $\int_0^1 \me^{-c u}\,\dif u = \frac{1- \me^{-c}}{c}$ for $c \ge 0$.
\end{prf}

The following lemma shows that Assumption~\ref{ass:alpha11}(v), a theoretical compatibility condition for $\Sigma_h$, implies
an empirical compatibility condition for $\tilde\Sigma_h$.

\begin{lem} \label{lem:emp-compat}
In the event $\Omega_{h1}$, Assumption \ref{ass:alpha11}(v) implies that
for any vector $b \in \bbR^{1+q_2}$ such that $\sum_{j\not\in S_{\bar\alpha_{11}}} |b_j|  \le \mu_2 \sum_{j \in S_{\bar\alpha_{11}}} |b_j| $, we have
\begin{align*}
 \nu_{21}^2 \left( \sum_{j \in S_{\bar\alpha_{11}}} |b_j| \right)^2 \le |S_{\bar\alpha_{11}}| \left( b^\T \tilde \Sigma_h b \right),
\end{align*}
where $\nu_{21} = \nu_2 \{1- \nu_2^{-2} (1+\mu_2)^2 \varrho_2 B_{h1} \}^{1/2} = \nu_2 (1-\varrho_3)^{1/2}$.
\end{lem}

\begin{prf}
In the event $\Omega_{h1}$, we have $ |b^\T (\tilde \Sigma_h - \Sigma_h) b | \le B_{h1} \lambda_2 \|b\|_1^2$ from Lemma~\ref{lem:prob-mat-h}.
Then Assumption~\ref{ass:alpha11}(v) implies that for any $b=(b_0,b_1,\ldots,b_p)^\T $ satisfying $\sum_{j\not \in S_{\bar\alpha_{11}}} |b_j| \le \mu_2 \sum_{j\in S_{\bar\alpha_{11}}} |b_j|$,
\begin{align*}
&\nu_2^2 \|b_{S_{\bar\alpha_{11}}} \|_1^2 \le |S_{\bar\alpha_{11}}| (b^\T \Sigma_h b) \le |S_{\bar\alpha_{11}}| \left(b^\T \tilde \Sigma_h b + B_{h1}\lambda_0 \|b\|_1^2 \right) \\
& \le |S_{\bar\alpha_{11}}| (b^\T \tilde \Sigma_h b ) + B_{h1} |S_{\bar\alpha_{11}}| \lambda_2 (1+\mu_2)^2 \|b_{S_{\bar\alpha_{11}}} \|_1^2 ,
\end{align*}
where $\|b_{S_{\bar\alpha_{11}}} \|_1=\sum_{j\in S_{\bar\alpha_{11}}} |b_j|$. The last inequality uses $\|b\|_1 \le (1+\mu_1) \| b_{S_{\bar\alpha_{11}}}\|_1$.
The desired result follows because $ |S_{\bar\alpha_{11}}| \lambda_2 \le \varrho_2$ and $\varrho_3=\nu_2^{-2} B_{h1} (1+\mu_2)^2 \varrho_2 < 1$ by Assumption~\ref{ass:alpha11}(vi).
\end{prf}

The final lemma completes the proof of Theorem~\ref{thm:alpha11}, because $P(\Omega_0 \cap\Omega_1  \cap \Omega_{h1} \cap\Omega_{f1} \cap \Omega_{f2}) \ge 1-(c_0+8)\epsilon$ by
probability Lemmas~\ref{lem:prob-grad}--\ref{lem:prob-mat-f2} in Section~\ref{sec:prob-lem}.

\begin{lem} \label{lem:invert}
For $A_2 > B_1 (\mu_2+1)/(\mu_2-1)$, inequality (\ref{eq:bound-alpha11}) holds in the event $\Omega_0 \cap\Omega_1  \cap \Omega_{h1} \cap\Omega_{f1} \cap \Omega_{f2}$:
\begin{align*}
& D^\dag_{11,\text{\tiny WL}} (  \hat \alpha_{11} , \bar \alpha_{11};  \bar \gamma_1, \bar\alpha_1) + (A_2 - B_1) \lambda_2 \|\hat\alpha_{11} - \bar\alpha_{11} \|_1 \\
& \le C_{g3}^{-1} C_{h3}^{-1} \left\{ 2 \mu_{21}^{-2} (M_{01}+M_{11}) |S_{\bar\gamma_1}| \lambda_0^2 + 2 \mu_{21}^{-2} M_{11} |S_{\bar\alpha_1}| \lambda_1^2
 + \mu_{22}^2 \nu_{21}^{-2}  |S_{\bar\alpha_{11}}| \lambda_2^2 \right\} .
\end{align*}
\end{lem}

\begin{prf}
Denote $b= \hat\alpha_{11} - \bar\alpha_{11}$,
$D^\dag_{11,\text{\tiny WL}} = D^\dag_{11,\text{\tiny WL}} (  \hat \alpha_{11} , \bar \alpha_{11};  \bar \gamma_1, \bar\alpha_1)$,
$Q = Q (  \hat \alpha_{11} , \bar \alpha_{11} ; \bar \gamma_1, \bar\alpha_1)$, and
$$
D^\ddag_{11,\text{\tiny WL}} =  D^\dag_{11,\text{\tiny WL}} +
(A_2 - B_1) \lambda_2 \| b \|_1 .
$$
In the event $\Omega_0 \cap \Omega_{h1} \cap\Omega_{f1} \cap \Omega_{f2}$, inequality (\ref{eq:basic-ineq-bar})
from Lemma~\ref{lem:basic-ineq-bar}  leads to two possible cases: either
\begin{align}
\mu_{21} D^\ddag_{11,\text{\tiny WL}}  \le \left\{ (M_{01} |S_{\bar\gamma_1}| \lambda_0^2)^{1/2} + M^{1/2}_{11} (|S_{\bar\gamma_1}| \lambda_0^2 + |S_{\bar\alpha_1}| \lambda_1^2)^{1/2} \right\}
Q ^{1/2}, \label{eq:invert-ineq1}
\end{align}
or $(1-\mu_{21}) D^\ddag_{11,\text{\tiny WL}} \le 2 A_2 \lambda_2 \sum_{j\in S_{\bar\alpha_{11}}} |b_j|$, that is,
\begin{align}
D^\ddag_{11,\text{\tiny WL}}
\le (\mu_2+1) (A_2-B_1) \lambda_2 \sum_{j\in S_{\bar\alpha_{11}}} |b_j| = \mu_{22} \lambda_2 \sum_{j\in S_{\bar\alpha_{11}}} |b_j| , \label{eq:invert-ineq2}
\end{align}
where $\mu_{21} = 1-2A_2 /\{ (\mu_2+1)(A_2-B_1)\} \in (0,1]$ because $A_2 > B_1 (\mu_2+1)/(\mu_2-1)$ and $\mu_{22}= (\mu_2+1)(A_2-B_1)$.
We deal with the two cases separately as follows.

If (\ref{eq:invert-ineq2}) holds, then $\sum_{j \not\in S_{\bar\alpha_{11}}} |b_j| \le \mu_2 \sum_{j\in S_{\bar\alpha_{11}}} |b_j|$,
which, by Lemma~\ref{lem:emp-compat}, implies
\begin{align}
\sum_{j\in S_{\bar\alpha_{11}}} |b_j| \le \nu_{21}^{-1} |S_{\bar\alpha_{11}}|^{1/2} \left( b^\T \tilde \Sigma_h b  \right)^{1/2}. \label{eq:invert-ineq3}
\end{align}
By Lemma~\ref{lem:local-quad}, we have
\begin{align}
D^\dag_{11,\text{\tiny WL}}
\ge C_{g3} C_{h3} \frac{1-\me^{-C_{h2} C_{h0} \|b\|_1 } }{C_{h2} C_{h0} \|b\|_1} \left( b^\T \tilde \Sigma_h b\right) . \label{eq:invert-ineq4}
\end{align}
Combining (\ref{eq:invert-ineq2}), (\ref{eq:invert-ineq3}), and (\ref{eq:invert-ineq4}) and using $D^\dag_{11,\text{\tiny WL}} \le D^\ddag_{11,\text{\tiny WL}}$ yields
\begin{align}
D^\ddag_{11,\text{\tiny WL}} \le \mu_{22}^2 \nu_{21}^{-2} C_{g3}^{-1} C_{h3}^{-1}|S_{\bar\alpha_{11}}|  \lambda_2^2  \frac{C_{h2} C_{h0} \|b\|_1}{1-\me^{-C_{h2} C_{h0} \|b\|_1 } }. \label{eq:invert-ineq5}
\end{align}
But $(A_2-B_1)\lambda_2 \|b\|_1 \le D^\ddag_{11,\text{\tiny WL}}$. Inequality (\ref{eq:invert-ineq5}) along with Assumption~\ref{ass:alpha11}(vi) implies that
$1 - \me^{- C_{h2} C_{h0} \|b\|_1} \le C_{h2} C_{h0} (A_2-B_1)^{-1} \mu_{22}^2 \nu_{21}^{-2} C_{g3}^{-1} C_{h3}^{-1} |S_{\bar\alpha_{11}}| \lambda_2 \le \varrho_5 \, (<1)$.
As a result, $C_{h2} C_{h0} \|b\|_1 \le - \log(1- \varrho_5)$ and hence
\begin{align}
\frac{1-\me^{-C_{h2} C_{h0} \|b\|_1 } }{C_{h2} C_{h0} \|b\|_1} = \int_0^1 \me^{-C_{h2} C_{h0} \|b\|_1 u} \dif u \ge \me^{-C_{h2} C_{h0} \|b\|_1} \ge 1-\varrho_5. \label{eq:invert-ineq6}
\end{align}
From this bound, inequality (\ref{eq:invert-ineq5}) then leads to $D^\ddag_{11,\text{\tiny WL}} \le \mu_{22}^2 \nu_{21}^{-2} C_{g3}^{-1} C_{h3}^{-1} |S_{\bar\alpha_{11}}| \lambda_2^2  $.

If (\ref{eq:invert-ineq1}) holds, then simple manipulation using $D^\dag_{11,\text{\tiny WL}} \le D^\ddag_{11,\text{\tiny WL}}$ and (\ref{eq:invert-ineq4}) together with
$Q =  b^\T \tilde \Sigma_h b$ gives
\begin{align}
D^\ddag_{11,\text{\tiny WL}}  \le \mu_{21}^{-2} C_{g3}^{-1} C_{h3}^{-1} \left\{ 2(M_{01}+M_{11}) |S_{\bar\gamma_1}| \lambda_0^2 + 2M_{11} |S_{\bar\alpha_1}| \lambda_1^2 \right\}
 \frac{C_{h2} C_{h0} \|b\|_1}{1-\me^{-C_{h2} C_{h0} \|b\|_1 } } . \label{eq:invert-ineq1b}
\end{align}
Similarly as above, using $(A_2-B_1)\lambda_2 \|b\|_1 \le D^\ddag_{11,\text{\tiny WL}}$ and inequality (\ref{eq:invert-ineq1b}) along with Assumption~\ref{ass:alpha11}(vi), we find
$1 - \me^{- C_{h2} C_{h0} \|b\|_1} \le C_{h2} C_{h0} (A_2-B_1)^{-1}\mu_{21}^{-2} C_{g3}^{-1} C_{h3}^{-1} \{ (M_{01}+M_{11}) |S_{\bar\gamma_1}| \lambda_0 + M_{11} |S_{\bar\alpha_1}| \lambda_1 \}
\le \varrho_6 \, (<1)$.
As a result, $C_{h2} C_{h0} \|b\|_1 \le - \log(1- \varrho_6)$ and hence
\begin{align}
\frac{1-\me^{-C_{h2} C_{h0} \|b\|_1 } }{C_{h2} C_{h0} \|b\|_1} = \int_0^1 \me^{-C_{h2} C_{h0} \|b\|_1 u} \dif u \ge \me^{-C_{h2} C_{h0} \|b\|_1} \ge 1-\varrho_6. \label{eq:invert-ineq1c}
\end{align}
From this bound, inequality (\ref{eq:invert-ineq1b}) then leads to $D^\ddag_{11,\text{\tiny WL}} \le \mu_{21}^{-2} C_{g3}^{-1} C_{h3}^{-1}
\{ 2(M_{01}+M_{11}) |S_{\bar\gamma_1}| \lambda_0^2 +  2 M_{11} |S_{\bar\alpha_1}| \lambda_1^2 \}$.
Therefore, (\ref{eq:bound-alpha11}) holds through (\ref{eq:invert-ineq1}) and (\ref{eq:invert-ineq2}) in the event $\Omega_0 \cap\Omega_1 \cap \Omega_{h1} \cap\Omega_{f1} \cap \Omega_{f2}$.
\end{prf}

From the preceding proof, inequality (\ref{eq:bound-alpha11-quad}) can also be deduced in the event $\Omega_0 \cap\Omega_1 \cap \Omega_{h1} \cap\Omega_{f1} \cap \Omega_{f2}$.
In fact, return to the two possible cases of (\ref{eq:invert-ineq1}) or (\ref{eq:invert-ineq2}).
If (\ref{eq:invert-ineq2}) holds, then we have, by (\ref{eq:invert-ineq4}) and (\ref{eq:invert-ineq6}), $b^\T \tilde \Sigma_h b \le C_{g3}^{-1} C_{h3}^{-1} (1-\varrho_5)^{-1} D^\dag_{11,\text{\tiny WL}} $.
If (\ref{eq:invert-ineq1}) holds, then we have, by (\ref{eq:invert-ineq4}) and (\ref{eq:invert-ineq1c}), $b^\T \tilde \Sigma_h b \le C_{g3}^{-1} C_{h3}^{-1} (1-\varrho_6)^{-1} D^\dag_{11,\text{\tiny WL}} $.
Hence $b^\T \tilde \Sigma_h b \le C_{g3}^{-1} C_{h3}^{-1} (1-\varrho_5 \vee \varrho_6)^{-1} D^\dag_{11,\text{\tiny WL}} $.

\subsection{Proof of Theorem \ref{thm:expansion-DY}} \label{sec:prf-thm-expansion-DY}

We split the proof into two lemmas. The first one deals with the convergence of the AIPW estimator
$\tilde E \{ \varphi_{\scriptscriptstyle D_1Y_{11}}(O;\hat{\pi}_1,\hat{m}_1, \hat{m}_{11} ) \}$ in (\ref{eq:expansion-bound-DY}).
Recall from (\ref{eq:phi-D1Y11}) that
$$
\varphi_{\scriptscriptstyle D_1Y_{11}}(O;\hat{\pi}_1,\hat{m}_1, \hat{m}_{11} )  =  \frac{Z}{\hat{\pi}_1(X)}  DY-\left\{\frac{Z}{\hat{\pi}_1 (X)}-1\right\} \hat{m}_1 (X)\hat{m}_{11}(X).
$$

\begin{lem} \label{lem:expansion-DY}
In the setting of Theorem~\ref{thm:alpha11}, take $r_1 = M_1 (|S_{\bar\gamma_1}| \lambda_0 + |S_{\bar\alpha_1}| \lambda_1) $
and $r_2 = M_2 (|S_{\bar\gamma_1}| \lambda_0 + |S_{\bar\alpha_1}| \lambda_1 + |S_{\bar\alpha_{11}}| \lambda_2) $.
Then we have in the event $(\Omega_0 \cap\Omega_1 \cap \Omega_{h1} \cap\Omega_{f1} \cap \Omega_{f2}) \cap \Omega_{f3} \cap \Omega_{2, r_1} \cap \Omega_{3, r_2}$,
\begin{align*}
& \left| \tilde E \left\{ \varphi_{\scriptscriptstyle D_1Y_{11}}(O;\hat{\pi}_1,\hat{m}_1, \hat{m}_{11} )\right\}-
\tilde E \left\{ \varphi_{\scriptscriptstyle D_1Y_{11}}(O;\bar{\pi}_1,\bar{m}_1, \bar{m}_{11} ) \right\} \right| \\
& \le M_{30} |S_{\bar\gamma_1}| \lambda_0(\lambda_1\vee\lambda_2) + M_{31} |S_{\bar\alpha_1}| \lambda_1 (\lambda_1\vee\lambda_2) + M_{32} |S_{\bar\alpha_{11}}| \lambda_2^2 ,
\end{align*}
where $M_{30} =  (B_1+1) M_0 + M_{02} + ( M_{11} + M_{21} )/2 + M_{12}+M_{22}$,
$M_{31} =  (M_{11}+M_{21})/2 + M_{12} + M_{22}$,
$M_{32} = M_{21}/2 + M_{22}$, depending on $(M_{02}, M_{11}, M_{21}, M_{12}, M_{22})$ in the proof.
\end{lem}

\begin{prf}
Consider the following decomposition
\begin{align}
& \varphi_{\scriptscriptstyle D_1Y_{11}}(O;\hat{\pi}_1,\hat{m}_1, \hat{m}_{11} )
=  \varphi_{\scriptscriptstyle D_1Y_{11}}(O;\bar{\pi}_1,\bar{m}_1, \bar{m}_{11} ) + \delta_1 + \delta_2 + \delta_3,  \label{eq:phi-decomp}
\end{align}
where
\begin{align*}
& \delta_1 = \{ \hat{m}_1 (X)\hat{m}_{11}(X) - \bar m_1(X) \bar m_{11}(X) \} \left\{1- \frac{Z}{\bar{\pi}_1 (X)} \right\} , \\
& \delta_2 = \{ D Y - \bar m_1(X) \bar m_{11}(X) \} \left\{\frac{Z}{\hat\pi_1 (X)} - \frac{Z}{\bar{\pi}_1 (X)} \right\} ,\\
& \delta_3 = \{ \hat{m}_1 (X)\hat{m}_{11}(X) - \bar m_1(X) \bar m_{11}(X) \} \left\{\frac{Z}{\bar\pi_1 (X)} - \frac{Z}{\hat{\pi}_1 (X)} \right\} .
\end{align*}
Denote $\Delta_1 = \tilde E(\delta_1)$, $\Delta_2 = \tilde E (\delta_2)$, and $\Delta_3 = \tilde E (\delta_3)$. Then
\begin{align*}
& \tilde E \left\{ \varphi_{\scriptscriptstyle D_1Y_{11}}(O;\hat{\pi}_1,\hat{m}_1, \hat{m}_{11} )\right\}
= \tilde E \left\{ \varphi_{\scriptscriptstyle D_1Y_{11}}(O;\bar{\pi}_1,\bar{m}_1, \bar{m}_{11} ) \right\} + \Delta_1 + \Delta_2 + \Delta_3,
\end{align*}
We handle the three terms $\Delta_1, \Delta_2, \Delta_3$ separately.

First, a Taylor expansion for $\Delta_2$ yields for some $u \in (0,1)$,
\begin{align*}
& \Delta_2 = - (\hat\gamma_1-\bar\gamma_1)^\T  \tilde E \left[ Z\{ D Y - \bar m_1(X) \bar m_{11}(X) \} \me^{-\bar\gamma_1^\T f} f  \right] \\
& \quad + (\hat\gamma_1-\bar\gamma_1)^\T  \tilde E \left[ Z\{ D Y - \bar m_1(X) \bar m_{11}(X) \} \me^{-\bar\gamma_1^\T f - u(\hat\gamma_1 -\bar\gamma_1)^\T f} f f^\T \right] (\hat\gamma_1-\bar\gamma_1) /2,
\end{align*}
denoted as $\Delta_{21} + \Delta_{22}$.
Because $f$ is a subvector of $h$, we have in the event $\Omega_0 \cap \Omega_1 $,
\begin{align*}
& | \Delta_{21} | \le \left\| \tilde E \left[ Z\{ D Y - \bar m_1(X) \bar m_{11}(X) \} \me^{-\bar\gamma_1^\T f} f  \right] \right\|_\infty \|\hat\gamma_1-\bar\gamma_1 \|_1 \\
& \le B_1 M_0 |S_{\bar\gamma_1}| \lambda_0 \lambda_2 .
\end{align*}
Moreover, we have in the event $\Omega_0 \cap \Omega_{f1} \cap \Omega_{f3} $,
\begin{align*}
& | \Delta_{22} | \le \me^{C_{f0} \|\hat\gamma_1-\bar\gamma_1\|_1} \tilde E \left[ Z | D Y - \bar m_1(X) \bar m_{11}(X) | \me^{-\bar\gamma_1^\T f } (\hat\gamma_1^\T f -\bar\gamma_1^\T f)^2 \right] /2
\le  M_{02} |S_{\bar\gamma_1}| \lambda_0^2 ,
\end{align*}
where $M_{02} = \me^{ C_{f0} M_0 \varrho_0 } [ (\sigma_0^2 + \sigma_1^2)^{1/2} M_0 +  \{B_{f3} + (\sigma_0^2 + \sigma_1^2)^{1/2} B_{f1}\} M_0^2 \varrho_0] /2$. The second step follows because
by (\ref{eq:bound-gamma1}) and (\ref{eq:ineq-mat-f3}) with $b=\hat\gamma_1 - \bar\gamma_1$,
\begin{align*}
& \tilde E \left[ Z  w_1(X;\bar\gamma_1) | DY - m_1(X; \bar{\alpha}_1) \psi_{\scriptscriptstyle Y} (\bar\alpha^\T _{11} h) |
 (\hat\gamma_1^\T f - \bar\gamma_1^\T f)^2 \right]  \\
& \le (\sigma_0^2 + \sigma_1^2)^{1/2} M_0 |S_{\bar\gamma_1}| \lambda_0^2 +  \{B_{f3} + (\sigma_0^2 + \sigma_1^2)^{1/2} B_{f1}\} M_0^2 \varrho_0 |S_{\bar\gamma_1}| \lambda_0^2 .
\end{align*}
Combining the preceding inequalities yields
\begin{align}
|\Delta_2 | \le | \Delta_{21} | + | \Delta_{22} | \le B_1 M_0 |S_{\bar\gamma_1}| \lambda_0 \lambda_2 +   M_{02} |S_{\bar\gamma_1}| \lambda_0^2 . \label{eq:Delta2-prf}
\end{align}

Second, the term $\Delta_3$ can be decomposed as
\begin{align*}
 \Delta_3 =  \tilde E \left[\bar m_{11} (X) \{ \hat m_1(X) - \bar m_1(X)\} \left\{\frac{Z}{\bar\pi_1 (X)} - \frac{Z}{\hat{\pi}_1 (X)} \right\} \right] \\
 \quad + \tilde E \left[\hat m_1(X) \{ \hat m_{11}(X) - \bar m_{11}(X)\}\left\{\frac{Z}{\bar\pi_1 (X)} - \frac{Z}{\hat{\pi}_1 (X)} \right\} \right],
\end{align*}
denoted as $\Delta_{31} + \Delta_{32}$.
Similarly as (\ref{eq:remove-hat-alpha1-b}) for $|\hat m_1(X) - \bar m_1(X)|$, we have
\begin{align}
| \hat m_{11}(X) - \bar m_{11}(X) | \le | (\hat\alpha_{11} - \bar\alpha_{11})^\T h| \psi_{\scriptscriptstyle Y}^\prime(0) \me^{C_{h2} (C_{h1} + C_{h0} \| \hat\alpha_{11}-\bar\alpha_{11}\|_1)} . \label{eq:remove-hat-alpha11}
\end{align}
Combining this inequality with (\ref{eq:remove-hat-gamma}) and (\ref{eq:remove-hat-alpha1-b}) and applying the Cauchy--Schwartz inequality shows that
in the event $\Omega_0 \cap\Omega_1 \cap \Omega_{h1} \cap\Omega_{f1} \cap \Omega_{f2}$,
\begin{align}
& | \Delta_{31} | \le \me^{C_{f0} \|\hat\gamma_1-\bar\gamma_1\|_1} \tilde E^{1/2} \left\{ Z w_1(X;\bar\gamma_1)  (\hat\gamma_1^\T f -\bar\gamma_1^\T f)^2 \right\} \nonumber \\
& \quad \times \tilde C_{h1} \tilde E^{1/2} \left\{  Z w_1(X;\bar\gamma_1) (\hat m_1 - \bar m_1  )^2 \right\} \nonumber \\
& \le (M_0 |S_{\bar\gamma_1}| \lambda_0^2 )^{1/2} M_{11}^{1/2} ( |S_{\bar\gamma_1}| \lambda_0^2 + |S_{\bar\alpha_1}| \lambda_1^2 )^{1/2} ,\label{eq:Delta3-prf-a}
\end{align}
and, with $\hat m_1(X) \in (0,1)$,
\begin{align}
& | \Delta_{32} | \le \me^{C_{f0} \|\hat\gamma_1-\bar\gamma_1\|_1} \tilde E^{1/2} \left\{ Z w_1(X;\bar\gamma_1)  (\hat\gamma_1^\T f -\bar\gamma_1^\T f)^2 \right\}
 \tilde E^{1/2} \left\{ Z w_1(X;\bar\gamma_1) (\hat m_{11} - \bar m_{11} )^2\right\}  \nonumber \\
& \le (M_0 |S_{\bar\gamma_1}| \lambda_0^2 )^{1/2}  M_{21}^{1/2}  ( |S_{\bar\gamma_1}| \lambda_0^2 + |S_{\bar\alpha_1}| \lambda_1^2 + |S_{\bar\alpha_{11}}| \lambda_2^2 )^{1/2} , \label{eq:Delta3-prf-b}
\end{align}
where $\tilde C_{h1} = \max\{|\psi_{\scriptscriptstyle Y}(-C_{h1})|,  |\psi_{\scriptscriptstyle Y}(C_{h1})|\}$ and
$M_{11} = \psi_{\scriptscriptstyle D}^{\prime 2} (0 ) \tilde C^2_{h1} \me^{ 2C_{f0} M_0 \varrho_0 +2 C_{g2} (C_{g1}+C_{g0} M_1 \varrho_1 ) } M_1$
as in Lemma~\ref{lem:remove-hat},
and $M_{21} = \psi_{\scriptscriptstyle Y}^{\prime 2} (0 ) \me^{ 2C_{f0} M_0 \varrho_0 +2 C_{h2} (C_{h1}+C_{h0} M_2 (\varrho_1+\varrho_2) ) } M_2$.
From these two inequalities, we obtain
\begin{align}
& | \Delta_3 | \le   (2 M_0 + M_{11} + M_{21} ) |S_{\bar\gamma_1}| \lambda_0^2 /2 + (M_{11}+M_{21})  |S_{\bar\alpha_1}| \lambda_1^2/2 + M_{21} |S_{\bar\alpha_{11}}| \lambda_2^2 /2. \label{eq:Delta3-prf}
\end{align}

Finally, the term $\Delta_1$ can be decomposed as
\begin{align*}
& \Delta_1 =  \tilde E \left[\bar m_{11} (X) \{ \hat m_1(X) - \bar m_1(X)\}\left\{1- \frac{Z}{\bar{\pi}_1 (X)} \right\} \right] \\
& \quad +  \tilde E \left[\hat m_1(X) \{ \hat m_{11}(X) - \bar m_{11}(X)\}\left\{1- \frac{Z}{\bar{\pi}_1 (X)} \right\} \right]  ,
\end{align*}
denoted as $\Delta_{11} + \Delta_{12}$. Recall that $\bar\pi_1(X) \equiv \pi^*(X)$ because model (\ref{eq:pi}) is correctly specified.
Take $r_1 = M_1 (|S_{\bar\gamma_1}| \lambda_0 + |S_{\bar\alpha_1}| \lambda_1) $ in Lemma~\ref{lem:prob-grad-pi-alpha1}.
Then in the event $ \Omega_0  \cap \Omega_{2, r_1}$, we have $\| \hat\alpha_1 - \bar\alpha_1 \|_1 \le r_1$ and hence
\begin{align*}
& |\Delta_{11}| \le 4 (\me^{-C_{f1}}+1) C_{g0} \{ |\psi_{\scriptscriptstyle Y}(-C_{h1})|, |\psi_{\scriptscriptstyle Y}(C_{h1})|\} \psi_{\scriptscriptstyle D}^\prime (0) \me^{C_{g2} (C_{g1} + C_{g0} r_1)} r_1 \lambda_1  \\
& \le M_{12} (|S_{\bar\gamma_1}| \lambda_0 + |S_{\bar\alpha_1}| \lambda_1) \lambda_1,
\end{align*}
where $M_{12} = 4 (\me^{-C_{f1}}+1) C_{g0} \tilde C_{h1} \psi_{\scriptscriptstyle D}^\prime (0) \me^{C_{g2} (C_{g1} + C_{g0} M_1 \varrho_1)} M_1$.
Take $r_2 = M_2 (|S_{\bar\gamma_1}| \lambda_0 + |S_{\bar\alpha_1}| \lambda_1 + |S_{\bar\alpha_{11}}| \lambda_2) $ in Lemma~\ref{lem:prob-grad-pi-alpha11}.
Then in the event $ (\Omega_0 \cap\Omega_1 \cap \Omega_{h1} \cap\Omega_{f1} \cap \Omega_{f2})  \cap \Omega_{3, r_2}$, we have
$\| \hat\alpha_{11} - \bar\alpha_{11} \|_1 \le r_2$ and hence
\begin{align*}
& |\Delta_{12}| \le 4 (\me^{-C_{f1}}+1) C_{h0} \psi_{\scriptscriptstyle Y}^\prime (0) \me^{C_{h2} (C_{h1} + C_{h0} r_2)} r_2 \lambda_2 \\
& \le M_{22} (|S_{\bar\gamma_1}| \lambda_0 + |S_{\bar\alpha_1}| \lambda_1 + |S_{\bar\alpha_{11}}| \lambda_2) \lambda_2,
\end{align*}
where $M_{22} =  4 (\me^{-C_{f1}}+1) C_{h0} \psi_{\scriptscriptstyle Y}^\prime (0) \me^{C_{h2} (C_{h1} + C_{h0} M_2(\varrho_1 + \varrho_2))} M_2$. Hence
\begin{align}
| \Delta_1 | \le M_{12} (|S_{\bar\gamma_1}| \lambda_0 + |S_{\bar\alpha_1}| \lambda_1) \lambda_1 +
M_{22} (|S_{\bar\gamma_1}| \lambda_0 + |S_{\bar\alpha_1}| \lambda_1 + |S_{\bar\alpha_{11}}| \lambda_2) \lambda_2. \label{eq:Delta1-prf}
\end{align}
Combining (\ref{eq:Delta2-prf}), (\ref{eq:Delta3-prf}), (\ref{eq:Delta1-prf}) yields the desired result.
\end{prf}

The second lemma deals with the convergence of the mean squared difference between
$\varphi_{\scriptscriptstyle D_1Y_{11}}(O;$ $\hat{\pi}_1,\hat{m}_1, \hat{m}_{11} ) $ and
$ \varphi_{\scriptscriptstyle D_1Y_{11}}(O;\bar{\pi}_1,\bar{m}_1, \bar{m}_{11} )$ in (\ref{eq:var-est-DY}) for variance estimation.

\begin{lem} \label{lem:var-est-DY}
In the setting of Theorem~\ref{thm:alpha11},
we have in the event $\Omega_0 \cap\Omega_1 \cap \Omega_{h1} \cap\Omega_{f1} \cap \Omega_{f2}$,
\begin{align*}
& \left| \tilde E \left[ \left\{ \varphi_{\scriptscriptstyle D_1Y_{11}}(O;\hat{\pi}_1,\hat{m}_1, \hat{m}_{11} ) -
 \varphi_{\scriptscriptstyle D_1Y_{11}}(O;\bar{\pi}_1,\bar{m}_1, \bar{m}_{11} ) \right\}^2 \right] \right| \\
& \le M_{40} (|S_{\bar\gamma_1}| \lambda_0^2 + |S_{\bar\alpha_1}| \lambda_1^2 + |S_{\bar\alpha_{11}}| \lambda_2^2) +
M_{41} (|S_{\bar\gamma_1}| \lambda_0 + |S_{\bar\alpha_1}| \lambda_1 + |S_{\bar\alpha_{11}}| \lambda_2)^2 ,
\end{align*}
where $M_{40} = 3  \me^{-C_{f1} } M_{01} + 6  \me^{-C_{f1} } ( 1 + \me^{ - C_{f0} M_0 \varrho_0}  )^2 (M_{11}+M_{21})$
and $M_{41} = 6 ( 1 +  \me^{-C_{f1} }  )^2 (M_{13}+M_{23})$, depending on $(M_{01}, M_{11}, M_{21}, M_{13}, M_{23})$ in the proof.
\end{lem}

\begin{prf}
Return to the decomposition (\ref{eq:phi-decomp}). Then
\begin{align*}
& \tilde E \left[ \left\{ \varphi_{\scriptscriptstyle D_1Y_{11}}(O;\hat{\pi}_1,\hat{m}_1, \hat{m}_{11} ) -
\varphi_{\scriptscriptstyle D_1Y_{11}}(O;\bar{\pi}_1,\bar{m}_1, \bar{m}_{11} ) \right\}^2 \right] = \tilde E \left\{ (\delta_1 + \delta_2 + \delta_3)^2 \right\} \\
& \le 3 \tilde E(\delta_1^2) + 3 \tilde E(\delta_2^2) + 3 \tilde E(\delta_3^2).
\end{align*}
We handle the three terms  $\tilde E(\delta_1^2), \tilde E(\delta_2^2) , \tilde E(\delta_3^2)$ separately.

First, we have by (\ref{eq:remove-hat-gamma}) and Assumption~\ref{ass:gamma1}(i)--(ii),
\begin{align}
& \tilde E(\delta_2^2) = \tilde E \left[ Z \{ D Y - \bar m_1(X) \bar m_{11}(X) \}^2 \left\{\frac{1}{\hat\pi_1 (X)} - \frac{1}{\bar{\pi}_1 (X)} \right\}^2 \right] \nonumber \\
& \le \me^{-C_{f1} + 2 C_{f0} \|\hat\gamma_1 - \bar\gamma_1\|_1} \tilde E \left[ Z w_1(X;\bar\gamma_1) \{  D Y - \bar m_1(X) \bar m_{11}(X) \}^2 (\hat\gamma_1^\T f-\bar\gamma_1^\T f)^2 \right] \nonumber \\
& \le \me^{-C_{f1} } M_{01} |S_{\bar\gamma_1}| \lambda_0^2 , \label{eq:delta2-sq-prf}
\end{align}
where $M_{01} = \me^{ 2 C_{f0} M_0 \varrho_0 } [ (\sigma_0^2 + \sigma_1^2) M_0 +  \{B_{f2} + (\sigma_0^2 + \sigma_1^2) B_{f1}\} M_0^2 \varrho_0]$ as in Lemma~\ref{lem:remove-hat}.

Second, writing $ \hat\pi_1^{-1} - \bar\pi_1^{-1} = \me^{-\bar\gamma_1^\T f} (\me^{- \hat\gamma_1^\T f + \bar\gamma_1^\T f} -1 )$
and using Assumption~\ref{ass:gamma1}(i)--(ii), we have
\begin{align*}
& \tilde E (\delta_3^2) = \tilde E \left[ \{ \hat{m}_1 (X)\hat{m}_{11}(X) - \bar m_1(X) \bar m_{11}(X) \}^2 \left\{\frac{Z}{\bar\pi_1 (X)} - \frac{Z}{\hat{\pi}_1 (X)} \right\}^2 \right] \\
& \le \me^{-C_{f1} } \left( 1 + \me^{ C_{f0} \|\hat\gamma_1 - \bar\gamma_1\|_1 } \right)^2 \tilde E \left[  Z w_1(X;\bar\gamma_1) \{ \hat{m}_1 (X)\hat{m}_{11}(X) - \bar m_1(X) \bar m_{11}(X) \}^2  \right] \\
& \le \me^{-C_{f1} } \left( 1 + \me^{ C_{f0} \|\hat\gamma_1 - \bar\gamma_1\|_1 } \right)^2 \\
& \quad \times 2 \left[ \tilde E \left\{  Z w_1(X;\bar\gamma_1) \bar m_{11}^2 (\hat{m}_1 -\bar m_1)^2 \right\} + \tilde E \left\{  Z w_1(X;\bar\gamma_1) \hat m_1^2 (\hat{m}_{11} -\bar m_{11})^2 \right\} \right].
\end{align*}
Then we obtain similarly as in  (\ref{eq:Delta3-prf-a})--(\ref{eq:Delta3-prf-b}),
\begin{align}
& \tilde E (\delta_3^2) \le 2 \me^{-C_{f1} } \left( 1 + \me^{ - C_{f0} M_0 \varrho_0} \right)^2  \nonumber \\
& \quad \times \left\{ M_{11} ( |S_{\bar\gamma_1}| \lambda_0^2 + |S_{\bar\alpha_1}| \lambda_1^2 ) +
 M_{21} ( |S_{\bar\gamma_1}| \lambda_0^2 + |S_{\bar\alpha_1}| \lambda_1^2 + |S_{\bar\alpha_{11}}| \lambda_2^2 ) \right\}, \label{eq:delta3-sq-prf}
\end{align}
where $M_{11}$ and $M_{21}$ are as in Lemma~\ref{lem:expansion-DY}.

Finally, using Assumption~\ref{ass:gamma1}(i)--(ii), we also have
\begin{align*}
& \tilde E (\delta_1^2) = \tilde E \left[ \{ \hat{m}_1 (X)\hat{m}_{11}(X) - \bar m_1(X) \bar m_{11}(X) \}^2 \left\{1- \frac{Z}{\bar{\pi}_1 (X)} \right\}^2 \right] \\
& \le \left( 1 +  \me^{-C_{f1} } \right)^2 \tilde E \left[  \{ \hat{m}_1 (X)\hat{m}_{11}(X) - \bar m_1(X) \bar m_{11}(X) \}^2 \right] \\
& \le 2 \left( 1 +  \me^{-C_{f1} } \right)^2 \left[ \tilde E \left\{\bar m_{11}^2 (\hat{m}_1 -\bar m_1)^2 \right\} + \tilde E \left\{\hat m_1^2 (\hat{m}_{11} -\bar m_{11})^2 \right\} \right].
\end{align*}
Then we obtain similarly as in  (\ref{eq:Delta3-prf-a})--(\ref{eq:Delta3-prf-b}) but using the bounds
$ | \hat\alpha_1^\T g - \bar\alpha_1^\T g| \le C_{g0} \|\hat\alpha_1 - \bar\alpha_1\|_1 $
and $| \hat\alpha_{11}^\T h - \bar\alpha_{11}^\T h | \le C_{h0} \|\hat\alpha_{11} - \bar\alpha_{11} \|_1 $,
\begin{align}
& \tilde E (\delta_1^2)  \le  2 \left( 1 +  \me^{-C_{f1} } \right)^2  \nonumber \\
& \quad \times \left\{ M_{13} ( |S_{\bar\gamma_1}| \lambda_0 + |S_{\bar\alpha_1}| \lambda_1 )^2 +
 M_{23} ( |S_{\bar\gamma_1}| \lambda_0 + |S_{\bar\alpha_1}| \lambda_1 + |S_{\bar\alpha_{11}}| \lambda_2 )^2 \right\}, \label{eq:delta1-sq-prf}
\end{align}
where $ M_{13} = C_{g0}^2 \psi_{\scriptscriptstyle D}^{\prime 2} (0 ) \tilde C_{h1}^2 \me^{ 2 C_{g2} (C_{g1}+C_{g0} M_1 \varrho_1 ) } M_1$
and $ M_{23} = C_{h0}^2 \psi_{\scriptscriptstyle Y}^{\prime 2} (0 ) \me^{ 2 C_{h2} (C_{h1}+C_{h0} M_2 (\varrho_1+\varrho_2) ) } M_2$.
Combining (\ref{eq:delta2-sq-prf})--(\ref{eq:delta1-sq-prf}) yields  the desired result.
\end{prf}

\vspace{.3in}
\centerline{\bf\Large References}

\begin{description}\addtolength{\itemsep}{-.1in}

\vspace{-.05in} \item Tan, Z. (2020a) Regularized calibrated estimation of propensity scores with model misspecification and high-dimensional data, {\em Biometrika}, 107, 137--158.

\vspace{-.05in} \item Tan, Z. (2020b) Model-assisted inference for treatment effects using regularized calibrated estimation with high-dimensional data, {\em Annals of Statistics}, 48, 811--837.

\end{description}


\end{document}